\documentclass[traditabstract]{aa} 
\usepackage{graphicx,natbib}
\usepackage{longtable,lscape,supertabular}
\usepackage{txfonts}
\begin{document}
   \title{Analysis of Far-UV Data of Central Stars of Planetary Nebulae: 
          Occurrence and Variability of Stellar Winds}

   \author{M.A.\ Guerrero
          \inst{1,2},
          \and
          O.\ De Marco
          \inst{2} 
          }

   \institute{
Instituto de Astrof\'\i sica de Andaluc\'\i a (IAA-CSIC), 
Glorieta de la Astronom\'\i a, E-18008, Granada, Spain
   \and
Department of Physics, Macquarie University, Sydney, NSW 2109, Australia \\
   \email{mar@iaa.es, orsola@science.mq.edu.au}
             }

\date{Received ; accepted }

 
\abstract{
The occurrence of stellar wind in the central star of a planetary nebula 
(CSPN) can be revealed by the presence of P~Cygni profiles of high-excitation 
lines overimposed on its stellar continuum.  
We have examined the entire Far-Ultraviolet Spectroscopic Explorer 
\emph{FUSE} archive and merged all useful spectroscopic observations 
of CSPNe to produce the highest quality spectra that can be used to 
assess the occurrence of stellar winds.  
Furthermore, the individual spectra of each CSPN have been compared to 
search for variability in the P~Cygni profile.  
P~Cygni profiles of high-excitation lines have been found in 44 CSPNe, 
with a clear correlation between the ionization potential of the lines 
and the effective temperature of the star.  
We introduce a prescription to derive the terminal wind velocity ($v_\infty$) 
from saturated and unsaturated P~Cygni profiles and provide new values of 
$v_\infty$ for these stars.   
Another 23 CSPNe do not show P~Cygni profiles or their data in the 
\emph{FUSE} archive are not conclusive to determine the occurrence 
of P~Cygni profiles.  
Variability in the P~Cygni profile of high-excitation far-UV 
lines is found for the first time in six CSPNe, namely Hen\,2-131, 
NGC\,40, NGC\,1535, NGC\,2392, Sp\,3, and SwSt\,1.  
This increases up to 13 the number of CSPNe with variable P~Cygni 
profiles in the UV, including those previously reported using 
\emph{IUE} or \emph{FUSE} observations.  
Variability is seen 
preferentially in the unsaturated P~{\sc v} and Si~{\sc iv} 
lines, but also in saturated C~{\sc iii} and O~{\sc vi} lines.  
The CSPNe with variable P~Cygni profiles have similar stellar 
properties (relatively low $\log(g)$ and $T_{\rm eff}$) that suggests 
they are less evolved CSPNe.  
Some of the CSPNe with variable P~Cygni profiles show O~{\sc vi} lines, 
while their effective temperature are insufficient to produce this ion.  
We suggest that this ion is produced by Auger ionization from X-rays 
associated to shocks in their stellar winds as is the case in massive 
OB stars of high ionization potential ions which cannot be abundantly 
produced by photoionizations.  
}

\keywords{Line: profiles -- 
          Stars: winds, outflows -- 
          ISM: planetary nebulae: general -- 
          Ultraviolet: stars
               }
   \authorrunning{Guerrero \& De Marco}
   \titlerunning{FUSE Survey of CSPNe}
   \maketitle
%

\section{Introduction}

Planetary nebulae (PNe) are a short stage in the stellar evolution of low- 
and intermediate-mass stars ($0.8-1.0\,M_\odot < M_{\rm i} < 8-10\,M_\odot$), 
when they have expelled most of their stellar envelopes at the end of 
the Asymptotic Giant Branch (AGB) phase and start their evolution 
towards white dwarfs (WD).  
The formation and evolution of a PN is believed to be dominated by 
the fast stellar wind that emanates from the central star of the PN 
(CSPN) as it sweeps-up the dense and slow AGB wind \citep{KPF78} 
and evacuates a central cavity \citep[see][for a review]{BF02}.
The interaction between this fast stellar wind and the slow AGB wind 
is so violent that the material in the stellar wind is shock-heated 
at high temperatures \citep{SSW08} resulting in a hot bubble filled 
with X-ray emission \citep[e.g.,][]{Kastner_etal00,Chu_etal01}.

Most models of PN formation consider the stellar winds of CSPNe to 
be isotropic, homogeneous, and only variable on long time-scales 
\citep[e.g.,][]{Villaver_etal02,Perinotto_etal04}.  
Since the radiatively-driven stellar winds of CSPNe and massive OB stars 
share many similarities \citep{Prinja90}, it can be expected that the 
stellar winds of CSPNe also present the clumping and time variability 
unveiled in the wind of massive OB stars by changes in the profile of 
optical lines and by discrete absorption features (DACs) in the P~Cygni 
profile of UV lines.  
The variability and clumping of the fast stellar wind can lead to 
insight on the interaction of the stellar wind with the nebular 
material, possibly requiring the reassessment of the wind-wind 
interaction in the models of PN formation and the revision of the 
mass-loss rate associated with these stellar winds.  
The hard X-ray emission observed in a few CSPNe, with NGC\,6543 
being the most prominent case \citep{Guerrero_etal01}, may be 
the smoking gun of shocks within the stellar winds of CSPNe as 
is seen in the strong stellar winds of OB stars \citep{LW80,GO95}.  
An investigation of the occurrence of wind variability and energies 
implied in the shocks in the limited sample of CSPNe with hard X-ray 
emission \citep[][Guerrero et al., in prep.]{Montez_etal10,Kastner_etal12} 
will help the disambiguation between late-type companions and shock-in 
winds as the origin of the hard X-ray emission.

More importantly, the physical processes behind the wind variability, 
such as magnetic fields, stellar rotation, and binary companions, are 
assumed to play important roles in the shaping of PNe.  
Indeed, episodic, collimated mass-ejections, as are seen, e.g., in the 
central star of N66 in the LMC \citep{Pena_etal08} or in the proto-PN 
M\,2-56 \citep{S-C_etal10}, are assumed to shape the most axysimmetric 
PNe \citep{ST98}.  
The importance of the stellar winds of CSPNe in their formation and 
evolution justifies detailed studies of their properties. 
For instance, wind variability may cause radial velocity (RV) RV 
variations that interfere the search for binary companions of CSPNe 
\citep{DeMarco_etal04}.  
The detection of variability on the P Cygni profiles of a CSPN would 
definitely make questionable the interpretation of RV variations as 
due to the reflex motion in a binary system.  
On the other hand, CSPNe that do not show variability on their P Cygni 
profiles will be cleared for future RV studies.  

Earlier studies of the variability of P~Cygni profiles of UV lines using 
\emph{International Ultraviolet Explorer} (\emph{IUE}) time-series suggested 
that the stellar winds of a few CSPNe (BD+30$^\circ$3639, IC\,4593, NGC\,40, 
NGC\,1535, NGC\,2392, NGC\,6543, and NGC\,6826) can be variable 
\citep{PP95,PP97}.  
The limited spectral resolution and sensitivity of \emph{IUE} and the 
lack of complete time-series data prevented the search for patterns 
in the variability of the winds of these CSPNe.  
More recently, detailed studies of the time-variability of 
the stellar winds of the CSPNe of IC\,418, IC\,2149, IC\,4593, 
NGC\,6543, NGC\,6826 and Hen\,2-138 have been obtained by 
\citet{Prinja_etal07,Prinja_etal10,Prinja_etal12} using 
\emph{Far Ultraviolet Spectroscopic Explorer} (\emph{FUSE}) 
observations.  
These studies indicate that the stellar wind of these CSPNe are 
variable and, in the case of NGC\,6543, this variability can be 
linked to the rotation of the star and has been used to measure 
the rotational velocity \citep{PMC12}.  
Photometric and spectroscopic changes in the CSPN of NGC\,6826 have 
also been related to the stellar rotation \citep{Handler_etal13}.

Inspired by these works and aiming at extending the results 
obtained in the 90's by \citet{PP95,PP97}, 
we present here a systematic study of all available \emph{FUSE} 
observations of CSPNe in the search for variability in the P Cygni 
profiles of far-UV lines of Galactic CPSNe.   
Since the available \emph{IUE} and \emph{FUSE} datasets are sparse in 
time, they are not suited to investigate in detail the time variability 
of these sources, but the detection of variability itself will aid in the 
selection of interesting targets for future detailed studies with with 
current 
and upcoming 
UV telescopes.  
In \S2, we describe the available \emph{FUSE} data and the sample 
of CSPNe that is been studied.  
In \S3, we present the data analysis and results, including an atlas 
of \emph{FUSE} P~Cygni profiles of CSPNe, and the evidence for the 
detection of time variability in the \emph{FUSE} P~Cygni profiles of 
CSPNe.  
In \S4, we discuss individually the CSPNe that show variable \emph{FUSE} 
P~Cygni profiles and describe into further details the observed variations.  
Finally, in \S5 we present a discussion of the results concerning the 
occurrence of wind variability, the time-scales and optical depths of 
the variations observed in the \emph{FUSE} P~Cygni profiles of CSPNe, 
and its correlation with hard X-ray emission.

\section{UV observations}

\subsection{The FUSE Observatory and FUSE Archive}

Launched in June 1999, \emph{FUSE} provides spectral coverage of the 
region 920--1190 \AA\ with a spectral resolving power $\sim$20,000 
\citep{Moos_etal00,Sahnow_etal00}.  
The \emph{FUSE} instruments consist of four telescopes and Rowland circle 
spectrographs and two photon-counting detectors: two telescopes and  
diffraction gratings are made from silicon carbide (SiC), providing 
coverage of the short-wavelength range, while the two other telescopes 
and gratings are lithium fluoride (LiF) coated and cover the spectral 
region long-ward of $\sim$1020 \AA.  
The combination of telescopes, gratings and detectors imply 
that a total of eight independent spectra are generated for 
each data set: two pairs of channels (SiC1 and SiC2, and LiF1 
and LiF2) which are actually recorded each of them by two 
detector segments (A and B).  
The channel and detector segment combination correspond to 
LiF1A  (987--1082 \AA), 
LiF1B (1094--1188 \AA), 
LiF2A (1086--1187 \AA), 
LiF2B  (979--1075 \AA), 
SiC1A (1004--1091 \AA), 
SiC1B  (905--993   \AA), 
SiC2A  (917--1006 \AA), and 
SiC2B (1016--1103 \AA).  
Therefore, flux at almost all wavelengths is recorded by at least two 
segments, with the short wavelength range (910--990 \AA) covered by the 
SiC1B and SiC2A spectra, the intermediate spectral range (1015--1075 \AA) 
covered by the LiF1A, LiF2B, SiC1A, and SiC2B spectra, and the 
long wavelength range (1095--1185 \AA) covered by the LiF1B and LiF2A 
spectra.

\begin{table*}
\centering
\scriptsize{
\caption{\footnotesize{
Sample of CSPNe with \emph{FUSE} observations}}

\label{FUSE_CSPN_sample}
\begin{tabular}{llrrrl|llrrrl}
\hline\hline

\multicolumn{1}{l}{Name}             & 
\multicolumn{1}{l}{Spectral Type}    & 
\multicolumn{1}{c}{$T_{\rm eff}$}     & 
\multicolumn{1}{c}{$\log (g)$}       & 
\multicolumn{1}{c}{$\dot M$}         & 
\multicolumn{1}{l}{Ref.}             & 
\multicolumn{1}{l}{Name}             & 
\multicolumn{1}{l}{Spectral Type}    & 
\multicolumn{1}{c}{$T_{\rm eff}$}     & 
\multicolumn{1}{c}{$\log (g)$}       & 
\multicolumn{1}{c}{$\dot M$}         & 
\multicolumn{1}{l}{Ref.}             \\
\multicolumn{1}{c}{}                 & 
\multicolumn{1}{c}{}                 & 
\multicolumn{1}{c}{[kK]}             & 
\multicolumn{1}{c}{[cgs]}            & 
\multicolumn{1}{c}{[$M_\odot$~yr$^{-1}$]} & 
\multicolumn{1}{l}{}                 & 
\multicolumn{1}{c}{}                 & 
\multicolumn{1}{c}{}                 & 
\multicolumn{1}{c}{[kK]}             & 
\multicolumn{1}{c}{[cgs]}            & 
\multicolumn{1}{c}{[$M_\odot$~yr$^{-1}$]} & 
\multicolumn{1}{l}{}                 \\
\hline
\multicolumn{10}{c}{~~~~~~~~~~~~~~~~~~~~~~~~~~~~~~CSPNe with P Cygni profiles} \\
\hline 

A\,30             & [WC]-PG1159   &     115~ & $\dots$~~~ & 5.0$\times$10$^{-8}$~ & 1 & NGC\,246          & PG1159        &     150~ &     5.7~~~ & 1.3$\times$10$^{-7}$~ & 2,12 \\  
A\,43             & Hybrid-PG1159 &     110~ &     5.7~~~ &        $\dots$~~~~~~~ & 2 & NGC\,1535         & O(H)          &      75~ &     5.0~~~ & 1.4$\times$10$^{-8}$~ & 3 \\  
A\,78             & [WC]-PG1159   &     113~ &     5.7~~~ & 4.7$\times$10$^{-8}$~ & 3 & NGC\,2371-2       & [WC]-PG1159   &     135~ &     6.3~~~ & 7.8$\times$10$^{-8}$~ & 3 \\  
BD+30$^\circ$3639 & [WC9]         &      47~ &     4.2~~~ & 1.6$\times$10$^{-6}$~ & 4 & NGC\,2392         & Of(H)         &      45~ &     3.7~~~ & 2.0$\times$10$^{-8}$~ & 10 \\  
Cn\,3-1           & WELS          & $\dots$~ & $\dots$~~~ &        $\dots$~~~~~~~ & $\dots$ & NGC\,2867         & [WO1-2]       &     141~ &     6.0~~~ & 5.8$\times$10$^{-7}$~ & 1 \\  
Hb\,7             & WELS          &      50~ & $\dots$~~~ & 2.9$\times$10$^{-8}$~ & 5 & NGC\,5882         & Of(H)         &      68~ &     3.8~~~ &         $\dots$~~~~~~~ & $\dots$ \\ 
Hen\,2-99         & [WC9]         &      49~ & $\dots$~~~ & 2.6$\times$10$^{-6}$~ & 6 & NGC\,6058         & O(H)          &      77~ &     5.0~~~ & 2.6$\times$10$^{-9}$~ & 10 \\ 
Hen\,2-131        & Of(H)         &      33~ &     3.1~~~ & 3.5$\times$10$^{-7}$~ & 7 & NGC\,6210         & O(H)          &      75~ &     5.0~~~ & 2.4$\times$10$^{-8}$~ & 10 \\  
Hen\,2-138        & Of(H)         &      29~ &     3.0~~~ & 1.2$\times$10$^{-7}$~ & 8 & NGC\,6543         & Of-WR(H)       &      60~ &    4.7~~~ & 1.1$\times$10$^{-7}$~ & 10 \\ 
IC\,418           & Of(H)         &      39~ &     3.7~~~ & 7.2$\times$10$^{-8}$~ & 7 & NGC\,6572         & WELS           &      60~ &    4.2~~~ & 3.0$\times$10$^{-8}$~  & 9,13 \\
IC\,2149          & Of(H)         &      42~ &     3.6~~~ & 2.0$\times$10$^{-8}$~ & 9 & NGC\,6826         & O3f(H)        &      44~ &     3.9~~~ & 1.8$\times$10$^{-7}$~ & 7 \\  
IC\,2448          & O(H)          &      95~ &     5.0~~~ & 1.1$\times$10$^{-8}$~ & 10 & NGC\,6891         & O(H)          &      50~ &     4.0~~~ & 1.2$\times$10$^{-7}$~ & 14 \\ 
IC\,2501          & Emission line & $\dots$~ & $\dots$~~~ &        $\dots$~~~~~~~ & 11 & NGC\,7009         & O(H)          &      87~ &     4.9~~~ & 2.9$\times$10$^{-9}$~ & 15,16 \\ 
IC\,2553          & Emission line & $\dots$~ & $\dots$~~~ &        $\dots$~~~~~~~ & 11 & NGC\,7094         & Hybrid PG1159 &     110~ &     5.7~~~ & 5.0$\times$10$^{-8}$~ & 12,17 \\ 
IC\,3568          & O5f           &      50~ &     4.0~~~ & 3.0$\times$10$^{-8}$~ & 9 & NGC\,7662         & $\dots$       &     119~ &     5.7~~~ & 2.5$\times$10$^{-9}$~ & 10 \\
IC\,4593          & O5f(H)        &      41~ &     4.0~~~ & 2.3$\times$10$^{-8}$~ & 10 & PB\,6             & [WO1]         &     158~ & $\dots$~~~ & 5.2$\times$10$^{-7}$~ & 1 \\  
IC\,4776          & WELS          &      65~ &     5.1~~~ & 1.4$\times$10$^{-8}$~ & 3 & PB\,8             & [WN/WC]       &      52~ & $\dots$~~~ & 8.5$\times$10$^{-8}$~ & 18 \\   
IC\,5217          & WELS          &      95~ & $\dots$~~~ &        $\dots$~~~~~~~ & $\dots$ & RX\,J2117.1+3412  & PG1159        &     170~ &     6.0~~~ & 1.0$\times$10$^{-7}$~ & 12 \\   
K\,1-16           & PG1159        &     140~ &     6.4~~~ & 2.5$\times$10$^{-8}$~ & 2,12 & Sp\,3             & WELS          &      50~ & $\dots$~~~ & 7.1$\times$10$^{-9}$~ & 5 \\
Lo\,4             & PG1159        &     120~ &     5.5~~~ & 2.5$\times$10$^{-8}$~ & 1 & SwSt\,1          & [WC9pec]      &      40~ & $\dots$~~~ & 1.9$\times$10$^{-7}$~ & 19 \\  
LSS\,1362         & O(H)          &     119~ &     5.0~~~ & 5.5$\times$10$^{-9}$~ & 10 & Tc\,1             & O(H)          &      35~ &     3.6~~~ & 2.1$\times$10$^{-8}$~ & 7 \\  
NGC\,40           & [WC8]         &      71~ &     4.8~~~ & 1.8$\times$10$^{-6}$~ & 4 & Vy\,2-3          & $\dots$       &      34~ & $\dots$~~~ &         $\dots$~~~~~~~ & $\dots$ \\  
\hline 
\multicolumn{10}{c}{~~~~~~~~~~~~~~~~~~~~~~~~CSPNe without P Cygni profiles} & \\
\hline 
A\,7              & hgO(H)        &      99~ &     7.0~~~ & $\dots$~~~~~~~ & $\dots$ & NGC\,1360   & O(H)    &     105~ &     5.6~~~ & 1.0$\times$10$^{-10}$~ & 10 \\
A\,31             & hgO(H)        &      85~ &     6.6~~~ & $\dots$~~~~~~~ & $\dots$ & NGC\,3132   & A2\,V   & $\dots$~ & $\dots$~~~ & $\dots$~~~~~~~ & $\dots$ \\
A\,35             & DAO           &      80~ &     7.7~~~ & $\dots$~~~~~~~ & 20      & NGC\,3211   & $\dots$ & $\dots$~ & $\dots$~~~ & $\dots$~~~~~~~ & $\dots$ \\
A\,39             & hgO(H)        &     117~ &     6.3~~~ & $\dots$~~~~~~~ & $\dots$ & NGC\,3587   & hgO(H)  &      94~ &     6.9~~~ & $\dots$~~~~~~~ & 24 \\
DeHt\,2           & O(H)          &     117~ &     5.6~~~ & $\dots$~~~~~~~ & $\dots$ & NGC\,3918   & $\dots$ &     140~ &     6.5~~~ & $\dots$~~~~~~~ & $\dots$ \\
GJJC\,1           & sdO           &      75~ &     5.0~~~ & $\dots$~~~~~~~ & 21      & NGC\,6720   & hgO(H)  &     101~ &     6.9~~~ & $\dots$~~~~~~~ & $\dots$ \\
HDW\,4            & hgO(H)        &      47~ &     7.9~~~ & $\dots$~~~~~~~ & $\dots$ & NGC\,7293   & hgO(H)  &     104~ &     7.0~~~ & $\dots$~~~~~~~ & 25 \\
Hen\,2-86         & [WC4]         &      68~ & $\dots$~~~ & $\dots$~~~~~~~ & 22      & Ps\,1       & sdO     &      39~ &     3.9~~~ & $\dots$~~~~~~~ & 26 \\
Hen\,3-1357       & $\dots$       & $\dots$~ & $\dots$~~~ & $\dots$~~~~~~~ & $\dots$ & PuWe\,1     & hgO(H)  &      94~ &     7.1~~~ & $\dots$~~~~~~~ & $\dots$ \\
K\,1-26           & hgO(H)        & $\dots$~ & $\dots$~~~ & $\dots$~~~~~~~ & $\dots$ & Sh\,2-174   & hgO(H)  &      69~ &     6.7~~~ & $\dots$~~~~~~~ & $\dots$ \\
K\,2-2            & hgO(H)        &      67~ &     6.1~~~ & $\dots$~~~~~~~ & $\dots$ & Sh\,2-216   & hgO(H)  &      83~ &     6.7~~~ & $\dots$~~~~~~~ & $\dots$ \\
M\,4-18           & [WC10]        &      31~ & $\dots$~~~ & $\dots$~~~~~~~ & 23      &             &         &          &            & & \\

\hline
\end{tabular}}
\tablebib{
(1)~\citet{Koesterke01}; 
(2) \citet{Miksa_etal02}; 
(3) \citet{HB04}; 
(4) \citet{Marcolino_etal07a}; 
(5) \citet{Gauba_etal01}; 
(6) \citet{LHJ96}; 
(7) \citet{PHM04}; 
(8) \citet{Prinja_etal10}; 
(9) \citet{MPP93}; 
(10) \citet{HB11}; 
(11) \citet{WG11}; 
(12) \citet{KW98}; 
(13) \citet{MHM90}; 
(14) \citet{McCarthy_etal90}; 
(15) \citet{C-SP89}; 
(16) \citet{Iping_etal06}; 
(17) \citet{KDR98}; 
(18) \citet{Todt_etal10}; 
(19) \citet{DeMarco_etal01}; 
(20) \citet{HB02}; 
(21) \citet{HP93}; 
(22) \citet{Gesicki_etal06}; 
(23) \citet{DMC99}; 
(24) \citet{N99}; 
(25) \citet{NS95}; 
(26) \citet{RHW02}. 
}
\end{table*}

Following the end of operations and decommissioning of \emph{FUSE} in 2007, 
the entire \emph{FUSE} science data were homogeneously reprocessed and 
archived at MAST, the Mikulski Archive for Space Telescopes, with the final 
version of the CalFUSE calibration pipeline software package, CalFUSE 3.2.3 
\citep{Dixon_etal07}.  
This calibration software processes \emph{FUSE} data to remove instrumental 
effects, extract spectra, and apply wavelength and flux calibrations.  
The final reprocessing of the \emph{FUSE} spectra permits the examination 
of a wealth of homogeneous, high-quality spectra covering the wavelength 
range from 920 \AA\ to 1190 \AA.  

\subsection{The sample}

We have searched the entire \emph{FUSE} archive at MAST for observations 
of Galactic CSPNe showing P~Cygni profiles of high-excitation UV lines.  
Following \citet{GR-LM10}, we have assembled a sample of 44 Galactic 
CSPNe with P Cygni profiles in \emph{FUSE} observations.  
The names of these PNe and spectral types of their CSPNe are listed in 
Table~\ref{FUSE_CSPN_sample}, together with their stellar parameters 
(effective temperature, surface gravity, and mass loss rate) and the 
references from which these data have been compiled.  
As listed in this same Table, another 23 Galactic CSPNe with data in the 
\emph{FUSE} archive are found to not have P~Cygni profiles, although the 
S/N ratio of the continua of five of them (Hen\,2-86, GJJC\,1, M\,4-18, 
NGC\,3132, and NGC\,3918) is too low to make a confident statement.  

%

We have then downloaded all available \emph{FUSE} data of the objects 
with P Cygni profiles and screened them to select only the useful 
exposures.  
Exposures that showed an anomalously low count rate in the stellar 
continuum, as compared to other exposures of the same object, were 
discarded.  
Likewise, LiF1B exposures affected by ``the worm'' in the 1140--1180 
\AA\ spectral range were disregarded for the analysis of the P Cygni 
profile of the C~{\sc iii} $\lambda$1176 multiplet.  
Table~\ref{FUSE_obs} summarizes the \emph{FUSE} observations of CSPNe, 
including the observation ID, UT starting time of the first exposure, 
number of useful exposures, and total useful exposure time.  
The cadence among exposures is typically the total exposure time 
divided by the number of exposures, although it may change notably 
among different exposures.   

\begin{table*}
\centering
\scriptsize{
\caption{\footnotesize{
Summary of FUSE observations of CSPNe with P Cygni profiles}} 
\label{FUSE_obs}
\begin{tabular}{lllrr|lllrr} 
\hline\hline
\multicolumn{1}{l}{Name}             & 
\multicolumn{1}{c}{Obs. ID}          & 
\multicolumn{1}{c}{Start time}       & 
\multicolumn{1}{c}{Number of}        & 
\multicolumn{1}{c}{Net exposure}     & 
\multicolumn{1}{l}{Name}             & 
\multicolumn{1}{c}{Obs. ID}          & 
\multicolumn{1}{c}{Start time}       & 
\multicolumn{1}{c}{Number of}        & 
\multicolumn{1}{c}{Net exposure}     \\ 
\multicolumn{1}{c}{}                 & 
\multicolumn{1}{c}{}                 & 
\multicolumn{1}{c}{date and UT}      & 
\multicolumn{1}{c}{exposures}        & 
\multicolumn{1}{c}{time}             & 
\multicolumn{1}{c}{}                 & 
\multicolumn{1}{c}{}                 & 
\multicolumn{1}{c}{date and UT}      & 
\multicolumn{1}{c}{exposures}        & 
\multicolumn{1}{c}{time}             \\
\multicolumn{1}{c}{}                 & 
\multicolumn{1}{c}{}                 & 
\multicolumn{1}{c}{}                 & 
\multicolumn{1}{c}{}                 & 
\multicolumn{1}{c}{[ks]}             & 
\multicolumn{1}{c}{}                 & 
\multicolumn{1}{c}{}                 & 
\multicolumn{1}{c}{}                 & 
\multicolumn{1}{c}{}                 & 
\multicolumn{1}{c}{[ks]}             \\
\hline
A\,30              & B0230101 & 2001-04-10 00:12:51 &  3~~~~~~~~~~ &  4.1~~~~~~ & Lo\,4              & B0230201 & 2001-05-26 09:00:03 & 12~~~~~~~~~~ & 24.1~~~~~~ \\
A\,43              & B0520201 & 2001-07-29 20:41:47 & 10~~~~~~~~~~ & 11.4~~~~~~ & LSS\,1362          & C1770301 & 2002-03-08 15:53:37 & 15~~~~~~~~~~ &  7.9~~~~~~ \\
                   & B0520202 & 2001-08-03 22:18:20 &  7~~~~~~~~~~ &  9.5~~~~~~ &                    & P3020701 & 2003-12-11 21:30:43 & 63~~~~~~~~~~ & 29.1~~~~~~ \\
A\,78              & B1100101 & 2000-11-13 03:56:14 & 12~~~~~~~~~~ &  7.9~~~~~~ & NGC\,40            & A0850101 & 2000-09-14 12:05:10 &  3~~~~~~~~~~ & 13.5~~~~~~ \\
                   & B1100102 & 2001-07-02 22:09:52 & 19~~~~~~~~~~ &  9.4~~~~~~ &                    & A0850102 & 2000-12-16 08:31:59 & 14~~~~~~~~~~ & 26.6~~~~~~ \\
                   & E1180101 & 2004-11-10 22:09:01 & 37~~~~~~~~~~ & 56.2~~~~~~ & NGC\,246           & E1180201 & 2004-07-12 17:01:48 & 15~~~~~~~~~~ &  6.5~~~~~~ \\
BD+30$^\circ$3639  & A0850301 & 2000-06-05 21:47:25 & 17~~~~~~~~~~ & 27.1~~~~~~ & NGC\,1535          & P1150808 & 2001-10-05 17:17:25 &  2~~~~~~~~~~ &  6.6~~~~~~ \\
Cn\,3-1            & D0890401 & 2004-05-15 07:45:43 &  9~~~~~~~~~~ & 21.9~~~~~~ &                    & C1770101 & 2003-01-01 20:27:40 & 15~~~~~~~~~~ &  7.9~~~~~~ \\
Hb\,7              & D0890201 & 2003-09-19 21:27:06 &  2~~~~~~~~~~ &  2.5~~~~~~ & NGC\,2371-2        & P1330301 & 2000-02-26 05:04:45 &  4~~~~~~~~~~ &  5.3~~~~~~ \\
Hen\,2-99          & D0890101 & 2003-02-17 13:53:09 &  6~~~~~~~~~~ & 24.7~~~~~~ & NGC\,2392          & B0320601 & 2001-02-21 16:27:25 &  2~~~~~~~~~~ &  3.9~~~~~~ \\
Hen\,2-131         & P1930301 & 2000-06-28 21:06:44 &  5~~~~~~~~~~ &  6.1~~~~~~ & NGC\,2867          & Z9110901 & 2003-02-22 12:07:50 &  4~~~~~~~~~~ & 14.9~~~~~~ \\
                   & U1091902 & 2006-06-28 06:50:16 &  4~~~~~~~~~~ &  7.6~~~~~~ &                    & U1072501 & 2006-07-13 06:06:40 &  2~~~~~~~~~~ &  4.3~~~~~~ \\
                   & U1091904 & 2006-06-29 22:53:21 &  3~~~~~~~~~~ &  9.9~~~~~~ &                    & U1072502 & 2007-06-30 12:31:47 &  5~~~~~~~~~~ & 12.6~~~~~~ \\
                   & U1091905 & 2007-04-16 20:02:45 &  2~~~~~~~~~~ &  1.0~~~~~~ & NGC\,5882          & B0690401 & 2002-08-14 15:01:12 &  2~~~~~~~~~~ &  3.3~~~~~~ \\
                   & U1091906 & 2007-06-15 00:22:39 &  2~~~~~~~~~~ &  8.0~~~~~~ &                    & B0690402 & 2002-08-15 09:20:04 &  8~~~~~~~~~~ & 14.9~~~~~~ \\
                   & U1091907 & 2007-06-16 04:14:03 &  9~~~~~~~~~~ & 11.9~~~~~~ &                    & B0690403 & 2002-06-16 13:41:18 &  5~~~~~~~~~~ & 10.0~~~~~~ \\
Hen\,2-138\tablefootmark{a}         & I8020201 & 1999-09-23 15:26:37 & 15~~~~~~~~~~ &  8.5~~~~~~ & NGC\,6058          & B0320401 & 2001-03-26 22:20:56 &  2~~~~~~~~~~ &  4.4~~~~~~ \\
                   & I8020202 & 1999-09-23 18:17:16 & 15~~~~~~~~~~ &  8.5~~~~~~ & NGC\,6210          & A0850201 & 2000-08-02 05:38:26 &  4~~~~~~~~~~ &  6.6~~~~~~ \\
                   & I8020203 & 1999-09-23 21:40:14 & 16~~~~~~~~~~ &  8.6~~~~~~ & NGC\,6543\tablefootmark{a}          & Q1080202 & 2001-10-01 21:52:01 &  9~~~~~~~~~~ &  3.8~~~~~~ \\
                   & I9040401 & 1999-09-21 21:15:39 &  2~~~~~~~~~~ &  8.2~~~~~~ &                    & F0340105 & 2007-01-13 03:01:24 & 26~~~~~~~~~~ & 11.8~~~~~~ \\
                   & P1042801 & 2000-03-29 22:25:04 & 11~~~~~~~~~~ & 20.2~~~~~~ &                    & F0340106 & 2007-01-14 02:38:02 & 18~~~~~~~~~~ &  6.9~~~~~~ \\
                   & S1010101 & 2000-07-13 09:06:18 &  8~~~~~~~~~~ &  2.1~~~~~~ &                    & F0340107 & 2007-01-15 02:14:23 & 19~~~~~~~~~~ &  8.2~~~~~~ \\
                   & M1140901 & 2000-07-12 23:06:34 &  1~~~~~~~~~~ &  3.3~~~~~~ &                    & F0340108 & 2007-01-16 01:37:52 & 10~~~~~~~~~~ &  3.6~~~~~~ \\
                   & S6013601 & 2002-04-21 06:55:48 &  2~~~~~~~~~~ &  7.7~~~~~~ & NGC\,6572          & B0320201 & 2001-09-15 10:48:58 &  4~~~~~~~~~~ &  6.9~~~~~~ \\
                   & M7276401 & 2007-02-16 05:05:15 &  2~~~~~~~~~~ &  1.8~~~~~~ & NGC\,6826          & P1930401 & 2000-08-07 22:48:04 & 12~~~~~~~~~~ &  5.8~~~~~~ \\
                   & M7276402 & 2007-02-17 03:05:27 &  2~~~~~~~~~~ &  1.7~~~~~~ &                    & D1200601 & 2003-06-23 19:56:26 &  2~~~~~~~~~~ &  4.3~~~~~~ \\
                   & M7276403 & 2007-02-18 23:37:42 &  2~~~~~~~~~~ &  2.4~~~~~~ &                    & D1201601 & 2003-10-16 11:01:28 &  1~~~~~~~~~~ &  0.8~~~~~~ \\
                   & M7276001 & 2007-04-19 00:30:46 &  2~~~~~~~~~~ &  2.7~~~~~~ &                    & F1600201 & 2006-11-06 23:12:27 &  5~~~~~~~~~~ &  2.1~~~~~~ \\
IC\,418            & P1151111 & 2001-12-02 02:23:33 &  9~~~~~~~~~~ &  4.1~~~~~~ &                    & F1600202 & 2006-11-09 22:15:00 &  3~~~~~~~~~~ &  1.9~~~~~~ \\
IC\,2149           & P1041402 & 1999-11-25 17:31:13 &  7~~~~~~~~~~ & 16.2~~~~~~ &                    & F1600209 & 2007-01-09 17:33:48 & 12~~~~~~~~~~ &  5.3~~~~~~ \\
                   & P1041401 & 1999-12-02 03:09:44 &  8~~~~~~~~~~ & 21.7~~~~~~ & NGC\,6891          & B0320301 & 2001-09-15 02:26:52 &  6~~~~~~~~~~ &  7.4~~~~~~ \\
                   & P1041403 & 2000-01-14 22:47:38 & 11~~~~~~~~~~ & 17.2~~~~~~ & NGC\,7009          & I8011001 & 1999-11-02 01:02:13 &  2~~~~~~~~~~ &  1.9~~~~~~ \\
IC\,2448           & B0320701 & 2002-03-07 00:32:47 &  3~~~~~~~~~~ & 11.5~~~~~~ &                    & X0140201 & 1999-11-02 07:36:07 & 12~~~~~~~~~~ &  5.8~~~~~~ \\
                   & S7011102 & 2005-03-30 15:50:37 &  6~~~~~~~~~~ & 12.4~~~~~~ &                    & I8011002 & 1999-11-02 17:26:29 &  2~~~~~~~~~~ &  2.7~~~~~~ \\
                   & S7011103 & 2005-03-31 05:42:49 &  5~~~~~~~~~~ & 18.1~~~~~~ & NGC\,7094          & P1043701 & 2000-11-13 08:53:28 & 14~~~~~~~~~~ & 22.6~~~~~~ \\
                   & U1072101 & 2006-01-20 01:34:40 &  7~~~~~~~~~~ & 23.9~~~~~~ & NGC\,7662          & B0690301 & 2001-07-20 20:08:48 &  4~~~~~~~~~~ & 10.2~~~~~~ \\
                   & U1072103 & 2007-06-30 00:10:28 &  5~~~~~~~~~~ & 14.8~~~~~~ & PB\,6              & Z9111101 & 2003-02-20 15:13:54 &  7~~~~~~~~~~ & 19.5~~~~~~ \\
IC\,2501           & Z9110101 & 2003-02-23 06:20:22 &  3~~~~~~~~~~ & 11.0~~~~~~ & PB\,8              & Z9111301 & 2003-06-17 12:35:50 &  5~~~~~~~~~~ & 16.7~~~~~~ \\
IC\,2553           & Z9110301 & 2003-04-24 07:06:05 &  3~~~~~~~~~~ &  8.5~~~~~~ & RX\,J2117.1+3412   & P1320501 & 2000-07-16 03:44:57 & 18~~~~~~~~~~ &  8.2~~~~~~ \\
IC\,3568           & P1042201 & 2000-03-02 00:49:26 &  7~~~~~~~~~~ & 20.0~~~~~~ &                    & D1800201 & 2003-06-21 13:50:31 & 13~~~~~~~~~~ &  6.4~~~~~~ \\
                   & U1031401 & 2006-02-03 20:18:19 &  1~~~~~~~~~~ &  1.2~~~~~~ &                    & D1800202 & 2003-06-22 11:09:45 & 23~~~~~~~~~~ & 10.7~~~~~~ \\
                   & U1031402 & 2006-02-04 23:58:51 &  1~~~~~~~~~~ &  1.2~~~~~~ & Sp\,3              & B0320801 & 2001-08-18 01:49:57 &  7~~~~~~~~~~ & 12.0~~~~~~ \\
                   & U1031404 & 2007-03-18 14:56:07 &  5~~~~~~~~~~ & 13.0~~~~~~ & SwSt\,1            & B0690101 & 2001-08-21 01:17:09 & 11~~~~~~~~~~ & 23.1~~~~~~ \\
IC\,4593           & B0320102 & 2001-08-03 15:28:55 &  4~~~~~~~~~~ &  1.9~~~~~~ &                    & B0690102 & 2001-08-22 10:34:39 &  7~~~~~~~~~~ & 15.3~~~~~~ \\
                   & D1200301 & 2004-04-16 13:25:30 &  3~~~~~~~~~~ &  7.3~~~~~~ & Tc\,1              & P1980304 & 2000-04-11 16:54:58 &  2~~~~~~~~~~ &  5.6~~~~~~ \\
IC\,4776           & P1330501 & 2000-05-21 00:02:15 &  2~~~~~~~~~~ &  6.2~~~~~~ & Vy\,2-3            & B0320501 & 2001-07-20 10:01:30 &  4~~~~~~~~~~ &  9.8~~~~~~ \\
IC\,5217           & Z9112401 & 2003-10-19 20:38:13 &  6~~~~~~~~~~ &  7.2~~~~~~ & & & & & \\
K\,1-16            & I8110310 & 1999-10-15 07:01:57 &  1~~~~~~~~~~ &  4.2~~~~~~ & & & & & \\
                   & M1031001 & 2000-07-23 18:59:07 &  9~~~~~~~~~~ &  9.9~~~~~~ & & & & & \\
                   & M1031002 & 2002-11-03 14:54:18 &  3~~~~~~~~~~ & 12.4~~~~~~ & & & & & \\
                   & M1031005 & 2002-12-25 19:05:12 &  3~~~~~~~~~~ &  8.5~~~~~~ & & & & & \\
                   & M7170101 & 2006-03-23 09:05:23 &  4~~~~~~~~~~ &  4.6~~~~~~ & & & & & \\
\hline
\end{tabular}
\tablefoot{
\tablefoottext{a}{
The \emph{FUSE} spectra of Hen\,2-138 and NGC\,6543 have been thoroughly 
investigated by \citet{Prinja_etal07,Prinja_etal10} and will not be 
discussed in the following sections. }
}}
\end{table*}

As it could be expected, the final dataset is highly inhomogeneous.  
To better describe this dataset, we need to distinguish 
explicitly between observation and exposure.  
Typically, each observation consists of several exposures 
taken consecutively along a period of time.  
There are many CSPNe that have only one observation consisting 
of different useful exposures (e.g., NGC\,2392), while there are 
some that have multiple observations taken at different epochs 
(e.g., NGC\,6826).  
For the latter, the observing epochs are either close in time, 
within a few days, or separated in time by several months or 
even years.  
Obviously, these ``sparse time-series'' are not always suited for the 
detailed study of time variability, but the availability of multiple 
exposures at least allows the search for the occurrence of changes in 
short-time scales of the P~Cygni profiles of the targeted UV lines.

\subsection{Data reduction}

Individual \emph{FUSE} spectra of a same object typically display 
small shifts in wavelength caused by the apparent motion of a source 
on the instrument aperture.  
Such motion may even move the source partially outside the aperture, 
resulting in differences in the flux registered by each segment and 
channel.  
To correct these effects, the individual \emph{FUSE} spectra of an 
object have been shifted by forcing narrow H$_2$, interstellar medium 
(ISM), and airglow (C~{\sc i}, O~{\sc i}, Ar\,{\sc i}, Fe~{\sc ii}, ...) 
absorptions to agree in wavelength with those of the higher S/N spectrum.  
Similarly, the flux of these spectra has been scaled so that 
the level of the stellar continuum agrees with that of the 
higher S/N spectrum.  
Since the \emph{FUSE} dispersion solution is highly non-linear and 
the vignetting of a source may affect the segments and channels 
defining different spectral ranges, this process has been applied 
locally to six spectral regions corresponding to the lines of 
(1) S~{\sc vi} $\lambda\lambda$933,944,  
(2) Ne~{\sc vii} $\lambda$973.3 and C~{\sc iii} $\lambda$977, 
(3) O~{\sc vi} $\lambda\lambda$1032,1038, 
(4) S~{\sc iv} $\lambda$1073, 
(5) Si~{\sc iii} $\lambda\lambda$1108,1110,1113, P~{\sc v} 
$\lambda\lambda$1118,1128 and Si~{\sc iv} $\lambda\lambda$1122,1128, 
and 
(6) C~{\sc iii} $\lambda\lambda$1176, 
implying that the shifts in wavelength and flux rescaling are fine 
tuned for each of these spectral ranges.  
We note that the shifts in wavelength between different spectra are 
typically small, $\lesssim$5~km~s$^{-1}$ (i.e., $\lesssim$1.5 pixels).

For most of the sources in our sample, the comparison among individual 
spectra, after registering them in wavelength and flux, resulted in 
differences within the noise of the individual spectra.  
This general consistency between different spectra wherever 
there is no variability supports the procedure for shifting 
and scaling individual spectra described above.  
Exposures taken at similar epochs and with different channels which were 
found to be not significantly different have been combined to increase 
the S/N ratio.  
Finally, all spectra have been merged together to build a ``master'' mean 
spectrum to which compare individual spectra taken at different epochs.

\begin{table*}
\centering
\scriptsize{
\caption{\footnotesize{
FUSE lines and variable P Cygni profiles in CSPNe\tablefootmark{a}}}

\label{CSPN_FUSE_lines}
\begin{tabular}{lcccccccccl}
\hline\hline

\multicolumn{1}{l}{CSPN}         & 
\multicolumn{1}{c}{S~{\sc vi}}   & 
\multicolumn{1}{c}{Ne~{\sc vii}} & 
\multicolumn{1}{c}{C~{\sc iii}}  & 
\multicolumn{1}{c}{O~{\sc vi}}   & 
\multicolumn{1}{c}{S~{\sc iv}}   & 
\multicolumn{1}{c}{Si~{\sc iii}} & 
\multicolumn{1}{c}{P~{\sc v}}    & 
\multicolumn{1}{c}{Si~{\sc iv}}  & 
\multicolumn{1}{c}{C~{\sc iii}}  & 
\multicolumn{1}{l}{Variability\tablefootmark{b,c}}  \\
\multicolumn{1}{l}{}             & 
\multicolumn{1}{l}{$\lambda\lambda$933,944 \AA} & 
\multicolumn{1}{l}{$\lambda$973 \AA} & 
\multicolumn{1}{l}{$\lambda$977 \AA} & 
\multicolumn{1}{l}{$\lambda\lambda$1032,1038 \AA} & 
\multicolumn{1}{l}{$\lambda$1073 \AA} & 
\multicolumn{1}{l}{$\lambda\lambda$1108,1110,1113 \AA} & 
\multicolumn{1}{l}{$\lambda$1118 \AA} & 
\multicolumn{1}{l}{$\lambda$1122 \AA} & 
\multicolumn{1}{l}{$\lambda$1175-1176 \AA} & 
\multicolumn{1}{l}{}             \\

\hline

A\,30              & $\dots$  & $\times$ & $\dots$  & $\times$ & $\dots$  & $\dots$  & $\dots$  & $\dots$  & $\dots$  & N: 1,2 \\
A\,43              & $\dots$  & $\times$ & $\dots$  & $\times$ & $\dots$  & $\dots$  & $\dots$  & $\dots$  & $\dots$  & N: 1,2 \\
A\,78              & $\dots$  & $\times$ & $\dots$  & $\times$ & $\dots$  & $\dots$  & $\dots$  & $\dots$  & $\dots$  & N: 2   \\
BD+30$^{\circ}$3639 & $\dots$  & $\dots$  & $\dots$ ?& $\dots$  & $\times$ & $\dots$  & $\times$ & $\times$ & $\times$ & N: 3 \\
Cn\,3-1            & $\dots$  & $\dots$  & $\dots$ ?& $\times$ & $\times$ & $\dots$  & $\times$ & $\times$ & $\times$ & N: 1   \\
Hb\,7              & $\times$ & $\dots$  & $\dots$  & $\times$ & $\dots$  & $\dots$  & $\dots$  & $\dots$  & $\dots$  & N: 1   \\
Hen\,2-99          & $\dots$  & $\dots$  & $\dots$  & $\times$ & $\times$ & $\dots$  & $\times$ & $\times$ & $\times$ & N: 1   \\
Hen\,2-131         & $\dots$  & $\dots$  & $\times$ & $\times$ & var      & var      & var      & var      & var      & Y: R,N \\
IC\,418            & $\dots$  & $\dots$  & $\times$ & var      & var      & $\dots$  & var      & var      & var      & Y: R,B,E \\ 
IC\,2149           & $\dots$  & $\dots$  & $\times$ & var      & $\dots$  & $\dots$  & var      & $\dots$  & var      & Y: R,E \\
IC\,2448           & $\dots$  & $\dots$  & $\dots$  & $\times$ & $\dots$  & $\dots$  & $\dots$  & $\dots$  & $\dots$  & N: 2   \\
IC\,2501           & $\times$ & $\dots$  & $\dots$  & $\times$ & $\dots$  & $\dots$  & $\times$ & $\dots$  & $\dots$  & N: 1   \\
IC\,2553           & $\dots$  & $\dots$  & $\dots$  & $\times$ & $\dots$  & $\dots$  & $\dots$  & $\dots$  & $\dots$  & N: 1,2 \\
IC\,3568           & $\times$ & $\dots$  & $\dots$  & $\times$ & $\dots$  & $\dots$  & $\dots$  & $\dots$  & $\dots$  & N: 3   \\
IC\,4593           & $\times$ & $\dots$  & $\times$ & var      & $\times$ & $\dots$  & var      & $\dots$  & var      & Y: R,B \\
IC\,4776           & $\times$ & $\dots$  & $\dots$  & $\times$ & $\dots$  & $\dots$  & $\dots$  & $\dots$  & $\dots$  & N: 1   \\
IC\,5217           & $\dots$  & $\dots$  & $\dots$  & $\times$ & $\dots$  & $\dots$  & $\dots$  & $\dots$  & $\dots$  & N: 1,2 \\
K\,1-16            & $\dots$  & $\times$ & $\dots$  & $\times$ & $\dots$  & $\dots$  & $\dots$  & $\dots$  & $\dots$  & N: 3   \\
Lo\,4              & $\dots$  & $\times$ & $\dots$  & $\times$ & $\dots$  & $\dots$  & $\dots$  & $\dots$  & $\dots$  & N: 1   \\
LSS\,1362          & $\dots$  & $\times$ & $\dots$  & $\times$ & $\dots$  & $\dots$  & $\dots$  & $\dots$  & $\dots$  & N: 3   \\
NGC\,40            & $\dots$  & $\dots$  & $\times$ & $\times$ & $\times$ & $\dots$  & var      & $\times$ & $\times$ & Y: R,B,E \\
NGC\,246           & $\dots$  & $\times$ & $\dots$  & $\times$ & $\dots$  & $\dots$  & $\dots$  & $\dots$  & $\dots$  & N: 3   \\
NGC\,1535          & $\dots$ ?& $\dots$  & $\dots$  & var      & $\dots$  & $\dots$  & $\dots$  & $\dots$  & $\dots$  & Y: N,B \\
NGC\,2371-2        & $\dots$  & $\times$ & $\dots$  & $\times$ & $\dots$  & $\dots$  & $\dots$  & $\dots$  & $\dots$  & N: 1,2 \\
NGC\,2392          & var      & $\dots$  & $\times$ & var      & $\dots$  & $\dots$  & var      & $\times$ & $\dots$  & Y: R,B \\
NGC\,2867          & $\dots$  & $\dots$ ?& $\dots$  & $\times$ & $\dots$  & $\dots$  & $\dots$  & $\dots$  & $\dots$  & N: 1,2 \\
NGC\,5882          & $\times$ & $\dots$  & $\dots$  & $\times$ & $\dots$  & $\dots$  & $\dots$  & $\dots$  & $\dots$  & N: 1,2 \\
NGC\,6058          & $\dots$  & $\dots$  & $\dots$  & $\times$ & $\dots$  & $\dots$  & $\dots$  & $\dots$  & $\dots$  & N: 1   \\
NGC\,6210          & $\times$ & $\dots$  & $\dots$  & $\times$ & $\dots$  & $\dots$  & $\dots$  & $\dots$  & $\dots$  & N: 3   \\
NGC\,6572          & $\times$ & $\dots$  & $\dots$  & $\times$ & $\dots$  & $\dots$  & $\times$ & $\dots$  & $\times$ & N: 1   \\
NGC\,6826          & var      & $\dots$  & var      & var      & $\dots$  & $\dots$  & var      & $\dots$  & var      & Y: R,N,B \\
NGC\,6891          & $\times$ & $\dots$  & $\dots$  & $\times$ & $\dots$  & $\dots$  & $\times$ & $\dots$  & $\dots$  & N: 1   \\
NGC\,7009          & $\dots$  & $\dots$  & $\dots$  & $\times$ & $\dots$  & $\dots$  & $\dots$  & $\dots$  & $\dots$  & N: 2   \\
NGC\,7094          & $\dots$  & $\times$ & $\dots$  & $\times$ & $\dots$  & $\dots$  & $\dots$  & $\dots$  & $\dots$  & N: 1,2 \\
NGC\,7662          & $\dots$  & $\dots$  & $\dots$  & $\times$ & $\dots$  & $\dots$  & $\dots$  & $\dots$  & $\dots$  & N: 3   \\
PB\,6              & $\dots$  & $\dots$ ?& $\dots$  & $\times$ & $\dots$  & $\dots$  & $\dots$  & $\dots$  & $\dots$  & N: 1,2 \\
PB\,8              & $\times$?& $\dots$ ?& $\dots$  & $\times$ & $\times$ & $\dots$  & $\times$ & $\times$ & $\times$ & N: 1   \\
RX\,J2117          & $\dots$  & $\times$ & $\dots$  & $\times$ & $\dots$  & $\dots$  & $\dots$  & $\dots$  & $\dots$  & N: 3   \\
Sp\,3              & var      & $\dots$  & $\dots$  & var      & $\dots$  & $\dots$  & $\times$ & $\dots$  & $\dots$  & Y: R,B \\
SwSt\,1            & $\dots$  & $\dots$  & $\dots$ ?& $\times$ & var      & var      & $\times$ & var      & var      & Y: R,N,B\\
Tc\,1              & $\dots$  & $\dots$  & $\times$ & $\times$ & $\times$ & $\dots$  & $\dots$  & $\dots$  & $\times$ & N: 1 \\
Vy\,2-3            & $\times$ & $\dots$  & $\dots$  & $\times$ & $\dots$  & $\dots$  & $\dots$  & $\dots$  & $\dots$  & N: 1 \\

\hline
\end{tabular}
\tablefoot{
\tablefoottext{a}{
The $\times$ sign indicates that the line shows a constant P~Cygni profile, 
while ``var'' means that the P~Cygni profile is variable. \\ }
\tablefoottext{b}{
When no variability (N) is detected, ``1'' means that the data may 
be insufficient or of low quality to search for variability, ``2'' 
that the P~Cygni profile is saturated or absorbed and the search 
for variability may be insensitive, and ``3'' that the data have 
good quality, but no variability was detected. 
\\ }
\tablefoottext{c}{
When variable P~Cygni profiles (Y), ``R'' means a ripple moving bluewards, 
``N'' a narrow absorption, ``E'' variability of the emission region of 
the profile, and ``B'' variability at velocities bluer than $v_\infty$.} }
}
\end{table*}

\subsection{Near-UV data}

While near-UV \emph{IUE} observations are available for a large 
sample of CSPNe, most of them have been already analyzed or 
compiled by \citet{PP91}.  
This study can be complemented with more recent analysis of \emph{IUE} data 
\citep{PP96,Feibelman97,Feibelman98,Feibelman99,Gauba_etal01,Marcolino_etal07b}. 
When necessary, we have adopted the information on the detection 
of P~Cygni profiles and the values of $v_\infty$ provided by these 
authors after an examination of the original \emph{IUE} data.  
We note that, in general, the spectral resolution and quality of 
the \emph{IUE} data is not comparable to those of \emph{FUSE} data.  

Near-UV \emph{HST} STIS observations are only available for a reduced 
sample of the CSPNe with \emph{FUSE} observations showing P~Cygni 
profiles (see below), namely Hen\,2-138, NGC\,1535, and NGC\,6543.

\section{Atlas of FUSE P-Cygni profiles of CSPNe}

\begin{figure*}
\centering
\includegraphics[bb=61 154 558 718,width=2.0\columnwidth,angle=0]{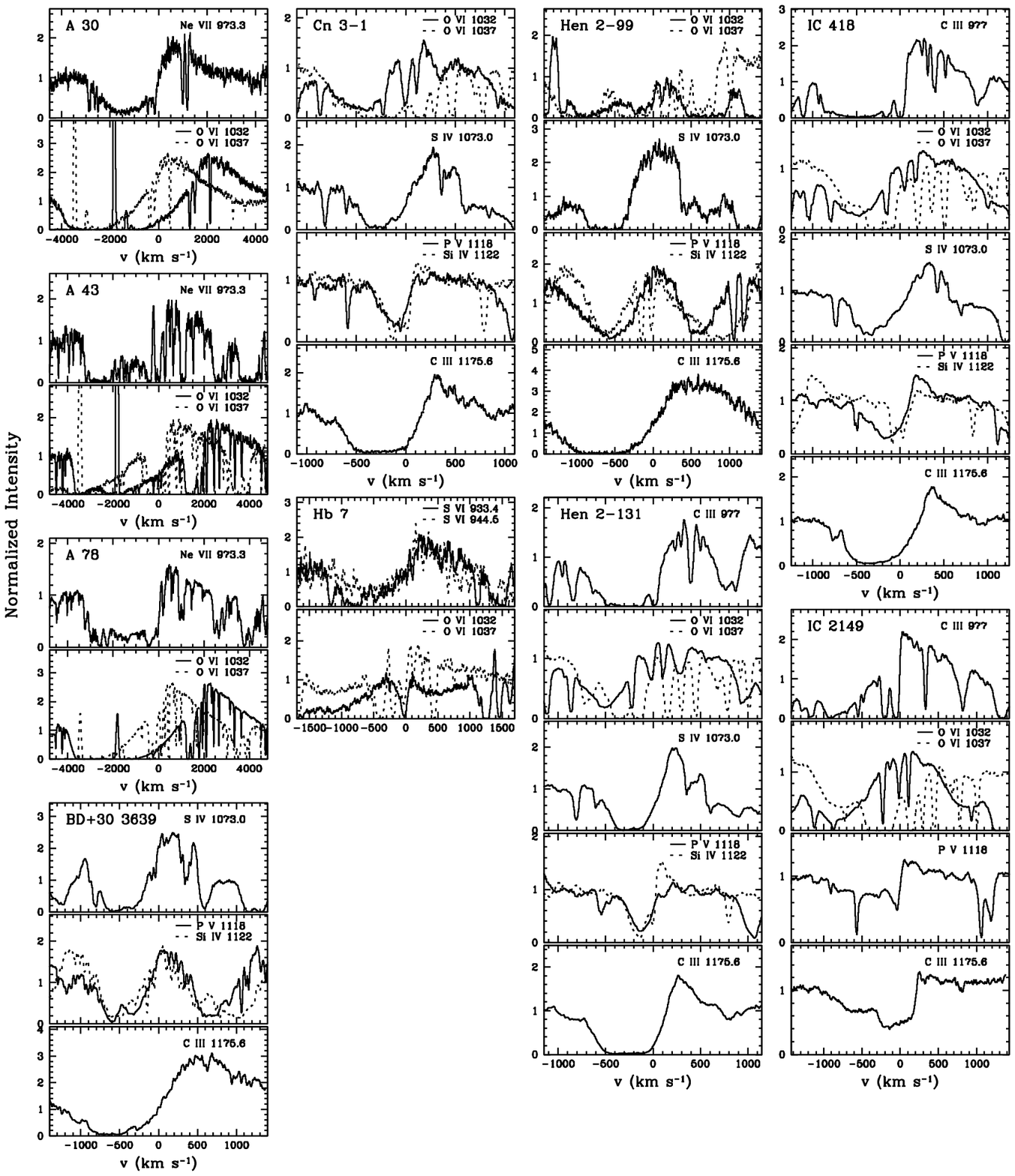}
\caption{
Normalized profiles of the lines in the \emph{FUSE} spectral range that show 
P~Cygni profiles.  
}
\label{fig_1a}
\end{figure*}

\begin{figure*}
\centering
\includegraphics[bb=61 174 558 718,width=2.00\columnwidth,angle=0]{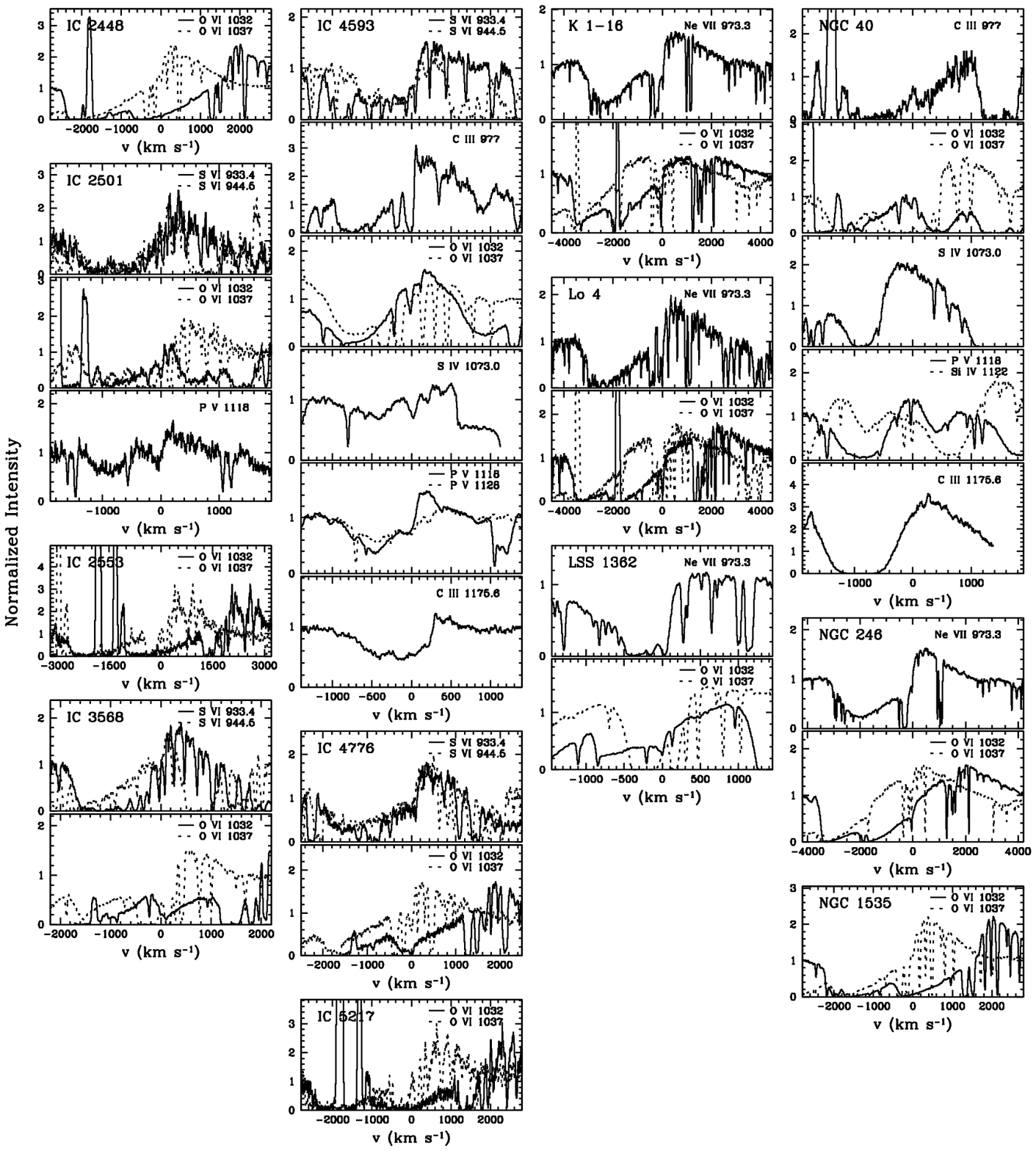}
\caption{Figure~\ref{fig_1a} continued.}
\label{fig_1b}
\end{figure*}

\begin{figure*}
\centering
\includegraphics[bb=61 144 558 718,width=2.00\columnwidth,angle=0]{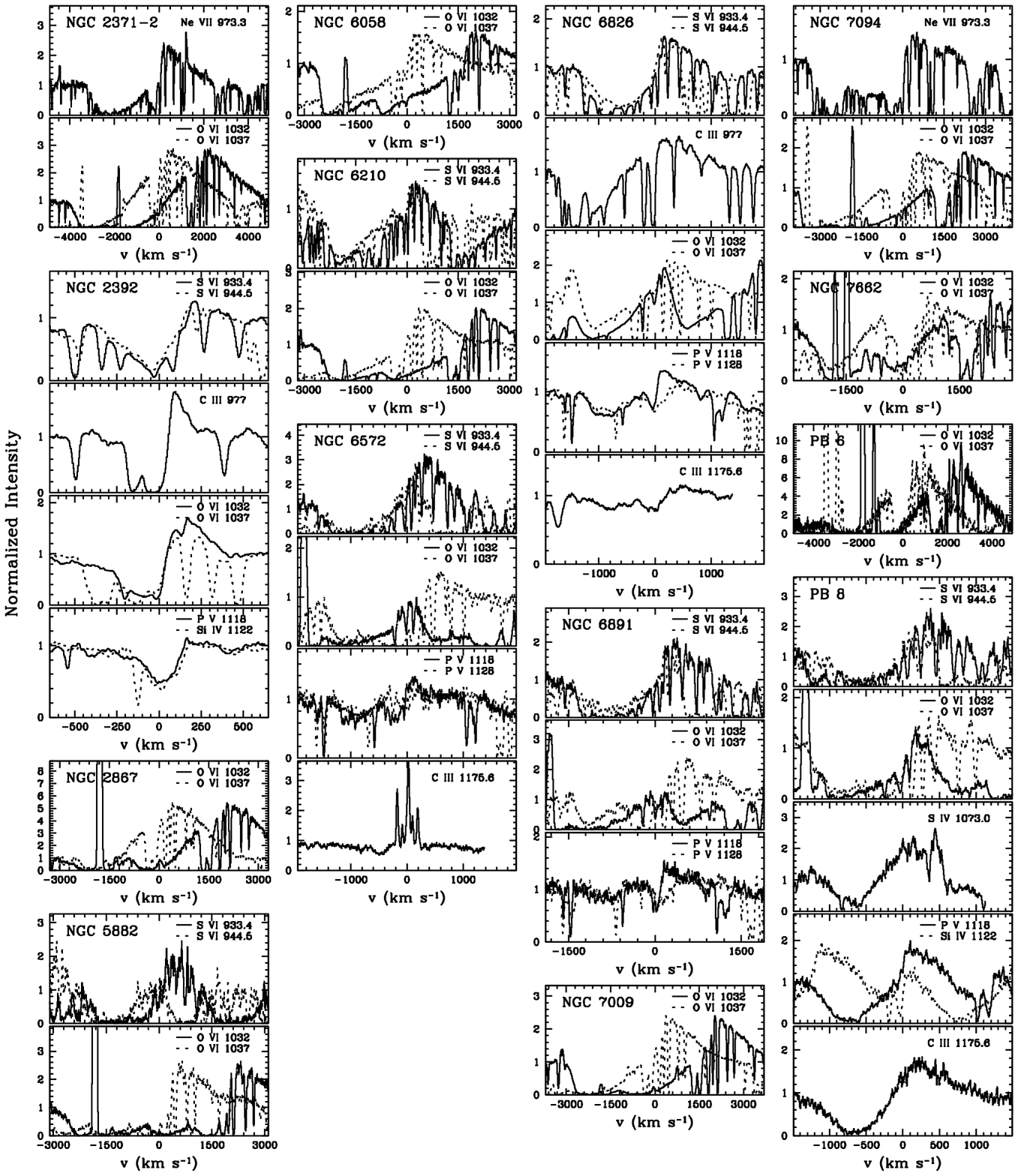}
\caption{Figure~\ref{fig_1a} continued.}
\label{fig_1c}
\end{figure*}

\begin{figure}
\centering
\includegraphics[bb=61 170 320 718,width=1.00\columnwidth,angle=0]{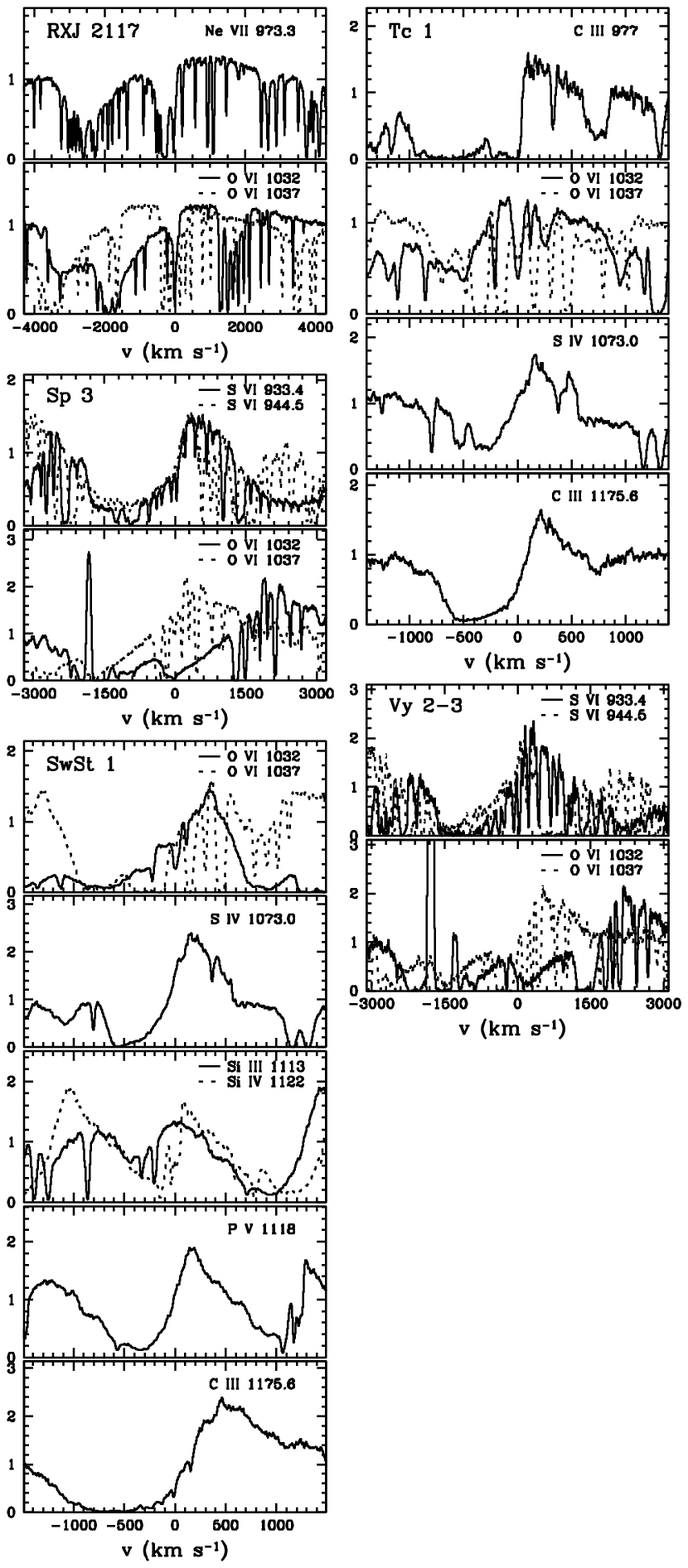}
\caption{Figure~\ref{fig_1a} continued.
}
\label{fig_1d}
\end{figure}

Table~\ref{CSPN_FUSE_lines} lists the CSPNe in our sample that show lines 
with P~Cygni profiles in their \emph{FUSE} spectra.  
In this table, the lines in the \emph{FUSE} spectral range are marked 
by a ``$\dots$'' sign if the P~Cygni profile is absent, a ``$\times$'' 
sign if a constant P~Cygni profile is present, or ``{\it var}'' if a 
variable P~Cygni profile is observed.  
The normalized spectra of the P Cygni profiles of these 
CSPNe, as obtained from the average of all useful spectra, 
are shown in Figures~\ref{fig_1a}-\ref{fig_1d}.  
Since we are combining all available \emph{FUSE} observations for every 
single object, we note that the resulting mean spectra presented here 
have the highest quality affordable by the \emph{FUSE} archive.

The \emph{FUSE} spectral range contains P~Cygni profiles of a variety of 
lines of different ions.  
The occurrence and shape of these line profiles carry information 
on the star and wind properties, including stellar effective 
temperature ($T_{\rm eff}$), wind ionization, mass-loss rate 
($\dot M$), terminal velocity ($v_\infty$),  chemical composition, 
and even binarity.  
The latter is expected if the companion is in a close orbit, within the 
line formation region, and its orbital motion modifies the line profiles 
in a periodic way.  
This mechanism, which has been reported in the massive O star plus WR star 
binary $\gamma$ Vel \citep{StLouis93,DeMarco02} may be expected in binary 
CSPNe as well, particularly those with an extended atmosphere such as [WC] 
CSPNe.  
Only one [WC] CSPN has been discovered to be a close binary 
using periodic light variability ascribed to irradiation 
effects \citep[PN\,G222.8$-$04.2,][]{HZG10}, but it has not 
been monitored for wind variability.  
Wind variability of the $\gamma$ Vel type may affect this and 
other [WC] CSPNe, even if other binary-induced effects are not 
prominent.

An examination of Figures~\ref{fig_1a}-\ref{fig_1d} reveals that, for a 
single CSPN, the P~Cygni profiles of different lines may be dramatically 
different among them.  
This is, for instance, the case of the C~{\sc iii} $\lambda$977, 
C~{\sc iii} $\lambda$1176, and S~{\sc iv} $\lambda$1073 lines 
of Hen\,2-131.  
The situation becomes more complicated when unsaturated line profiles, such as 
the O~{\sc vi} $\lambda\lambda$1031.9,1037.6, P~{\sc v} $\lambda$1118, and 
Si~{\sc iv} $\lambda$1122 lines of Hen\,2-131, are considered.  
Since the distinct physical processes that dominate a line determine 
its shape, and therefore the observational indicators of $v_\infty$ 
(lines formed farther from the star tend to have higher $v_\infty$), 
it is worthwhile to describe in detail the properties and 
characteristics of the ions and lines observed in CSPNe by 
\emph{FUSE}.

\subsection{Ions in the stellar wind of CSPNe in the FUSE spectral range}

Table~\ref{UV_lines} provides the identifications and rest wavelengths 
for typical P Cygni line profiles present in the \emph{FUSE} spectra 
of CSPNe.  
This table includes also information on the electronic 
transition, the nature of the line, and its transition 
probability.  
The relative transition probability of multiplet lines provides 
an indication of the relative strength of the lines in a same 
multiplet.  
We describe these lines in decreasing ionization potential (IP).

\begin{figure}
\centerline{
\includegraphics[bb=18 174 592 718,width=1.00\columnwidth]{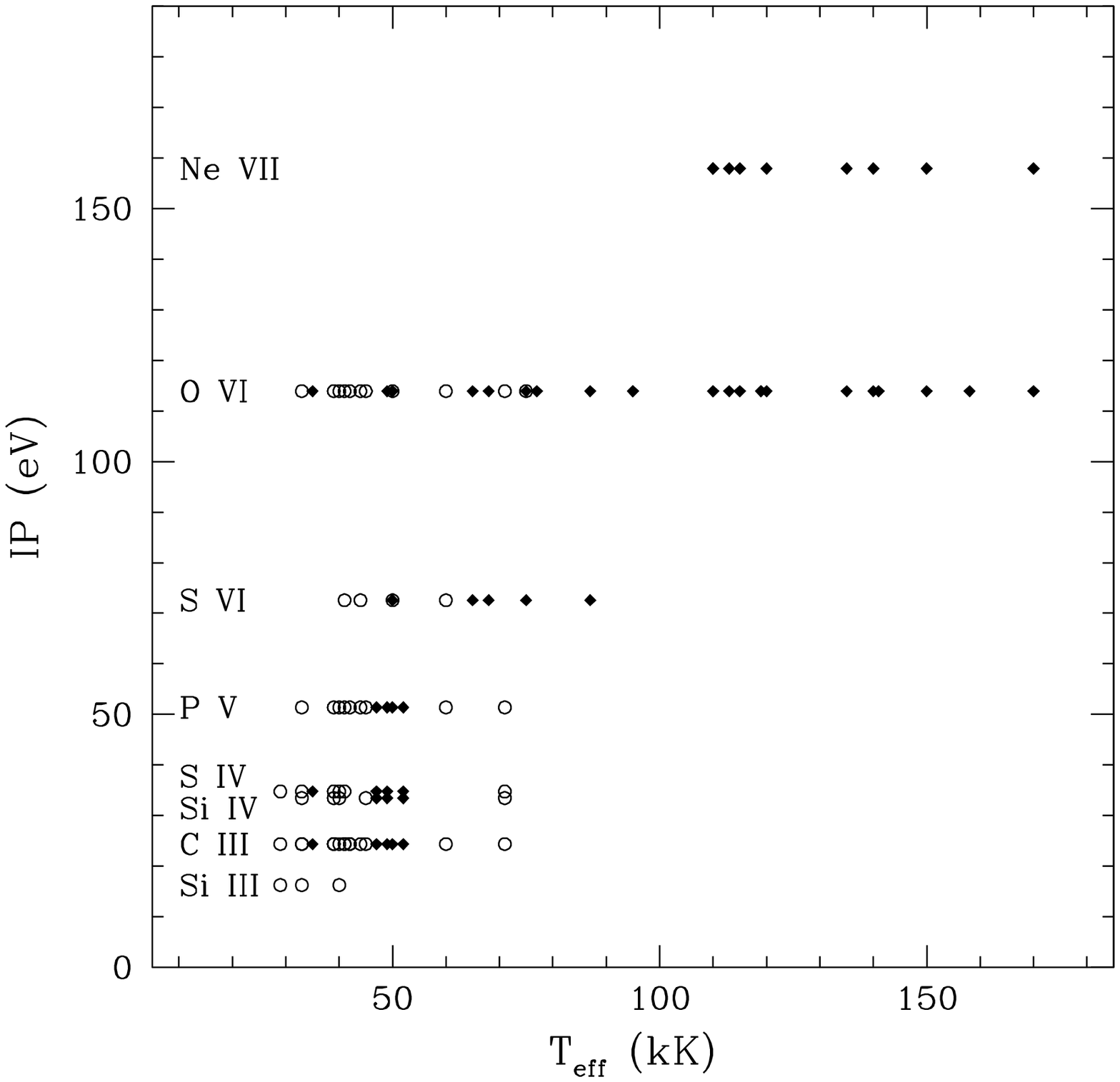}
}
\caption{
Ionization potential of the lines with P~Cygni profiles seen in the 
\emph{FUSE} spectrum of a CSPN vs.\ its effective temperature. 
Objects that show line variability are shown as open circles 
and those that do not display variability as filled diamonds.  
}
\label{teff_line}
\end{figure}

The highest ionization species present in the \emph{FUSE} spectra 
of CSPNe is revealed by the Ne~{\sc vii} $\lambda$973.3 line (IP: 
157.9--207.3 eV).  
This line has only recently been identified in the atmosphere of a number 
of extremely hot hydrogen-deficient WDs \citep{Werner_etal04}, and in the 
stellar wind of the hot 
CSPNe A\,78, K\,1-16, and NGC\,2371-2 \citep{HBH05}.  
We have also identified this feature in the \emph{FUSE} spectra of A\,30, 
A\,43, Lo\,4, LSS\,1362, NGC\,246, NGC\,7094, and RX\,J2117.1+3412, thus 
increasing up to 10 the number of CSPNe that show this line.  
Only the Ne~{\sc vii} P~Cygni profiles of NGC\,2371-2 and LSS\,1362 are 
saturated, while it is unclear as for A\,43 and Lo\,4 due to the large 
number of narrow absorptions over-imposed on the profile;  
the remaining objects show unsaturated P~Cygni profiles.  
All the CSPNe with P~Cygni profile in the Ne~{\sc vii} line have 
$T_{\rm eff}>$100,000~K (Figure~\ref{teff_line}), as such a large 
effective temperature is needed to produce a significant flux of 
photons with energies above the IP of 0.16 keV, i.e., in the X-ray 
domain.  
The photospheric emission of K\,1-16, NGC\,246, NGC\,2371-2, 
NGC\,7094, and RX\,J2117.4+3412 is detected in soft X-rays 
\citep{MWP93,Hoare_etal95,Kastner_etal12}, whereas the central 
star of A\,30 also exhibits soft X-ray emission, although its 
nature is uncertain \citep{Guerrero_etal12}.  
We also note that all the stars that show Ne~{\sc vii} P~Cygni profiles 
are CSPNe of the type PG\,1159, with LSS\,1362 being the only exception.  
Otherwise, several stars hotter than 100,000~K do not show 
clear evidences of the Ne~{\sc vii} line: A\,7, A\,39, 
DeHt\,2, NGC\,1360, NGC\,2867, NGC\,7293, NGC\,7662, and 
PB\,6.  
This may suggest that the occurrence of the Ne~{\sc vii} feature in the 
spectrum of a CSPN requires both high temperature (ionisation effect) 
and hidrogen deficiency (abundance effect) conditions at the surface of 
the star.  
A high neon abundance linked to hydrogen deficiency would provide 
support to our recent suggestion that [WC] stars and born-again CSPNe 
may be related with Ne novae \citep{LdML11}. 
%
%
%
%

\begin{table*}
\centering
\scriptsize{
\caption{\footnotesize{
FUSE lines with P~Cygni profiles in CSPNe}} 

\label{UV_lines}
\begin{tabular}{lrccrl}
\hline\hline

\multicolumn{1}{l}{Ion}         & 
\multicolumn{1}{c}{$\lambda_0$} & 
\multicolumn{1}{c}{Transition}  & 
\multicolumn{1}{c}{Term}        & 
\multicolumn{1}{c}{Transition Probability} & 
\multicolumn{1}{l}{Note}        \\
\multicolumn{1}{c}{}            & 
\multicolumn{1}{c}{[\AA]}       & 
\multicolumn{1}{c}{}            & 
\multicolumn{1}{c}{}            & 
\multicolumn{1}{c}{[s]}         & 
\multicolumn{1}{c}{}            \\

\hline

C~{\sc iii}  &977.03\,\,& 1s$^2$2s$^2$ -- 1s$^2$2s2p & $^1$S -- $^1$P$^0$     & $1.7\times 10^9$~~~~~~~~ & Resonance line \\
C~{\sc iii}  & 1174.933 & 1s$^2$2s2p -- 1s$^2$2p$^2$ & $^3$P$^1$ -- $^3$P$^2$ & $3.3\times 10^8$~~~~~~~~ & Multiplet \\ 
C~{\sc iii}  & 1175.263 & 1s$^2$2s2p -- 1s$^2$2p$^2$ & $^3$P$^0$ -- $^3$P$^1$ & $4.4\times 10^8$~~~~~~~~ & Multiplet \\ 
C~{\sc iii}  & 1175.590 & 1s$^2$2s2p -- 1s$^2$2p$^2$ & $^3$P$^1$ -- $^3$P$^1$ & $3.3\times 10^8$~~~~~~~~ & Multiplet \\ 
C~{\sc iii}  & 1175.711 & 1s$^2$2s2p -- 1s$^2$2p$^2$ & $^3$P$^2$ -- $^3$P$^2$ & $9.9\times 10^8$~~~~~~~~ & Multiplet \\ 
C~{\sc iii}  & 1175.987 & 1s$^2$2s2p -- 1s$^2$2p$^2$ & $^3$P$^1$ -- $^3$P$^0$ & $1.3\times 10^9$~~~~~~~~ & Multiplet \\
C~{\sc iii}  & 1176.370 & 1s$^2$2s2p -- 1s$^2$2p$^2$ & $^3$P$^2$ -- $^3$P$^1$ & $5.5\times 10^8$~~~~~~~~ & Multiplet \\ 
%
%
%
Ne~{\sc vii} &  973.302 & 1s$^2$2s2p -- 1s$^2$2p$^2$ & $^1$P$^0$ -- $^1$D     & $\dots$~~~~~~~~~~~~~~  & Resonance line          \\ 
O~{\sc vi}   & 1031.926 & 1s$^2$2s -- 1s$^2$2p       & $^2$S -- $^2$P$^0$     & $4.2\times 10^8$~~~~~~~~ & Resonance doublet \\
O~{\sc vi}   & 1037.617 & 1s$^2$2s -- 1s$^2$2p       & $^2$S -- $^2$P$^0$     & $4.1\times 10^8$~~~~~~~~ & Resonance doublet \\
P~{\sc v}    & 1117.977 & 2p$^6$3s -- 2p$^6$3p       & $^2$S -- $^2$P$^0$     & $1.2\times 10^9$~~~~~~~~ & Resonance doublet \\
P~{\sc v}    & 1128.008 & 2p$^6$3s -- 2p$^6$3p       & $^2$S -- $^2$P$^0$     & $1.2\times 10^9$~~~~~~~~ & Resonance doublet \\
S~{\sc iv}   & 1072.973 & 3s$^2$3p -- 3s3p$^2$       & $^2$P$^0$ -- $^2$D     & $2.0\times 10^9$~~~~~~~~ & Fine-structure line \\
S~{\sc iv}   & 1073.518 & 3s$^2$3p -- 3s3p$^2$       & $^2$P$^0$ -- $^2$D     & $3.3\times 10^8$~~~~~~~~ & Fine-structure line \\
%
S~{\sc vi}   &  933.376 & 2p$^6$3s -- 2p$^6$3p       & $^2$S -- $^2$P$^0$     & $1.7\times 10^9$~~~~~~~~ & Resonance doublet \\
S~{\sc vi}   &  944.525 & 2p$^6$3s -- 2p$^6$3p       & $^2$S -- $^2$P$^0$     & $1.6\times 10^9$~~~~~~~~ & Resonance doublet \\
%
%
Si~{\sc iii} & 1108.358 & 3s3p -- 3s3d              & $^3$P$^0$ -- $^3$D     & $1.5\times 10^9$~~~~~~~~ & Triplet \\
Si~{\sc iii} & 1109.970 & 3s3p -- 3s3d              & $^3$P$^0$ -- $^3$D     & $2.1\times 10^9$~~~~~~~~ & Triplet \\
Si~{\sc iii} & 1113.230 & 3s3p -- 3s3d              & $^3$P$^0$ -- $^3$D     & $2.7\times 10^9$~~~~~~~~ & Triplet \\
Si~{\sc iv}  & 1122.485 & 2p$^6$3p -- 2p$^6$3d      & $^2$P$^0$ -- $^2$D     & $2.1\times 10^9$~~~~~~~~ & Doublet \\
Si~{\sc iv}  & 1128.340 & 2p$^6$3p -- 2p$^6$3d      & $^2$P$^0$ -- $^2$D     & $2.5\times 10^9$~~~~~~~~ & Doublet \\

\hline
\end{tabular}}
\end{table*}

The most common P Cygni profiles in the \emph{FUSE} spectral range 
among CSPNe are these of the O~{\sc vi} $\lambda\lambda$1031.9,1037.6 
(IP: 113.9--138.1 eV) resonance doublet, i.e., lines resulting from 
transitions to the ground level.  
While the occurrence of this doublet is widespread among our sample of 
CSPNe (Table~\ref{CSPN_FUSE_lines}), with BD+30$^\circ$3639 being the 
only one that does not display a P~Cygni profile of these lines, 
there is still a dependence of this line with stellar effective 
temperature:  
not a single CSPN in our sample with $T_{\rm eff}<50,000$~K shows 
saturated O~{\sc vi} P~Cygni profiles, while it is saturated for most 
CSPNe with $T_{\rm eff}>50,000$~K.  
The chemical composition of the star may also play a role; [WR] 
CSPNe, with larger abundances of oxygen in their stellar winds, 
show a significant fraction of saturated O~{\sc vi} profiles.  
We note, however, that their effective temperatures are greater than 
50,000~K in most of the cases and so the possible effects of oxygen 
abundances are masked by the prevalence of effective temperature.  
Interestingly, the O~{\sc vi} P~Cygni profiles of four hot CSPNe 
(K\,1-16, LSS\,1362, NGC\,7662, and RX\,J2117.1$+$3412) are not 
saturated.  
These trends suggest that O~{\sc vi} is the dominant ionization stage 
for CSPNe with effective temperatures in the range between 50,000~K 
and 100,000~K.  
This is unclear for CSPNe with effective temperatures above 100,000~K 
because the unsaturated O~{\sc vi} lines may result either from 
ionization stages above O~{\sc vi} being dominant (the IP of O~{\sc vii} 
is 138.12 eV) or from the reduced optical thickness of the wind resulting 
from their low $\dot M$.  
Finally, we also note that the measurement of terminal velocities greater 
than $\sim$1000 km~s$^{-1}$ from the O~{\sc vi} lines is complicated by 
the separation of the doublet, $\sim$1650 km~s$^{-1}$, and by the blend 
with the interstellar Ly$\beta$ line, $\sim$1800 km~s$^{-1}$ bluewards 
of the O~{\sc vi} blue component at 1031.6 \AA.   

%
%

The S~{\sc vi} $\lambda\lambda$933.4,944.5 (IP: 72.6--88.1 eV) resonance 
doublet is present in the shortward region of the \emph{FUSE} spectra of 
15 CSPNe.  
Since the IP of S~{\sc vi} is smaller than that of O~{\sc vi}, the 
S~{\sc vi} doublet is present in CSPNe with effective temperatures 
lower ($T_{\rm eff}<87,000$~K) than those of the CSPNe with O~{\sc vi} 
lines (Figure~\ref{teff_line}).  
We note that the spectral region of the S~{\sc vi} doublet is affected by a 
myriad of H~{\sc i} Lyman, He~{\sc ii} Balmer, and very particularly nebular 
and interstellar H$_2$ lines.  
Despite this difficulty, the similarities between the P~Cygni 
profiles of the two components complement well each other, so 
that it lends weight to the determination of the line profile.  
Owing very likely to the sulfur abundance in the stellar wind being lower 
than that of oxygen, the line is saturated just in three CSPNe (IC\,2501, 
NGC\,5882, and NGC\,6572) out of the 15 displaying this line.

It is worth mentioning that these three ions, Ne~{\sc vii}, O~{\sc vi} 
and S~{\sc vi}, have IPs well above that of He~{\sc ii} (IP: 54.4 eV).
Because the large bound-free absorption of He~{\sc i} of the 
ionizing stellar continuum flux shortward of 228 \AA\ (i.e., 
with energies greater than 54.4 eV), the so-called ``super 
ions'' \citep{Massa_etal03} with IPs above 54.4 eV are not 
expected to be abundantly produced by photoionizations from 
the ground state in massive OB stars.  
Thus, when ``super-ions'' are present in these stars, it is believed that 
they are produced by Auger ionizations of X-ray photons generated by shocks 
in the stellar wind \citep{CO79}.  
Note, however, that CSPNe may have much larger effective temperatures 
than OB stars, and thus they can produce a large fraction of ionizing 
photons with energies above the IP of He~{\sc ii}.  
Indeed, the occurrence of the Ne~{\sc vii} or S~{\sc vi} lines 
is clearly limited to CSPNe with effective temperatures above 
or below 100,000~K, respectively.  
On the other hand, the O~{\sc vi} line is present over the entire 
range of effective temperatures.  
This issue will be discussed into more detail in section $\S$6.3.

The P~Cygni profile of the P~{\sc v} $\lambda\lambda$1118.0,1128.0 (IP: 
51.4--65.0 eV) resonance doublet is typically unsaturated, given the 
relatively low phosphorus abundances.  
The red component of the doublet is usually blended with the Si~{\sc iv} 
$\lambda$1128.3 line, and thus we have focused on the blue component.  
The P~Cygni profiles of unsaturated P~{\sc v} $\lambda$1118,1128 lines 
have been singled out as a sensitive diagnostic of variable conditions 
in the wind of CSPNe \citep{Prinja_etal07}.

The S~{\sc iv} $\lambda\lambda$1073.0,1073.5 lines (IP: 34.8--47.2 eV) are 
two fine-structure lines.  
The strength of the blue component is $\sim$9 times that of the 
red component (Table~\ref{UV_lines}), and thus the blue component 
is expected to dominate the line profile.  
Although the occurrence of S~{\sc iv} and S~{\sc vi} lines overlaps over 
a wide range of stellar temperatures (Figure~\ref{teff_line}), there are 
only two CSPNe, IC\,4593 and PB\,8, that display P~Cygni profiles of both 
species.

The Si~{\sc iv} $\lambda\lambda$1122.5,1128.3 doublet (IP: 33.5--45.1 eV) 
is found among a significant number of CSPNe, all of them also presenting 
the P~{\sc v} doublet.  
As referred above, the red components of the Si~{\sc iv} and P~{\sc v} 
doublets are blended and these have not been used in our study.  
The Si~{\sc iv} $\lambda$1122 is a line from a radiatively excited state.  
As the lower level of an excited state line is the upper level of a 
resonance line transition, its population depends strongly on the 
local radiation field and decreases rapidly with radial distance 
\citep{O81}.

The C~{\sc iii} $\lambda$1176 (IP: 24.4--47.9 eV) is a triplet which 
has 5 separate, closely spaced levels, resulting in a multiplet composed 
of at least 6 bright lines.  
This line can act much like a resonance line in dense winds, 
scattering radiation in any region wherever C$^{++}$ is present.  
A prominent P~Cygni feature is detected at the bluewards edge of the 
line in [WR] central stars of BD+30$^\circ$3639, NGC\,40 and PB\,8, and 
the WELS star of Sp\,3.  
The origin of this line is uncertain, but it can be attributed to 
the C~{\sc iv} doublet $\lambda\lambda$1168.9,1169.0, or to the 
N~{\sc iv} $\lambda\lambda$1168.6,1169.1,1169.5 multiplet.  
The C~{\sc iii} $\lambda$977 resonance line is also present in some 
CSPNe, mostly (but not always) when the C~{\sc iii} $\lambda$1176 
is detected.

Finally, the \emph{FUSE} spectra of CSPNe also include the Si~{\sc iii} 
$\lambda\lambda$1108.4,1110.0,1113.2 triplet (IP: 16.3--33.5 eV).  
These lines are only found in the CSPNe of Hen\,2-131, Hen\,21-138, 
and SwSt\,1, stars with $T_{\rm eff}\leq$40,000~K, in correspondence 
with the low IP of this ion.  
Interestingly, these CSPNe also show Si~{\sc iv} P~Cygni profiles, 
thus indicating that both species are simultaneously present in 
their stellar winds.

\section{Terminal wind velocities}

The normalized spectra shown in Figures~\ref{fig_1a}-\ref{fig_1d} can be 
used to investigate the wind terminal velocities, $v_\infty$, of the CSPNe 
with \emph{FUSE} P~Cygni profiles.  
Following the prescriptions by \citet{PBH90}, we have determined 
the values of two observational indicators of $v_\infty$: 
(1) the black velocity, $v_{\rm black}$, the velocity of the blueward 
edge of the absorption region that has zero intensity, and 
(2) the edge velocity, $v_{\rm edge}$, the velocity at which the profile 
rejoins the continuum level.  
In many cases, $v_{\rm black}$ could not be determined because the 
line profile was not saturated.  
For these cases, we have defined a new observational indicator of 
$v_\infty$, the grey velocity, $v_{\rm grey}$, as the velocity of the 
blueward edge of the absorption region that has the lowest intensity.

The values of these velocities are listed in Table~\ref{FUSE_vel}.  
From the comparison between velocities derived from lines of a 
doublet (e.g., the O~{\sc vi} $\lambda$1031.9 and $\lambda$1037.6 lines, 
or the S~{\sc vi} $\lambda$933.4 and $\lambda$944.5 lines), the typical 
uncertainty in these velocities is estimated to be 10--40 km~s$^{-1}$, 
depending on the signal-to-noise ratio, precise line shape, and blending 
with narrow telluric, H$_2$, and atomic line absorptions.  
In certain cases, the blue edge of the line is heavily affected by these 
absorption lines, hampering the determination of the edge velocity.  
When this was the case, either an upper or a lower limit 
has been provided for the edge velocity.

\begin{table*}
\centering
\scriptsize{
\caption{\footnotesize{
Stellar wind velocities for CSPNe}} 

\label{FUSE_vel}
\begin{tabular}{llrrr|llrrr}
\hline\hline

\multicolumn{1}{c}{Name}             & 
\multicolumn{1}{c}{Line}             & 
\multicolumn{1}{c}{$v_{\rm edge}$}  & 
\multicolumn{1}{c}{$v_{\rm grey}$}  & 
\multicolumn{1}{c}{$v_{\rm black}$} & 
\multicolumn{1}{c}{Name}             & 
\multicolumn{1}{c}{Line}             & 
\multicolumn{1}{c}{$v_{\rm edge}$}  & 
\multicolumn{1}{c}{$v_{\rm grey}$}  & 
\multicolumn{1}{c}{$v_{\rm black}$} \\ 
\multicolumn{1}{c}{}                 & 
\multicolumn{1}{c}{}                 & 
\multicolumn{1}{c}{[km~s$^{-1}$]}   & 
\multicolumn{1}{c}{[km~s$^{-1}$]}   & 
\multicolumn{1}{c}{[km~s$^{-1}$]}   & 
\multicolumn{1}{c}{}                 & 
\multicolumn{1}{c}{}                 & 
\multicolumn{1}{c}{[km~s$^{-1}$]}   & 
\multicolumn{1}{c}{[km~s$^{-1}$]}   & 
\multicolumn{1}{c}{[km~s$^{-1}$]}   \\

\hline

A\,30              & Ne~{\sc vii} &    2975~~~ &    1950~~~ & $\dots$~~~ & NGC\,40            & C~{\sc iii} $\lambda$977  & $\dots$~~~ & $\dots$~~~ &   910~~~ \\   
                   & O~{\sc vi}   &    4025~~~ & $\dots$~~~ &    2675~~~ &                    & C~{\sc iii} $\lambda$1176 &    1400~~~ & $\dots$~~~ &  1150~~~ \\   
A\,43              & Ne~{\sc vii} & $<$3275~~~ & $\dots$~~~ &    3000~~~ &                    & O~{\sc vi}   &    1600~~~ & $\dots$~~~ &    1280:~~ \\              
                   & O~{\sc vi}   &    3825~~~ & $\dots$~~~ &    3600~~~ &                    & P~{\sc v}    & $>$1400~~~ &     980~~~ & $\dots$~~~ \\              
A\,78              & Ne~{\sc vii} & $<$3125~~~ &    3025~~~ & $\dots$~~~ &                    & S~{\sc iv}   &    1330~~~ & $\dots$~~~ &    1000~~~ \\              
                   & O~{\sc vi}   &    4025~~~ & $\dots$~~~ &    3450~~~ &                    & Si~{\sc iv}  &     970~~~ &     800~~~ & $\dots$~~~ \\              
BD+30$^\circ$3639  & C~{\sc iii} $\lambda$1176 & $\dots$~~~ & $\dots$~~~ &     710~~~ & NGC\,246           & Ne~{\sc vii} &    3170~~~ &    2150~~~ & $\dots$~~~ \\
                   & P~{\sc v}    &     800~~~ &     420~~~ & $\dots$~~~ &                    & O~{\sc vi}   &    3580~~~ & $\dots$~~~ &    3390~~~ \\
                   & Si~{\sc iv}  &     770~~~ &     640~~~ & $\dots$~~~ & NGC\,1535          & O~{\sc vi}   & $>$2080~~~ &    1825~~~ & $\dots$~~~ \\
                   & S~{\sc iv}   &  $>$740~~~ & $\dots$~~~ &     630~~~ & NGC\,2371-2        & Ne~{\sc vii} &    3030~~~ & $\dots$~~~ &    2550~~~ \\
Cn\,3-1            & C~{\sc iii} $\lambda$1176 &     690~~~ &     370~~~ & $\dots$~~~ &       & O~{\sc vi}   &    3990~~~ & $\dots$~~~ &    3500~~~ \\
                   & P~{\sc v}    &     370~~~ &     180~~~ & $\dots$~~~ & NGC\,2392          & C~{\sc iii} $\lambda$977  &  $>$100~~~ & $\dots$~~~ &  65~~~ \\   
                   & Si~{\sc iv}  &     370~~~ & $\dots$~~~ & $\dots$~~~ &                    & O~{\sc vi}   &  $>$270~~~ & $\dots$~~~ &      90~~~ \\             
                   & S~{\sc iv}   &  $>$480~~~ & $\dots$~~~ &     380~~~ &                    & P~{\sc v}    &     140~~~ & $\dots$~~~ & $\dots$~~~ \\            
Hb\,7              & O~{\sc vi}   & $>$1540~~~ &    1360~~~ & $\dots$~~~ &                    & S~{\sc vi}   &     330~~~ & $\dots$~~~ & $\dots$~~~ \\            
                   & S~{\sc vi}   &    1170~~~ &     860~~~ & $\dots$~~~ &                    & Si~{\sc iv}  &  $<$160~~~ & $\dots$~~~ & $\dots$~~~ \\  
Hen\,2-99          & C~{\sc iii} $\lambda$1176 &    1230~~~ & $\dots$~~~ &    910~~~ & NGC\,2867          & O~{\sc vi}   &    2560~~~ & $\dots$~~~ &    2110~~~ \\
                   & O~{\sc vi}   & $\dots$~~~ & $\dots$~~~ &     980:~~ & NGC\,5882          & O~{\sc vi}   & $<$2630~~~ & $\dots$~~~ &    1720~~~ \\
                   & P~{\sc v}    &    1070~~~ &     590~~~ & $\dots$~~~ &                    & S~{\sc vi}   &    2250~~~ & $\dots$~~~ &    1870~~~ \\ 
                   & S~{\sc iv}   &  $>$910~~~ & $\dots$~~~ &     830~~~ & NGC\,6058          & O~{\sc vi}   &    2740~~~ & $\dots$~~~ &    2600~~~ \\  
                   & Si~{\sc iv}  &     810~~~ &     720~~~ & $\dots$~~~ & NGC\,6210          & O~{\sc vi}   &    2530~~~ &    2100~~~ & $\dots$~~~ \\
Hen\,2-131         & C~{\sc iii} $\lambda$977  & $\dots$~~~ & $\dots$~~~ &    340~~~ &        & S~{\sc vi}   &    2290~~~ & $\dots$~~~ &    1910~~~ \\
                   & C~{\sc iii} $\lambda$1176 &  $>$680~~~ & $\dots$~~~ &    390~~~ & NGC\,6572          & C~{\sc iii} $\lambda$1176 &     600~~~ &     520~~~ & $\dots$~~~ \\

                   & O~{\sc vi}   & $\dots$~~~ &     560~~~ & $\dots$~~~ &                    & O~{\sc vi}   &    1420~~~ & $\dots$~~~ &    1190~~~ \\
                   & P~{\sc v}    &     330~~~ &     170~~~ & $\dots$~~~ &                    & P~{\sc v}    &    1210~~~ &    1160~~~ & $\dots$~~~ \\
                   & S~{\sc iv}   &  $>$480~~~ & $\dots$~~~ &     340~~~ &                    & S~{\sc vi}   &    1600~~~ & $\dots$~~~ &    1140~~~ \\
                   & Si~{\sc iii} &  $>$290~~~ &     210~~~ & $\dots$~~~ & NGC\,6826          & C~{\sc iii} $\lambda$977  &    1200~~~ &     760~~~ & $\dots$~~~ \\ 
                   & Si~{\sc iv}  &     330~~~ &     180~~~ & $\dots$~~~ &                    & O~{\sc vi}   &    1230~~~ & $\dots$~~~ &    1100~~~ \\   
IC\,418            & C~{\sc iii} $\lambda$977  & $\dots$~~~ & $\dots$~~~ &    510~~~ &        & P~{\sc v}    &    1290~~~ &    1080~~~ & $\dots$~~~ \\
                   & C~{\sc iii} $\lambda$1176 &     630~~~ & $\dots$~~~ &    390~~~ &        & S~{\sc vi}   & $>$1210~~~ &     710~~~ & $\dots$~~~ \\
                   & O~{\sc vi}   &  $>$900~~~ &     530~~~ & $\dots$~~~ & NGC\,6891          & O~{\sc vi}   &    1420~~~ &    1230~~~ & $\dots$~~~ \\
                   & P~{\sc v}    &  $>$310~~~ &     190~~~ & $\dots$~~~ &                    & P~{\sc v}    &    1260~~~ &     910~~~ & $\dots$~~~ \\
                   & S~{\sc iv}   &  $>$480~~~ &     420~~~ & $\dots$~~~ &                    & S~{\sc vi}   &  $>$990~~~ & $\dots$~~~ &     950~~~ \\
                   & Si~{\sc iv}  &     120~~~ &     100~~~ & $\dots$~~~ & NGC\,7009          & O~{\sc vi}   & $>$2870~~~ & $\dots$~~~ &    2540~~~ \\
IC\,2149           & C~{\sc iii} $\lambda$977  & $>$1030~~~ & $\dots$~~~ &     930~~~ & NGC\,7094          & Ne~{\sc vii} &    3330~~~ &    1510~~~ & $\dots$~~~ \\
                   & C~{\sc iii} $\lambda$1176 &    1030~~~ &     180~~~ & $\dots$~~~ &                    & O~{\sc vi}   &    3790~~~ & $\dots$~~~ &    3610~~~ \\
                   & O~{\sc vi}   &    1190~~~ & $\dots$~~~ &     890~~~ & NGC\,7662          & O~{\sc vi}   & $<$2580~~~ &    1640~~~ & $\dots$~~~ \\
                   & P~{\sc v}    &     870~~~ & $\dots$~~~ &     800~~~ & PB\,6              & O~{\sc vi}   &    3140~~~ & $\dots$~~~ &    2980~~~ \\
IC\,2448           & O~{\sc vi}   &    2470~~~ & $\dots$~~~ &    2280~~~ & PB\,8              & C~{\sc iii} $\lambda$1176 & $\dots$~~~ &    1140~~~ &  720~~~ \\  
IC\,2501           & O~{\sc vi}   &    1400~~~ & $\dots$~~~ &    1200~~~ &                    & O~{\sc vi}   &    1300~~~ & $\dots$~~~ &    1000~~~ \\            
                   & P~{\sc v}    &    1320~~~ &    1080~~~ & $\dots$~~~ &                    & P~{\sc v}    &    1240~~~ & $\dots$~~~ &     790~~~ \\            
                   & S~{\sc vi}   &    1430~~~ & $\dots$~~~ &    1220~~~ &                    & S~{\sc iv}   &    1240~~~ &     700~~~ & $\dots$~~~ \\            
IC\,2553           & O~{\sc vi}   &    2700~~~ & $\dots$~~~ &    2450~~~ &                    & S~{\sc vi}   &    1010~~~ &     750~~~ & $\dots$~~~ \\            
IC\,3586           & O~{\sc vi}   & $>$1700~~~ &    1640~~~ & $\dots$~~~ &                    & Si~{\sc iv}  & $\dots$~~~ &     470~~~ & $\dots$~~~ \\ 
                   & S~{\sc vi}   &    1780~~~ & $\dots$~~~ &    1580~~~ & RX\,J2117.1+3412   & Ne~{\sc vii} &    3490~~~ &    3030~~~ & $\dots$~~~ \\
IC\,4593           & C~{\sc iii} $\lambda$977  &     960~~~ & $\dots$~~~ &    820~~~ &                    & O~{\sc vi}   &    3650~~~ &    3140~~~ & $\dots$~~~ \\
                   & C~{\sc iii} $\lambda$1176 &     720~~~ &     420~~~ & $\dots$~~~ & Sp\,3              & O~{\sc vi}   & $>$1850~~~ & $\dots$~~~ &    1680~~~ \\
                   & O~{\sc vi}   &    1150~~~ &     860~~~ & $\dots$~~~ &                    & S~{\sc vi}   &    1940~~~ &    1680~~~ & $\dots$~~~ \\
                   & P~{\sc v}    &     920~~~ &     480~~~ & $\dots$~~~ & SwSt\,1            & C~{\sc iii} $\lambda$1176 &    1500~~~ & $\dots$~~~ &  770~~~ \\
                   & S~{\sc iv}   &    1020~~~ &     440~~~ & $\dots$~~~ &                    & O~{\sc vi}   &    1220~~~ &     880~~~ & $\dots$~~~ \\
                   & S~{\sc vi}   &     750~~~ &     585~~~ & $\dots$~~~ &                    & P~{\sc v}    &  $>$760~~~ &     400~~~ & $\dots$~~~ \\
IC\,4776           & O~{\sc vi}   &    2060~~~ & $\dots$~~~ &    1840~~~ &                    & S~{\sc iv}   &     770~~~ & $\dots$~~~ &     600~~~ \\
                   & S~{\sc vi}   &    1960~~~ &    1820~~~ & $\dots$~~~ &                    & Si~{\sc iii} &     610~~~ &     450~~~ & $\dots$~~~ \\
IC\,5217           & O~{\sc vi}   &    2530~~~ & $\dots$~~~ &    2360~~~ &                    & Si~{\sc iv}  &  $>$550~~~ &     330~~~ & $\dots$~~~ \\
K\,1-16            & Ne~{\sc vii} & $>$3000~~~ &    2410~~~ & $\dots$~~~ & Tc\,1              & C~{\sc iii} $\lambda$977  &     900~~~ & $\dots$~~~ &    1050~~~ \\
                   & O~{\sc vi}   &    3670~~~ &    3550~~~ & $\dots$~~~ &                    & C~{\sc iii} $\lambda$1176 &     790~~~ &     540~~~ & $\dots$~~~ \\
Lo\,4              & Ne~{\sc vii} & $>$3100~~~ &    3000~~~ & $\dots$~~~ &                    & O~{\sc vi}   &     950~~~ &     510~~~ & $\dots$~~~ \\
                   & O~{\sc vi}   &    3830~~~ &    3430~~~ & $\dots$~~~ &                    & S~{\sc iv}   &     440~~~ &     390~~~ & $\dots$~~~ \\ 
LSS\,1362          & Ne~{\sc vii} &  $>$520~~~ & $\dots$~~~ &     480~~~ & Vy\,2-3            & O~{\sc vi}   &    1740~~~ &    1460~~~ & $\dots$~~~ \\
                   & O~{\sc vi}   &  $>$500~~~ & $\dots$~~~ & $\dots$~~~ &                    & S~{\sc vi}   &    1700~~~ & $\dots$~~~ &    1460~~~ \\

\hline
\end{tabular}}
\end{table*}

An inspection of Table~\ref{FUSE_vel} reveals that the estimates 
of $v_{\rm edge}$, $v_{\rm grey}$, and $v_{\rm black}$ implied from 
different lines can differ significantly from one line to another 
for a given CSPN, although these differences tend to be smaller 
for estimates of $v_{\rm black}$ from different lines.  
This is not completely unexpected as the atlas of \emph{FUSE} P~Cygni 
profiles of CSPNe in Figures~\ref{fig_1a}-\ref{fig_1d} reveals a 
variety of P~Cygni profiles for the same CSPN.  
It is clear that these differences can result in uncertain estimates 
of the wind velocity.

As discussed by \citet{Abbott78} and \citet{PBH90}, $v_{\rm black}$ can 
be expected to provide a value to $v_\infty$ closer than $v_{\rm edge}$.   
According to Table~\ref{FUSE_vel}, $v_{\rm edge}$ is typically 150-200 
km~s$^{-1}$ bluer than $v_{\rm black}$, with some objects and lines for 
which the deviations are much greater.  
If we exclude objects with deviations greater than 1,000 km~s$^{-1}$, 
the best linear fit between these two velocities is: 
\begin{equation}
v_{\rm black} = 
(0.939\pm0.017) \times v_{\rm edge} - (175\pm32)  \;\; [{\rm km~s}^{-1}]. 
\end{equation} 
which allows us to obtain an accurate estimate of $v_{\rm black}$ and 
$v_\infty$ from $v_{\rm edge}$.

We note, however, that the prescription that equals $v_{\rm black}$ to 
$v_\infty$ holds in the assumption that the P~Cygni profile is saturated.  
Unfortunately, this is not the case for many of the P~Cygni 
profiles shown by CSPNe in the spectral range of \emph{FUSE}.  
In such a case, it is pertinent to investigate the correlation 
between $v_{\rm grey}$ and $v_{\rm edge}$.  
By definition, $v_{\rm grey}$ is always redder than $v_{\rm edge}$, 
as it is also the case for $v_{\rm black}$.  
A linear fit also excluding objects with deviations greater than 
1,000 km~s$^{-1}$ provides the following relation between these 
two velocities: 
\begin{equation}
v_{\rm grey} = 
(0.942\pm0.027) \times v_{\rm edge} - (210\pm50)  \;\; [{\rm km~s}^{-1}]. 
\end{equation} 
Such relationship is identical, within the uncertainties, to the 
relation between $v_{\rm black}$ and $v_{\rm edge}$, suggesting that 
$v_{\rm grey}$ can be used as a proxy of $v_{\rm black}$ in those cases 
when the line is not saturated.

In the following, we will adopt the terminal velocity of the wind 
to be $v_{\rm black}$ of saturated P~Cygni profiles when they are 
available.  
In these cases, the values of $v_{\rm grey}$ derived from 
unsaturated P~Cygni profiles, such as those of the 
Si~{\sc iv} $\lambda\lambda$1122.5,1128.3, and P~{\sc v} 
$\lambda\lambda$1118.0,1128.0, have not been used to 
determine $v_\infty$.  
Typically, the O~{\sc vi} $\lambda\lambda$1031.9,1037.6 lines, when 
saturated, have been used to derive $v_{\rm black}$ and these values 
have been assigned to $v_\infty$.  
In a number of cases, other saturated lines have been used instead: 
S~{\sc iv}\footnote{
There is, however, a caveat as the wavelength of the line has been 
assumed to be that of the dominant blue component.  } 
(BD+30$^\circ$3639, Cn\,3-1, Hen\,2-99, Hen\,2-131, IC\,2501, and 
NGC\,40), 
C~{\sc iii} $\lambda$977 (IC\,418, IC\,2149, IC\,4593, and NGC\,2392), 
S~{\sc vi} $\lambda\lambda$933.4,944.5 (IC\,2501, IC\,3568, NGC\,6572, 
and Vy\,2-3), 
and Ne~{\sc vii} $\lambda$973 (LSS\,1362).  
The C~{\sc iii} $\lambda$1176 multiplet is typically saturated, but 
$v_{\rm black}$ is not easy to estimate because the contribution to 
the profile of line components bluewards of the brightest component 
at $\lambda$1175.711~\AA.  
To derive the value of $v_{\rm black}$ for this line, we have computed the 
value of $v_{\rm black}$ for each component of the multiplet and taken the 
weighted average using the transition probabilities as weights.  

The \emph{FUSE} spectra of Hb\,7, K\,1-16, Lo\,4, NGC\,1535, 
NGC\,7662, and RX\,J2117.1+3412 did not show any saturated 
P~Cygni profile.  
For those cases, we assumed $v_\infty$ to be $v_{\rm grey}$ of the 
most reliable line.  
The values of $v_\infty$ adopted from \emph{FUSE} $v_{\rm black}$ values 
are listed in column 2 of Table~\ref{FUSE_IUE_vel} together with those 
from \emph{IUE} data (column 3) available in the literature as compiled 
by \citet{PP91} or computed more recently by others 
\citep{PP96,Feibelman97,Feibelman98,Feibelman99,Gauba_etal01,Marcolino_etal07b} 
that assumed $v_\infty$ to be $v_{\rm edge}$. 
The terminal wind velocity derived from spectral fits is also 
listed in column 4, together with the references (column 5) 
from which those values have been compiled.

\begin{table}
\centering
\scriptsize{
\caption{\footnotesize{
CSPNe wind terminal velocities}}

\label{FUSE_IUE_vel}
\begin{tabular}{lrrrc}
\hline\hline

\multicolumn{1}{c}{}                        & 
\multicolumn{1}{c}{\emph{FUSE}}             &
\multicolumn{1}{c}{\emph{IUE}}              & 
\multicolumn{2}{c}{Model fit}               \\
\multicolumn{1}{c}{Name~~~~~~~~~~~~~~~~~~~} & 
\multicolumn{1}{c}{$v_{\rm black}$}          & 
\multicolumn{1}{c}{$v_{\rm edge}$}           & 
\multicolumn{1}{c}{$v_{\rm model}$}          & 
\multicolumn{1}{c}{Ref.}                   \\
\multicolumn{1}{c}{}                       & 
\multicolumn{1}{c}{[km~s$^{-1}$]}           & 
\multicolumn{1}{c}{[km~s$^{-1}$]}           & 
\multicolumn{1}{c}{[km~s$^{-1}$]}           & 
\multicolumn{1}{c}{}                       \\

\hline

A\,30              &    3000~~~ &    3400~~~ & 4000~~~ & 1 \\ 
A\,43              &    3000~~~ &$\dots$~~~~ & $\dots$~~~ & $\dots$ \\
A\,78              &    3450~~~ &    3500~~~ & 3200~~~ & 2 \\ 
BD+30$^\circ$3639  &     670~~~ &    1000~~~ &  700~~~ & 3 \\  
Cn\,3-1            &     380~~~ &     330~~~ & $\dots$~~~ & 3 \\  
Hb\,7              &    1360:~~ &    1320~~~ & $\dots$~~~ & 4 \\  
Hen\,2-99          &     870~~~ & 1000--2000 &  900~~~ & 5 \\  
Hen\,2-131         &     370~~~ &     850~~~ &  450~~~ & 6 \\  
Hen\,2-138         &   $\dots$  & $\dots$    &  300~~~ & 7 \\  
IC\,418            &     450~~~ &    1050~~~ &  700~~~ & 8 \\ 
IC\,2149           &     930~~~ &    1300~~~ & 1000~~~ & 8 \\ 
IC\,2448           &    2280~~~ &$\dots$~~~~ & 2000~~~ & 9 \\ 
IC\,2501           &    1210~~~ &    1280~~~ & $\dots$~~~ & $\dots$ \\
IC\,2553           &    2450~~~ &$\dots$~~~~ & $\dots$~~~ & $\dots$ \\
IC\,3568           &    1580~~~ &    1850~~~ & 1730~~~ & 10 \\ 
IC\,4593           &     820~~~ &    1100~~~ &  750~~~ & 9 \\  
IC\,4776           &    1840~~~ &    1760~~~ & 2300~~~ & 2,6 \\  
IC\,5217           &    2360~~~ &$\dots$~~~~ & $\dots$~~~ & $\dots$ \\
K\,1-16            &    3550:~~ &    3800~~~ & 4000~~~ & 11,12 \\  
Lo\,4              &    3430:~~ &$\dots$~~~~ & 4000~~~ & 1,13 \\  
LSS\,1362          &     480~~~ &$\dots$~~~~ & 2400~~~ & $\dots$ \\
NGC\,40            &    1030~~~ &    1600~~~ & 1000~~~ & 3 \\ 
NGC\,246           &    3390~~~ & $>$3300~~~ & 4000~~~ & 11 \\ 
NGC\,1535          &    1825:~~ &    2150~~~ & 1950~~~ & 2 \\ 
NGC\,2371-2        &    3500~~~ &    4500~~~ & 3700~~~ & 2,14 \\ 
NGC\,2392          &      65~~~ &     600~~~ &  300~~~ & 9 \\
NGC\,2867          &    2110~~~ &$\dots$~~~~ & 1800~~~ & 1,15 \\  
NGC\,5882          &    1720~~~ &    1525~~~ & $\dots$~~~ & $\dots$ \\
NGC\,6058          &    2600~~~ &$\dots$~~~~ & 2300~~~ & 9 \\
NGC\,6210          &    2100~~~ &    2350~~~ & 2150~~~ & 9 \\
NGC\,6572          &    1140~~~ &    1550~~~ & 1190~~~ & 10 \\  
NGC\,6543          &    $\dots$ & $\dots$    & 1500~~~ & 8 \\ 
NGC\,6826          &    1100~~~ &    1600~~~ & 1200~~~ & 6 \\ 
NGC\,6891          &     950~~~ &    1950~~~ & $\dots$~~~ & 16 \\ 
NGC\,7009          &    2540~~~ &    2750~~~ & 2770~~~ & 17,18 \\ 
NGC\,7094          &    3610~~~ &    3600~~~ & 3500~~~ & 11 \\ 
NGC\,7662          &    1640:~~ &$\dots$~~~~ & 2250~~~ & 9 \\
PB\,6              &    2980~~~ &$\dots$~~~~ & 3000~~~ & 1 \\ 
PB\,8              &    1000~~~ &    1060~~~ & 1000~~~ & 19 \\ 
RX\,J2117.1+3412   &    3140:~~ &$\dots$~~~~ & 3500~~~ & 11 \\ 
Sp\,3              &    1680~~~ &    1600~~~ & $\dots$~~~ & 4 \\ 
SwSt\,1            &     880~~~ &    1580~~~ &  900~~~ & 20 \\ 
Tc\,1              &    1050~~~ &    1530~~~ &  900~~~ & 6 \\ 
Vy\,2-3            &    1460~~~ & $\dots$~~~ & $\dots$~~~ & $\dots$ \\

\hline
\end{tabular}
\tablefoot{
The ``:'' sign indicates that these velocities have been derived 
from P~Cygni profiles that are not completely saturated.  
}
\tablebib{
(1) \citet{Koesterke01}; 
(2) \citet{HB04}; 
(3) \citet{Marcolino_etal07a}; 
(4) \citet{Gauba_etal01}; 
(5) \citet{LHJ96}; 
(6) \citet{PHM04}; 
(7) \citet{Prinja_etal10}; 
(8) \citet{Prinja_etal12}; 
(8) \citet{MPP93};
(9) \citet{HB11}; 
(10) \citet{MPP93};
(11) \citet{KW98}; 
(12) \citet{PP96}; 
(13) \citet{Feibelman99}; 
(14) \citet{Feibelman97}; 
(15) \citet{Feibelman98}; 
(16) \citet{McCarthy_etal90};
(17) \citet{C-SP89}; 
(18) \citet{Iping_etal06};
(19) \citet{Todt_etal10}; 
(20) \citet{DeMarco_etal01}. 
}
}
\end{table}

\begin{figure}
\centerline{
\includegraphics[bb=18 200 592 718,width=1.0\columnwidth,angle=0]{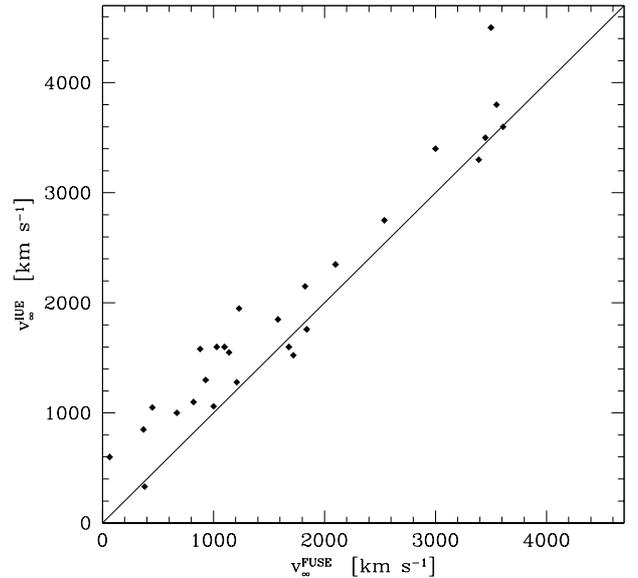}
}
\caption{
Comparison of \emph{FUSE} and \emph{IUE} terminal velocities for the CSPNe 
listed in Table~\ref{FUSE_IUE_vel}.  
The solid line corresponds to the 1:1 correspondence.  
}
\label{IUE_vs_FUSE}
\end{figure}

\begin{figure*}
\centerline{
\includegraphics[bb=18 350 592 718,width=1.75\columnwidth,angle=0]{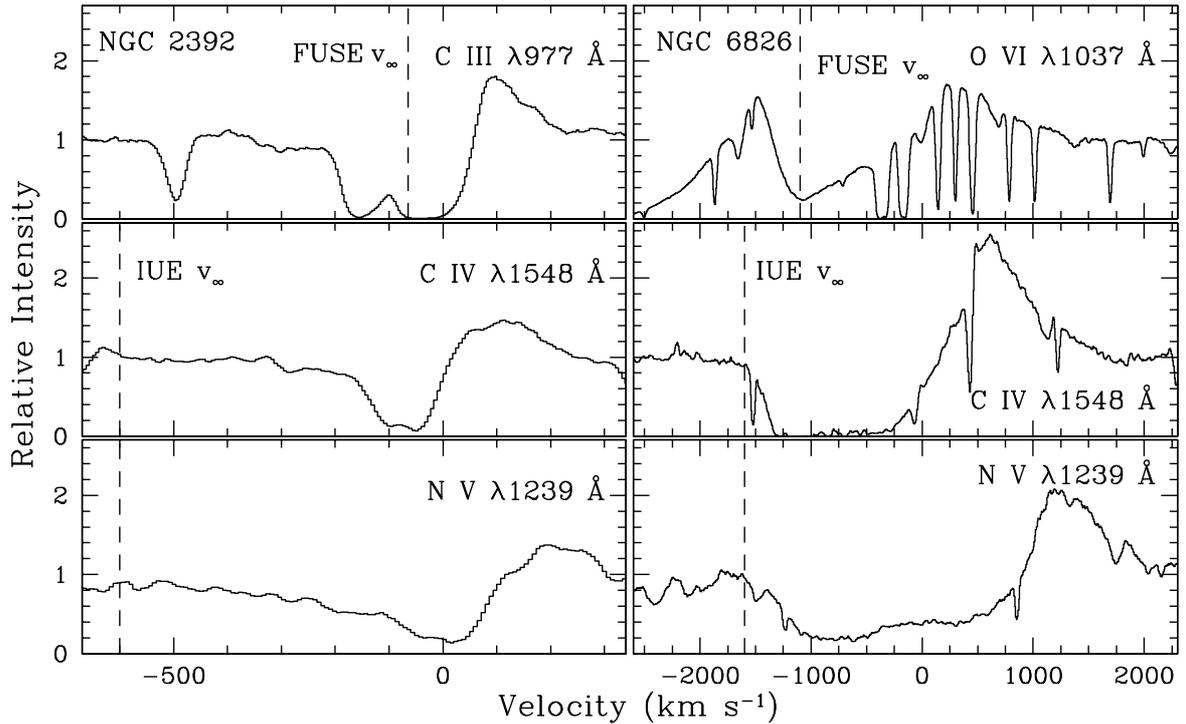}
}
\caption{
Comparison of \emph{FUSE} and \emph{IUE} terminal velocities in the CSPNe 
of NGC\,2392 ({\it left}) and NGC\,6826 ({\it right}).  
The top panels show the normalized P~Cygni profiles of the 
NGC\,2392 C~{\sc iii} $\lambda$977.0 and NGC\,6826 O~{\sc vi} 
$\lambda$1037.6 lines derived from \emph{FUSE} data.  
The middle and bottom panels show respectively the C~{\sc iv} 
$\lambda$1548.2 and N~{\sc v} $\lambda$1238.8 normalized P~Cygni 
profiles of NGC\,2392 ({\it left}) and NGC\,6826 ({\it right}) 
derived from \emph{IUE} data.  
The \emph{FUSE} and \emph{IUE} \citep{PP91} terminal velocities 
are shown by dashed vertical lines.  
}
\label{IUE_FUSE}
\end{figure*}

The comparison between \emph{FUSE} and \emph{IUE} terminal velocities 
(Table~\ref{FUSE_IUE_vel} and Figure~\ref{IUE_vs_FUSE}) implies that 
the latter are generally larger.  
The velocity difference is notably large, $\sim$500~km~s$^{-1}$, for objects 
with terminal velocities lower than $\sim$1000~km~s$^{-1}$ (Cn\,3-1 being the 
only notable exception), suggesting that the spectral resolution of 
\emph{IUE} is not sufficient to determine values of $v_\infty$ below 
1000~km~s$^{-1}$.  
The typical difference between \emph{FUSE} and \emph{IUE} terminal velocities 
is similar to that found between \emph{FUSE} black and edge velocities, as it 
can be expected because the \emph{IUE} velocities compiled by \citet{PP91} 
correspond to $v_{\rm edge}$ rather than to $v_{\rm black}$.  
This suggestion is further illustrated by Figure~\ref{IUE_FUSE} which 
compares our determination of $v_\infty$ for NGC\,2392 (65 km~s$^{-1}$) 
and NGC\,6826 (1100 km~s$^{-1}$) with those derived from \emph{IUE} data 
(600 and 1600 km~s$^{-1}$, respectively).  
It is clear that the \emph{IUE} velocities correspond to $v_{\rm edge}$ 
and therefore they must be discounted.  
This figure also sheds light on the large discrepancy between 
the \emph{IUE} and \emph{FUSE} terminal velocities of NGC\,2392, 
as the terminal velocity in the former data has been associated 
to a shallow trough extending up to $\sim$600 km~s$^{-1}$ in the 
N~{\sc v} $\lambda\lambda$1238.8,1242.8 and C~{\sc iv} 
$\lambda$1548.2 lines (but not in the C~{\sc iv} red component 
at $\lambda$1150.8 \AA).  

\section{Wind variability}

\subsection{Detecting variability in FUSE P Cygni profiles of CSPNe}

The search for time variability in the P~Cygni profiles of high-excitation UV 
lines in the \emph{FUSE} spectra of a CSPN has been performed by subtracting 
its ``master'' mean spectrum (Figures~\ref{fig_1a}-\ref{fig_1d}) from 
individual exposures and visually inspecting the residuals to search for 
meaningful differences.  
This process has resulted in the identification of definite 
variations
in the P~Cygni profiles of the 10 CSPNe listed as 
variable in Table~\ref{CSPN_FUSE_lines}, of which 3, namely 
Hen\,2-131, Sp\,3 and SwSt\,1, are found variable for the 
first time in this work.  
Since the line profile variations are not present in all of their P~Cygni 
profiles, the lines that are found to be variable have been especifically 
listed in this table as ``var''.

Table~\ref{CSPN_FUSE_lines} also includes information on 
the CSPNe that do not show variability.  
These CSPNe can be classified into three different groups.  
The first group (N=1 in Table~\ref{CSPN_FUSE_lines}) is composed 
of sources for which the number of individual spectra may be 
insufficient and/or the S/N ratio may be inadequate for the 
investigation of variability. 
The second group (N=2 in Table~\ref{CSPN_FUSE_lines}) includes objects 
whose P~Cygni profiles are strongly saturated or highly absorbed by 
multiple narrow lines, and thus they may be insensitive to line variations.  
It must be noted that the P~Cygni profiles of some variable 
sources, while similarly affected by saturation and/or narrow 
line absorption, showed undoubtful evidence of variability 
(e.g., NGC\,1535).  
Finally, the third group (N=3 in Table~\ref{CSPN_FUSE_lines}) consists 
of sources that, having suitable P~Cygni profiles for their study and a 
sufficient number of spectra with adequate S/N ratio, lack variability 
in the particular \emph{FUSE} observations presented in this paper.  
The latter are ``bona-fide'' candidates to non-variable winds, 
although the fragmented datasets here presented cannot absolutely 
preclude their variability.

\begin{figure*}
\centerline{
\includegraphics[bb=21 150 585 710,width=1.90\columnwidth,angle=0]{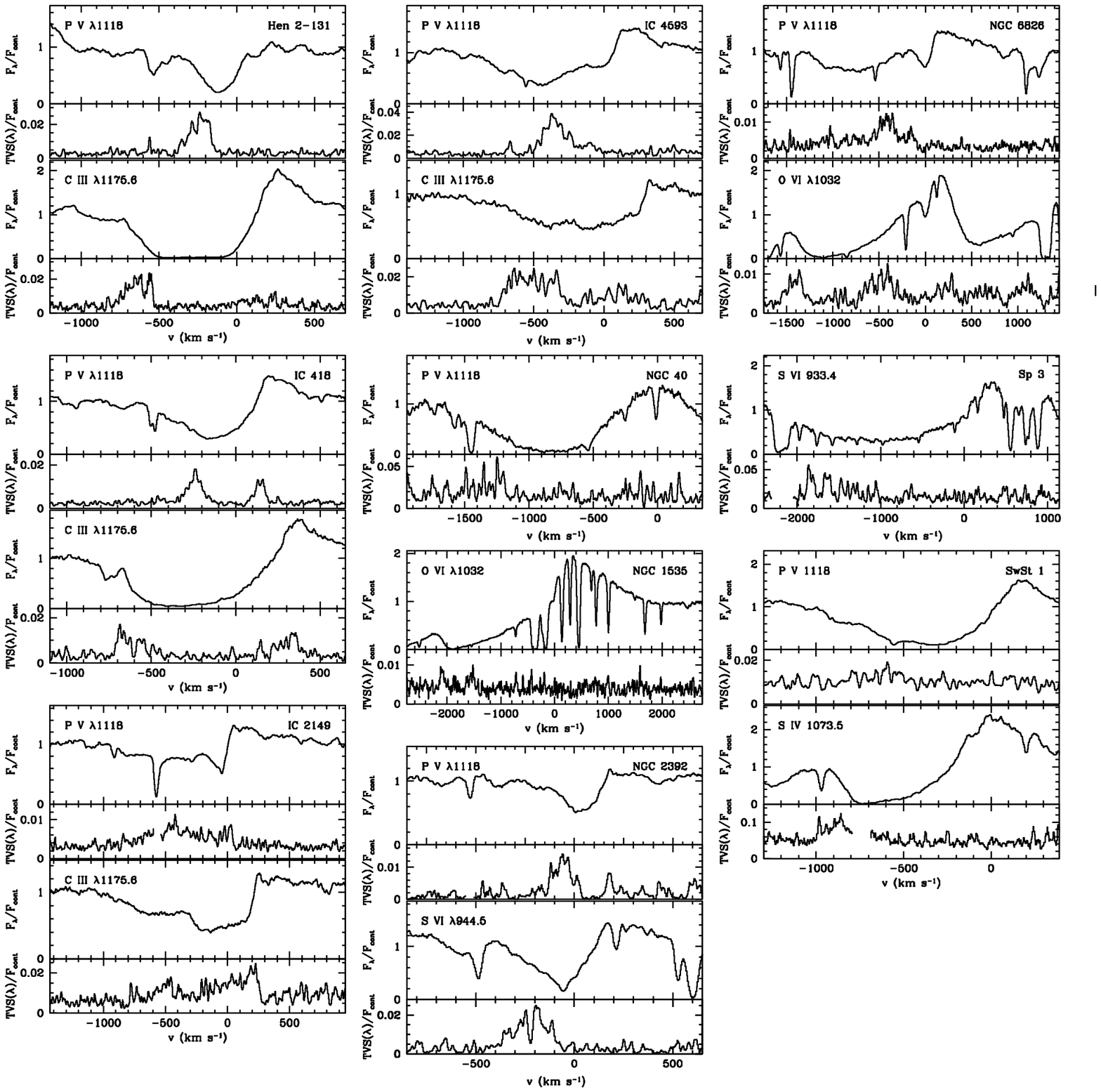}
}
\caption{
Selected P Cygni profiles of the CSPNe showing \emph{FUSE} variability.  
For each line, the top panel shows the averaged spectrum and the lower 
panel the temporal variation spectrum, TVS, both normalized by the 
continuum level at the spectral range of the line.  
Note that we have omitted the TVS at specific wavelength ranges where the 
continuum level of the averaged spectrum falls to zero, either in the trough 
of some lines (e.g., the S~{\sc vi} $\lambda$933.4 \AA\ line of Sp\,3) or at 
the location of narrow, deep absorption lines (e.g., the P~{\sc v} 
$\lambda$1118 \AA\ line of IC\,2149).  
}
\label{spec_dif}
\end{figure*}

The variability detected in the different lines for every CSPN is 
illustrated in Figure~\ref{spec_dif}.  
The upper panels show the normalized averaged spectrum of selected 
individual P Cygni profiles in velocity space, while the lower 
panels display the temporal variance spectrum (TVS) that we define 
as the scaled unweighted variance spectrum: 
\begin{equation}
TVS(\lambda) = \frac{1}{N} \frac{\sum [ F_i(\lambda) - \bar{F}(\lambda) ]^2}{\bar{F}(\lambda)}  
\end{equation}
where $F_i(\lambda)$ is the flux at wavelength $\lambda$ of the $ith$ 
spectrum, 
$\bar{F}(\lambda)$ is the averaged spectrum at wavelength $\lambda$, and 
$N$ is the number of spectra \citep[e.g.,][]{Howarth_etal93,FGB96}.  
As described by \citet{Howarth_etal93}, the division by the averaged 
spectrum, $\bar{F}(\lambda)$, scales TVS so that it does not vary with 
wavelength in the absence of variability, regardless of the signal 
level.   
Actually, the plots in the lower panels of Figure~\ref{spec_dif} 
have been divided by the continuum level at the spectral range of 
the line, $F_{\rm cont}$, to provide a normalized quantity.  
These plots illustrate the different variability of these CSPNe and even 
the different variability of P~Cygni profiles of different lines for the 
same CSPN.  
For instance, Hen\,2-131 and IC\,4593 show variability at 
different velocities in the P~{\sc v} and C~{\sc iii} lines.  
The variability can affect a significant fraction of the 
line profile (e.g., IC\,2149) or be restricted to a narrow 
velocity range (e.g., NGC\,1535).  
%
%
%

It is interesting to compare the samples of \emph{FUSE} (this paper) 
and \emph{IUE} variable CSPNe \citep{PP95,PP97}: 
IC\,4593, NGC\,40, NGC\,1535, NGC\,2392, NGC\,6826, and NGC\,6543 
are both \emph{FUSE} and \emph{IUE} variable CSPNe, while NGC\,246, 
NGC\,6210, NGC\,6572, and NGC\,7009 are found to be non-variable.  
This agreement strengthens the statements about the variability 
of these CSPNe.  
Meanwhile, the limitations of the \emph{IUE} and \emph{FUSE} datasets 
used by these studies is evidenced by BD+30$^\circ$3639, which was found 
to be variable by \emph{IUE} \citep{PP97}, but non-variable by our 
\emph{FUSE} study.  
On the other hand, the \emph{FUSE} data reveal the variability of 
IC\,418 and IC\,2149 that was missed by \emph{IUE}.

\subsection{Characteristics of the variable P Cygni profiles}

In order to investigate into more detail the variability of the 
\emph{FUSE} P~Cygni profiles of CSPNe, we have plotted in 
Figures~\ref{spec_time_Hen2_131} to \ref{spec_time_SwSt1} 
spectra derived from different individual exposures (bottom-panels) 
normalized by the mean spectra (top-panels).  
The P~Cygni profiles shown in these figures exhibit different types of 
line variability that are indicated in Table~\ref{CSPN_FUSE_lines} 
such as ripples (R), variability in the emission region ($\lambda > 
\lambda_0$) of the profile (E), and narrow features detected on 
the line profile trough (N) or bluewards of $v_\infty$ (B).  
The most common variable features are broad ripples (R) moving bluewards 
that are in many aspects similar to the DACs reported in the P~Cygni UV 
profiles of massive OB stars.  
The velocities and time-scales of individual features are discussed 
below for each object.  


\subsubsection{Hen\,2-131}

\begin{figure*}
\centerline{
\includegraphics[bb=31 160 588 718,width=1.60\columnwidth,angle=0]{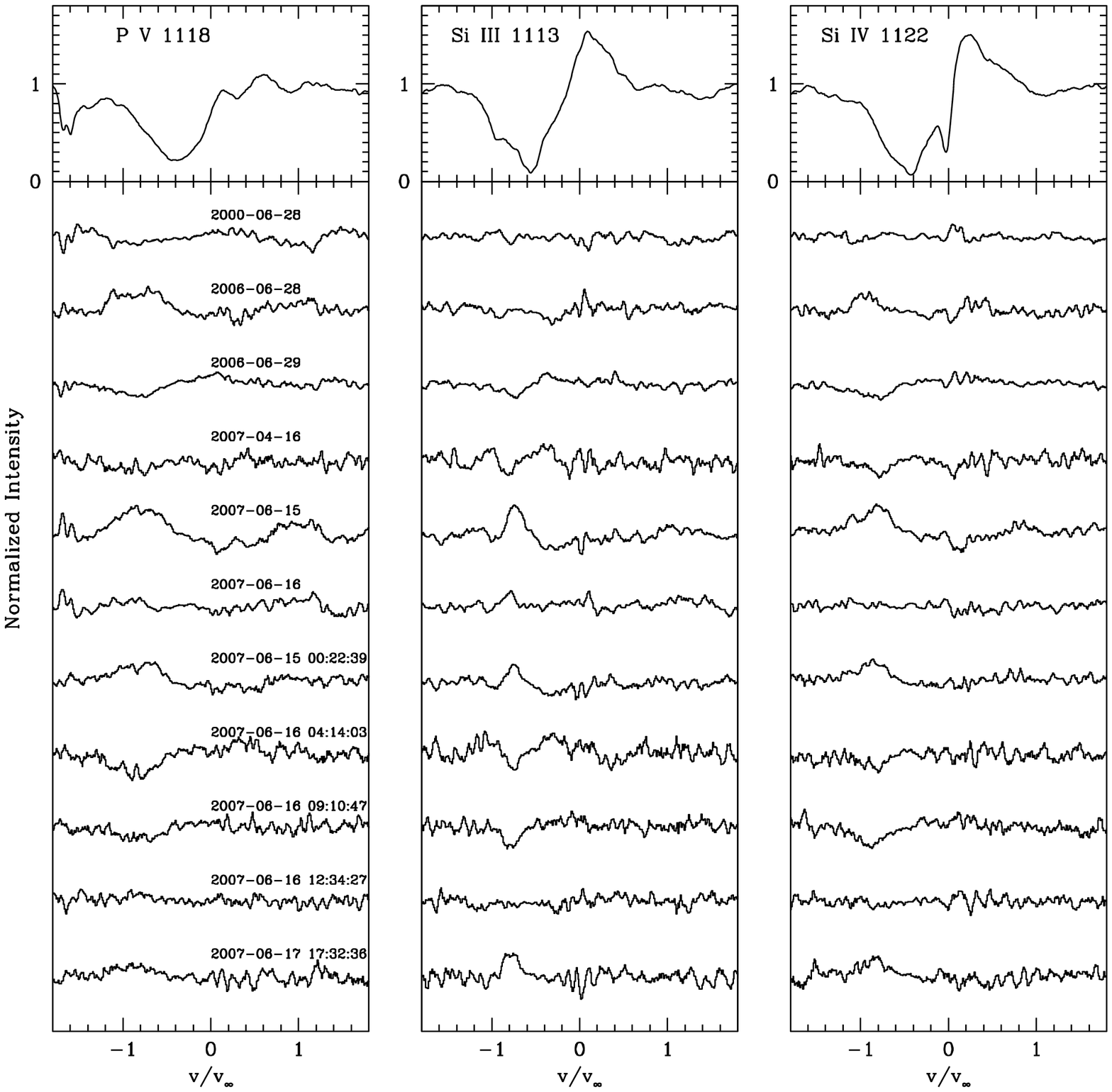}
}
\caption{
(top-panels) 
Averaged P~{\sc v} $\lambda$1118 (left), Si~{\sc iii} $\lambda$1113 (center), and 
Si~{\sc iv} $\lambda$1122 (right) P~Cygni profiles of Hen\,2-131.  
(bottom-panels) 
Normalized differences between individual spectra and the corresponding averaged 
spectrum shown in the top-panel.  
The epoch of each observation is indicated on the P~{\sc v} spectra.  
}
\label{spec_time_Hen2_131}
\end{figure*}

Figure~\ref{spec_time_Hen2_131} plots the profiles of the P~{\sc v} 
$\lambda$1118, Si~{\sc iii} $\lambda$1113, and Si~{\sc iv} $\lambda$1122 
lines of Hen\,2-131 (bottom-panels) normalized by its mean spectrum 
(top-panels).  
The normalized spectra taken on 2000-06-28, 2006-06-28, 2006-06-29, 
2007-04-15, 2007-06-15, and 2007-06-16 illustrate the long-term 
variability of these lines in Hen\,2-131, while the individual 
exposures obtained on 2007-06-15 and 2007-06-16 illustrate its 
short-term variability.

All the individual normalized spectra of the P~{\sc v} and Si~{\sc iv} lines 
show a broad DAC that is persistently observed in the trough of the line at 
a velocity up to 90\% of $v_\infty$.  
The Si~{\sc iii} line also shows a feature at this velocity, but it is 
narrower (N) than in the two other lines.  
The first two spectra taken on 2007-06-16 (${\Delta}t\sim$5 hours) show 
the absorption moving bluewards.  
The feature is also clearly seen in the spectra taken on 2007-06-15 
UT00:22:39 and 2007-06-17 UT17:32:36, i.e., $\sim$40 hours apart.  
If we assume this corresponds to the same DAC moving bluewards, the observed 
shift in velocity imply an acceleration $\sim$1$\times$10$^{-4}$ km~s$^{-2}$ 
for a terminal wind velocity of 370 km~s$^{-1}$.

Although there are \emph{IUE} high-dispersion SWP data available for 
Hen\,2-131 (SWP07653, SWP10854, and SWP10855), their S/N ratio is 
inadequate to search for time variability of the P~Cygni profiles of 
the C~{\sc iv} $\lambda\lambda$1548,1551 lines.  
Otherwise, the \emph{FUSE} variability detected in this study can 
be linked to the occurrence of variable optical lines reported by 
\citet{Mendez89}, although these variations were attributed to 
binary motions rather than to wind variability.  
This claim on the binary status of this object is not confirmed yet.


\subsubsection{IC\,418}

\begin{figure*}
\centerline{
\includegraphics[bb=31 470 588 718,width=1.90\columnwidth,angle=0]{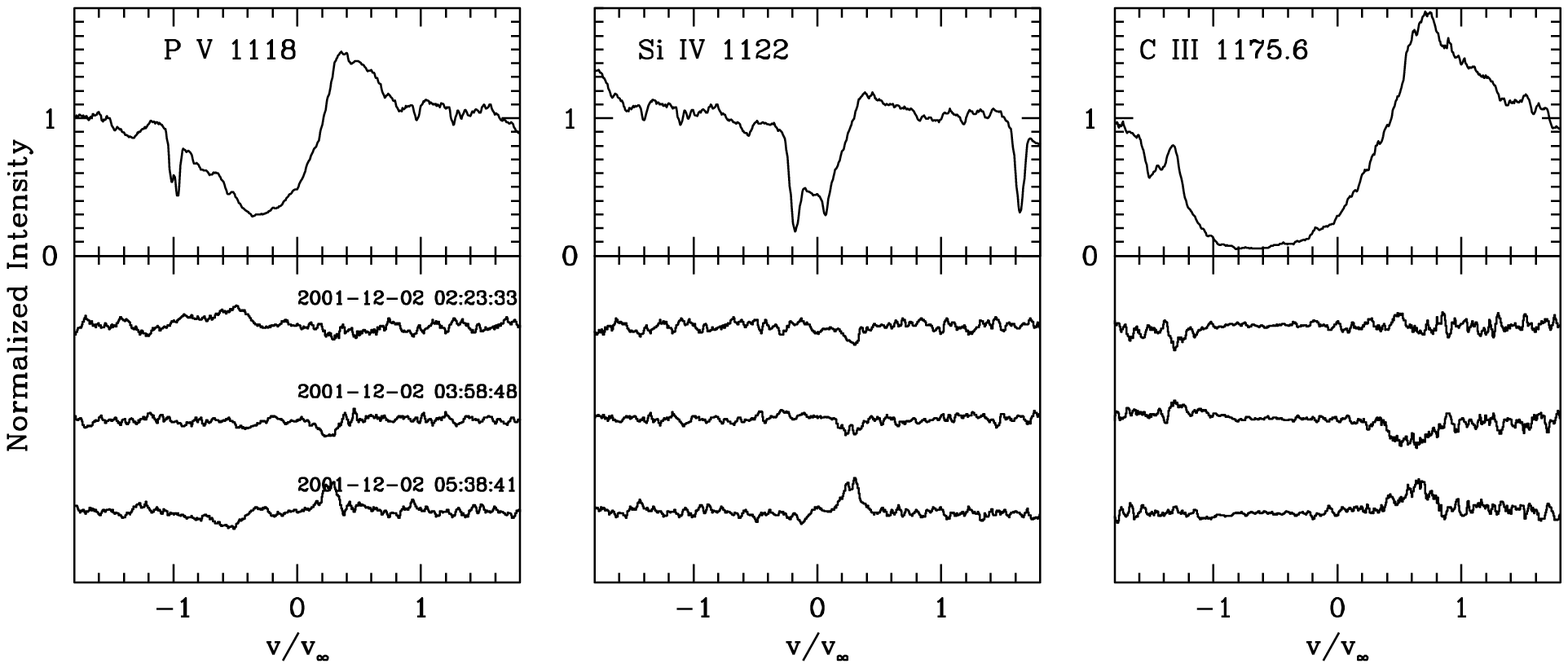}
}
\caption{
Same as Figure~\ref{spec_time_Hen2_131} for the P~{\sc v} $\lambda$1118, 
Si~{\sc iv} $\lambda$1122, and C~{\sc iii} $\lambda$1176 P~Cygni 
profiles of IC\,418.  
}
\label{spec_time_IC418}
\end{figure*}

Figure~\ref{spec_time_IC418} plots the profiles of the P~{\sc v} 
$\lambda$1118, Si~{\sc iv} $\lambda$1122, and C~{\sc iii} $\lambda$1176 
lines of IC\,418 (bottom-panels) normalized by its mean spectrum 
(top-panels).  
All the three normalized spectra were taken on 2001-12-02 with a 
cadence $\sim$1.6 hours.

As for Hen\,2-131, the P~Cygni profile of the P~{\sc v} line 
of IC\,418 shows a broad DAC that seems to move bluewards.  
For this feature, \citet{Prinja_etal12} reports a velocity shift 
$\sim$100 km~s$^{-1}$.  
Contrary to Hen\,2-131, the ripples in the P~{\sc v} and C~{\sc iii} 
P~Cygni profiles of IC\,418 extend to velocities bluer than $v_\infty$ 
(B).  
Moreover, these two lines, as well as the Si~{\sc iv} line, show 
variability of the emission region of the profile (E).


The \emph{IUE} observations of IC\,418 detected P~Cygni profiles of the 
Si~{\sc iv} $\lambda\lambda$1393,1401, C~{\sc iv} $\lambda\lambda$1548,1551, 
and N~{\sc iv} $\lambda$1719 lines whose possible variability was discussed 
in detail and rejected by \citet{PP95}.  
We have examined the P~Cygni profile of the C~{\sc iv} lines 
based on \emph{IUE} SWP data obtained on December 8, 1989.  
The line shows some evidence of variability in the emission region (E), 
as well as a broad ripple at velocities above $v_\infty$ (B), although 
the quality of the \emph{IUE} data is not sufficient to claim a firm 
detection of variability.  
We note that \citet{Mendez89} detected stellar emission and absorption 
lines moving in anti-phase that can be attributed to binary motion or 
to wind variability.

\subsubsection{IC\,2149}

\begin{figure}
\centerline{
\includegraphics[bb=1 240 240 720,width=0.95\columnwidth,angle=0]{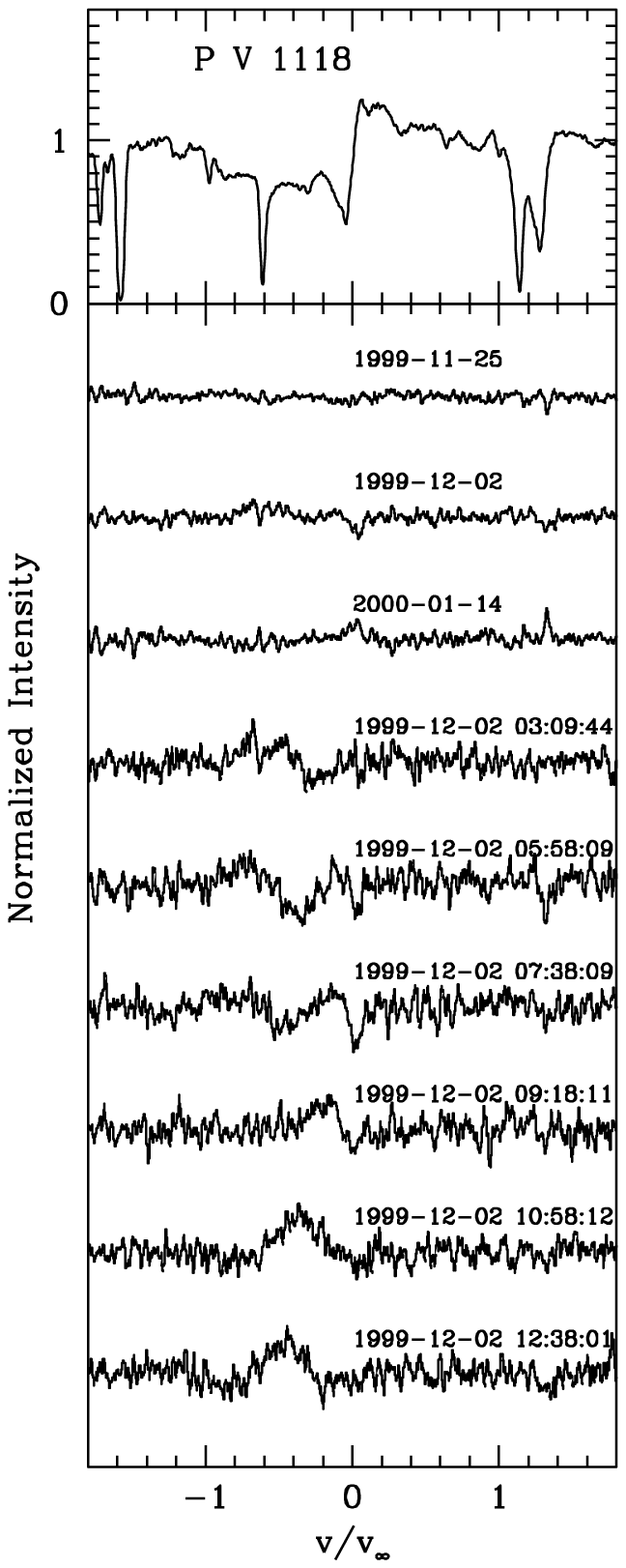}
}
\caption{
Same as Figure~\ref{spec_time_Hen2_131} for the P~{\sc v} $\lambda$1118 P~Cygni 
profile of IC\,2149.  
}
\label{spec_time_IC2149}
\end{figure}

Figure~\ref{spec_time_IC2149} plots profiles of the P~{\sc v} 
$\lambda$1118 line of IC\,2149 (bottom-panels) normalized by 
its mean spectrum (top-panels).  
The normalized spectra taken on 1999-11-26, 1999-12-02, and 2000-01-14 
show long-term variability, while the individual exposures obtained on 
1999-12-02 with a cadence $\sim$1.9 hours illustrate the short-term 
variability.

The P~Cygni profile of the P~{\sc v} line of IC\,2149 show large 
variations within time-scales as short as the spectra cadence.  
There is a notable broad DAC that seems to be traveling bluewards 
for which \citet{Prinja_etal12} reports an acceleration $\sim$0.01 
km~s$^{-2}$.  
Other spectra show also a broad feature at velocities up to $\sim$80\% 
of $v_\infty$.  
There is some evidence of variability in the emission region 
of the profile (E).


\emph{IUE} SWP observations were obtained on August 1980, April 1982, and 
August 1983 and examined by \citet{PP95}.  
P~Cygni profiles of the N~{\sc v}, Si~{\sc iv}, and C~{\sc iv} lines 
are detected, but no variability could be claimed from those data.

\subsubsection{IC\,4593}

\begin{figure*}
\includegraphics[bb=31 230 588 718,width=1.90\columnwidth,angle=0]{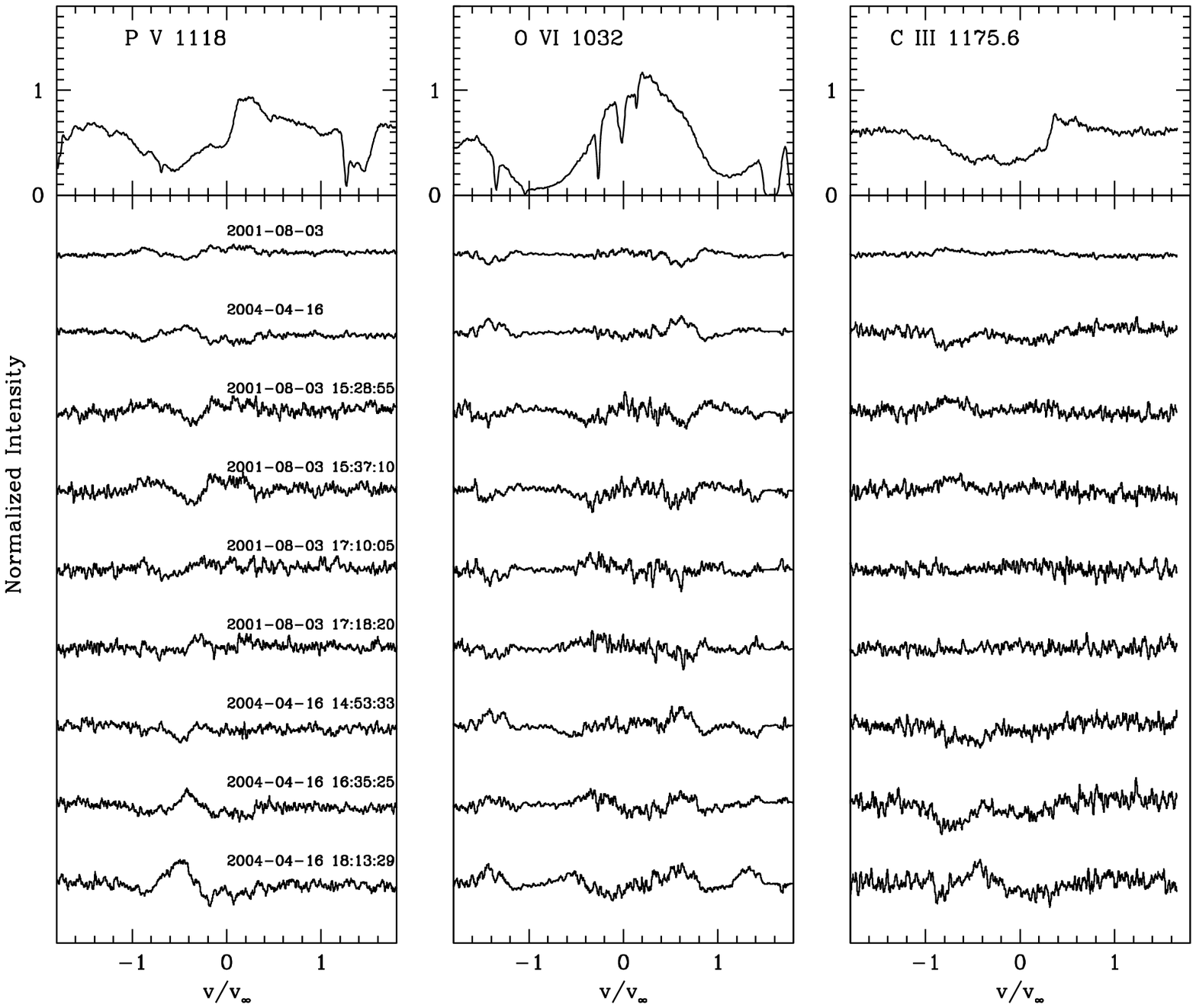}
\caption{
Same as Figure~\ref{spec_time_Hen2_131} for the P~{\sc v} $\lambda$1118, 
O~{\sc vi} $\lambda$1032, and C~{\sc iii} $\lambda$1176  P~Cygni 
profile of IC\,4593.  
}
\label{spec_time_IC4593}
\end{figure*}

Figure~\ref{spec_time_IC4593} plots the profiles of the P~{\sc v} 
$\lambda$1118, O~{\sc vi} $\lambda$1032, and C~{\sc iii} $\lambda$1176 
lines of IC\,4593 (bottom-panels) normalized by its mean spectrum 
(top-panels).  
The two top profiles correspond to the mean spectra obtained on 
2001-08-03 and 2004-04-16, respectively, whereas the following 
spectra are individual exposures obtained during these dates.

The P~Cygni profiles of the three lines display noticeable DACs that 
move bluewards.  
On 2004-04-16, the DAC in the P~{\sc v} profile moved by $\sim$20\% 
of $v_\infty$ over a period of 1.5 hours, implying an acceleration 
$\sim$0.03 km~s$^{-2}$, comparable to the measurement of 0.042 km~s$^{-2}$ 
reported by \citet{Prinja_etal12}.  
On these same observations, the ripple in the O~{\sc vi} lines extends 
beyond $v_\infty$ (B).  
No obvious variations of the emission region of the profile is detected.


IC\,4593 was one of the CSPNe claimed to be variable by \emph{IUE} 
\citep{PP95}.  
The \emph{IUE} C~{\sc iv}, O~{\sc iv}, Si~{\sc iv}, N~{\sc iv}, and N~{\sc v} 
lines show a behavior similar to this of the \emph{FUSE} lines described 
above, i.e., little or no changes in the emission region of the profile and 
variable troughs at velocities near or above $v_\infty$.  
\citet{DeMarco_etal07} detected a variability in the emission component 
of the He~{\sc ii} $\lambda$4686~\AA\ line, whereby the entire emission 
line shifted to the red and the P~Cygni trough of the line is also 
variable.  
This behaviour remains unexplained.

\subsubsection{NGC\,40}

\begin{figure}
\includegraphics[bb=1 430 240 720,width=0.95\columnwidth,angle=0]{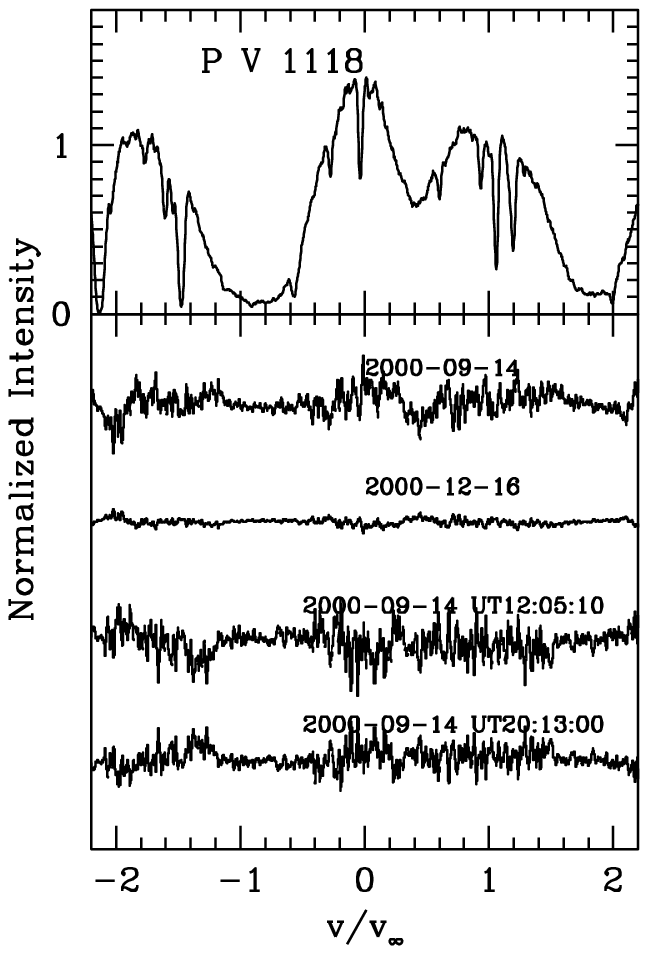}
\caption{
Same as Figure~\ref{spec_time_Hen2_131} for the P~{\sc v} $\lambda$1118 P~Cygni 
profile of NGC\,40.  
}
\label{spec_time_NGC40}
\end{figure}

Figure~\ref{spec_time_NGC40} plots profiles of the P~{\sc v} 
$\lambda$1118 line of NGC\,40 (bottom-panel) normalized by 
its mean spectrum (top-panel).  
The two top normalized profiles correspond to the mean spectra obtained 
on 2000-09-14 and 2000-12-16, respectively, whereas the following two 
are individual exposures obtained on 2000-09-14.  
The separation in time between these two spectra is $\sim$8 hours.

The P~Cygni profile of the P~{\sc v} line of NGC\,40 is rather flat, 
but it shows noticeable ripples above the terminal velocity (R,B) and 
on the emission region of the profile (E).


NGC\,40 was claimed to have non variable P~Cygni profiles by \citet{PP95}, 
based on \emph{IUE} data acquired on January 1983 and August 1991.  
However, new \emph{IUE} high resolution observations obtained on 
1994-08-22 and 1994-11-06 revealed the variability of the P~Cygni 
profiles of the C~{\sc iv} lines, while the N~{\sc v}, O~{\sc v}, 
and Si~{\sc iv} P~Cygni profiles also present in the \emph{IUE} 
spectral range were found to be constant \citep{PP97}.  
As for the \emph{FUSE} P~{\sc v} line, the P~Cygni profile of the \emph{IUE} 
C~{\sc iv} line exhibits ripples at and above $v_\infty$ (R,B) and on the 
emission region of the profile (E).

\subsubsection{NGC\,1535}

\begin{figure}
\includegraphics[bb=1 500 240 720,width=0.95\columnwidth,angle=0]{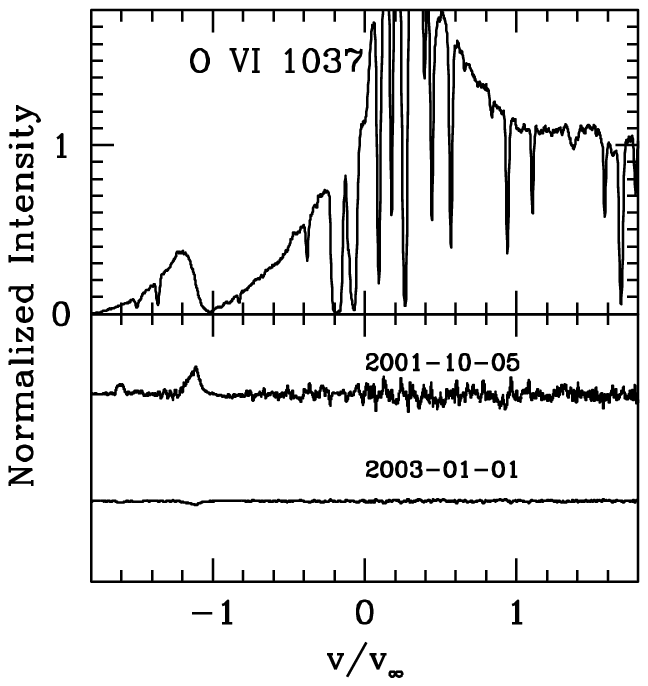}
\caption{
Same as Figure~\ref{spec_time_Hen2_131} for the O~{\sc vi} $\lambda$1037 
P~Cygni profile of NGC\,1535.  
}
\label{spec_time_NGC1535}
\end{figure}

Figure~\ref{spec_time_NGC1535} plots profiles of the O~{\sc vi} 
$\lambda$1037 line of NGC\,1535 (bottom-panel) normalized by its 
mean spectrum (top-panel).  
The normalized profiles correspond to the mean spectra obtained 
on 2001-10-05 and 2003-01-01, respectively.

The P~Cygni profile of the O~{\sc vi} line of NGC\,1535 shows a 
narrow feature at a velocity $\sim$10\% above its wind terminal 
velocity of 1825~km~s$^{-1}$ (N,B).  
No noticeable changes are observed in the emission region of the profile.  
The variability detected by \emph{FUSE} does not match completely 
the changes observed in the N~{\sc v} P~Cygni profiles by \emph{IUE} 
between December 1980 and March 1981, 
as this line also shows variability of the emission region \citep{PP95}.


\subsubsection{NGC\,2392}

\begin{figure*}
\includegraphics[bb=-71 515 300 718,width=1.250\columnwidth,angle=0]{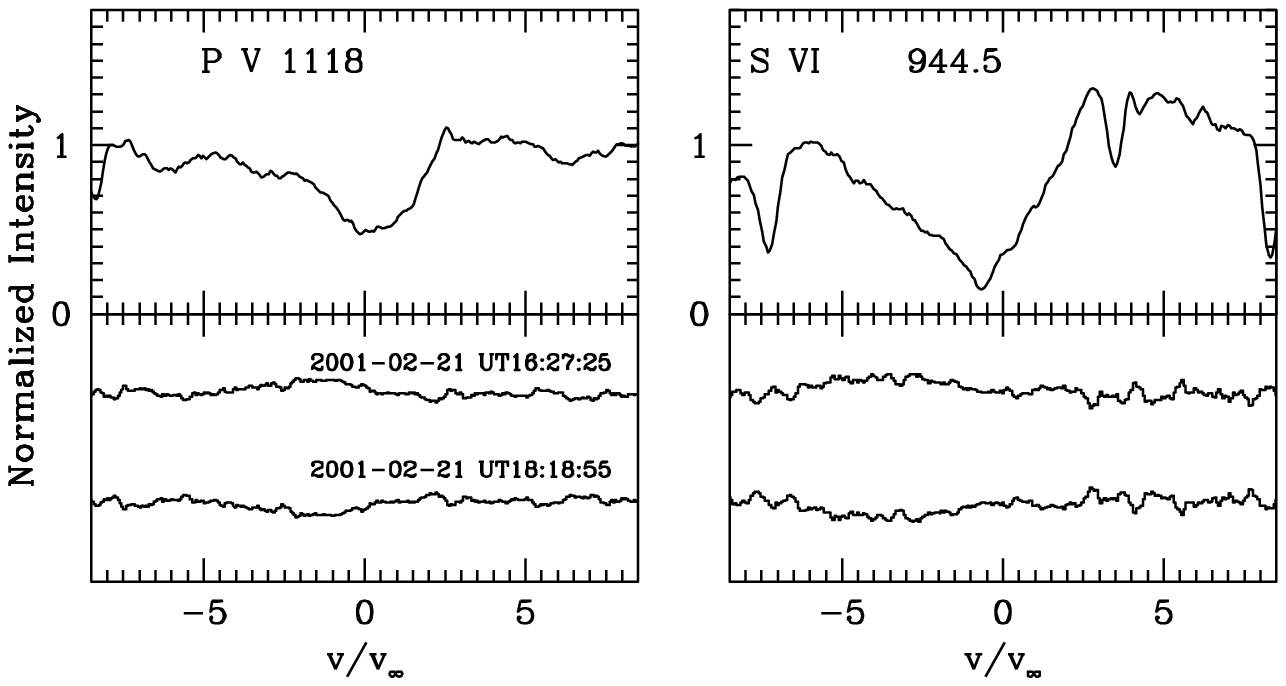}
\caption{
Same as Figure~\ref{spec_time_Hen2_131} for the P~{\sc v} $\lambda$1118 and 
S~{\sc vi} $\lambda$944.5 P~Cygni profiles of NGC\,2392.  
}
\label{spec_time_NGC2392}
\end{figure*}

Figure~\ref{spec_time_NGC2392} plots profiles of the P~{\sc v} $\lambda$1118 
and S~{\sc vi} $\lambda$944.5 lines of NGC\,2392 (bottom-panels) normalized 
by the mean spectrum (top-panels).  
The normalized profiles correspond to individual exposures obtained on 
2001-02-21 taken with a difference in time $\sim$1.85 hours.

The P~Cygni profiles of both the P~{\sc v} and S~{\sc vi} lines of NGC\,2392 
show a broad ripple (R) that extends far bluewards of $v_\infty$ (B) due to 
its low value in this CSPN ($v_\infty=65$~km~s$^{-1}$).  
We note that the ripple of these two lines are not completely in 
phase, with the peak of the P~{\sc v} line in the profile taken 
at UT16:27:25 at $\sim$3~$\times\,v_\infty$, while that of the 
S~{\sc vi} line is at $\sim$7~$\times\,v_\infty$.


The variability of the P~Cygni profiles of NGC\,2392 has already been 
reported by \citet{PP95} based on \emph{IUE} high-dispersion observations 
obtained on May 1983, and October and November 1990.  
The variability of the \emph{IUE} C~{\sc iv} and N~{\sc v} lines, 
however, affects the emission region of the profiles (E), while 
the \emph{FUSE} variability is better described as a broad ripple 
(R) at velocities well above $v_\infty$ (B), but with no variable 
emission region of the profile.

\subsubsection{NGC\,6826}

\begin{figure*}
\includegraphics[bb=-71 245 300 718,width=1.25\columnwidth,angle=0]{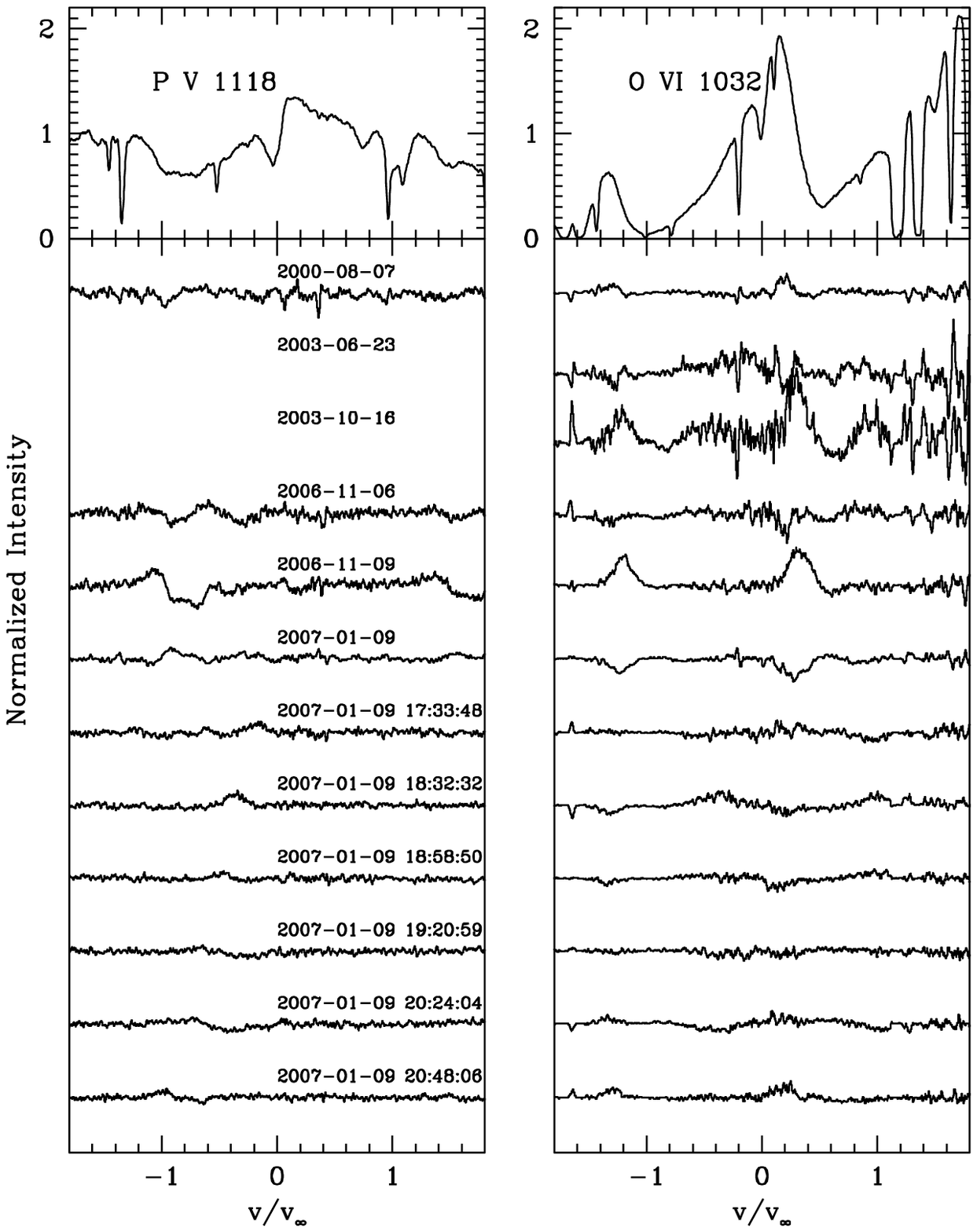}
\caption{
Same as Figure~\ref{spec_time_Hen2_131} for the P~{\sc v} $\lambda$1118 and 
O~{\sc vi} $\lambda$1032 P~Cygni profiles of NGC\,6826.  
}
\label{spec_time_NGC6826}
\end{figure*}

Figure~\ref{spec_time_NGC6826} plots profiles of the P~{\sc v} 
$\lambda$1118 and O~{\sc vi} $\lambda$1032 lines of NGC\,6826 
(bottom-panels) normalized by the mean spectrum (top-panels).  
The normalized profiles correspond to mean spectra 
obtained on 2000-08-07, 2003-06-23, 2003-10-16, 
2006-11-06, 2006-11-09, and 2007-11-09, but we 
need to note that no suitable observations of the 
P~{\sc v} line are available for the observations 
obtained on 2003 (2003-06-23 and 2003-10-16).  
The spectra obtained on the 2007-11-09 has further been split into 
6 individual exposures acquired with a cadence between 0.5 and 1.0 
hours.

The P~Cygni profile of the O~{\sc vi} line of NGC\,6826 show persistent 
narrow features (N) that are found at velocities above the terminal 
velocity (B).  
Broad DACs are also traveling across the profile trough as 
seen in the P~Cygni profile of the P~{\sc v} line.  
In particular, the individual exposures acquired on 2007-11-09 show 
a feature that moves from a velocity $\sim$0.2~$\times\,v_\infty$ 
(i.e., $\sim$200~km~s$^{-1}$) up to $\sim$1.0~$\times\,v_\infty$ 
(i.e., 1100~km~s$^{-1}$), thus implying a velocity change of 
$\sim$900~km~s$^{-1}$ in a period of 5.25 hours, or an averaged 
acceleration $\sim$0.05~km~s$^{-2}$ in perfect agreement with 
the measurement by \citet{Prinja_etal12}.


The variability of the P~Cygni profiles of the \emph{IUE} C~{\sc iv}, 
N~{\sc iv}, and N~{\sc v} lines of NGC\,6826 was reported by 
\citet{PP95,PP97} based on high-resolution observations acquired on July 
and August 1982, February 1986, August 1994, and September 1995.  
All the variations observed in the profiles of the C~{\sc iv} and 
N~{\sc v} lines affect to the trough at velocities close to or above 
$v_\infty$, while the changes observed in the unsaturated N~{\sc iv} 
line would be better described as a broad ripple (R).  
Additional photometric and spectroscopic variations have been interpreted 
in terms of wind variability \citep{Handler_etal13}.

\subsubsection{Sp\,3}

\begin{figure*}
\includegraphics[bb=-71 470 300 718,width=1.25\columnwidth,angle=0]{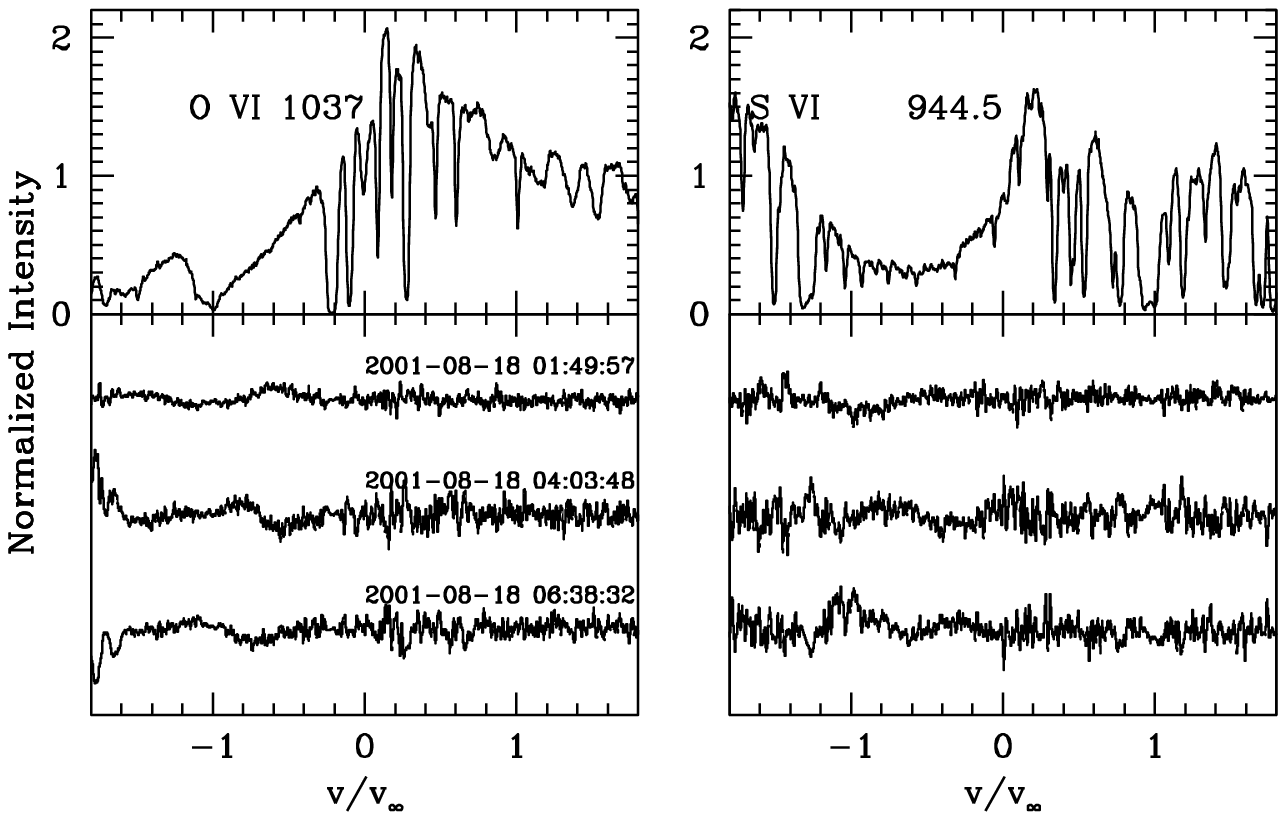}
\caption{
Same as Figure~\ref{spec_time_Hen2_131} for the O~{\sc vi} $\lambda$1032 and 
S~{\sc vi} $\lambda$944.5 P~Cygni profiles of Sp\,3.  
}
\label{spec_time_Sp3}
\end{figure*}

Figure~\ref{spec_time_Sp3} plots profiles of the O~{\sc vi} $\lambda$1037 
and S~{\sc vi} $\lambda$944.5 lines of Sp\,3 (bottom-panels) normalized 
by the mean spectrum (top-panels).  
The normalized profiles correspond to individual exposures obtained on 
2001-08-18 taken with a cadence $\sim$2.4 hours.

The P~Cygni profiles of both the O~{\sc vi} and S~{\sc vi} lines of Sp\,3 
show a broad ripple (R) traveling bluewards through the profile trough up 
to velocities bluewards of $v_\infty$ (B).  
The peak of these ripples are not completely in phase, as the 
O~{\sc vi} ripple is (0.1--0.2)$\times v_\infty$ bluewards of 
the S~{\sc vi} ripple, but they move coherently, with a change 
of velocity $\sim$0.5~$\times v_\infty$ (i.e., $\sim$800~km~s$^{-1}$) 
in a period $\sim$4.8 hours.  
This implies an averaged acceleration of $\sim$0.05~km~s$^{-2}$, very 
similar to that found in NGC\,6826.


\subsubsection{SwSt\,1}

\begin{figure*}
\includegraphics[bb=31 430 588 718,width=1.90\columnwidth,angle=0]{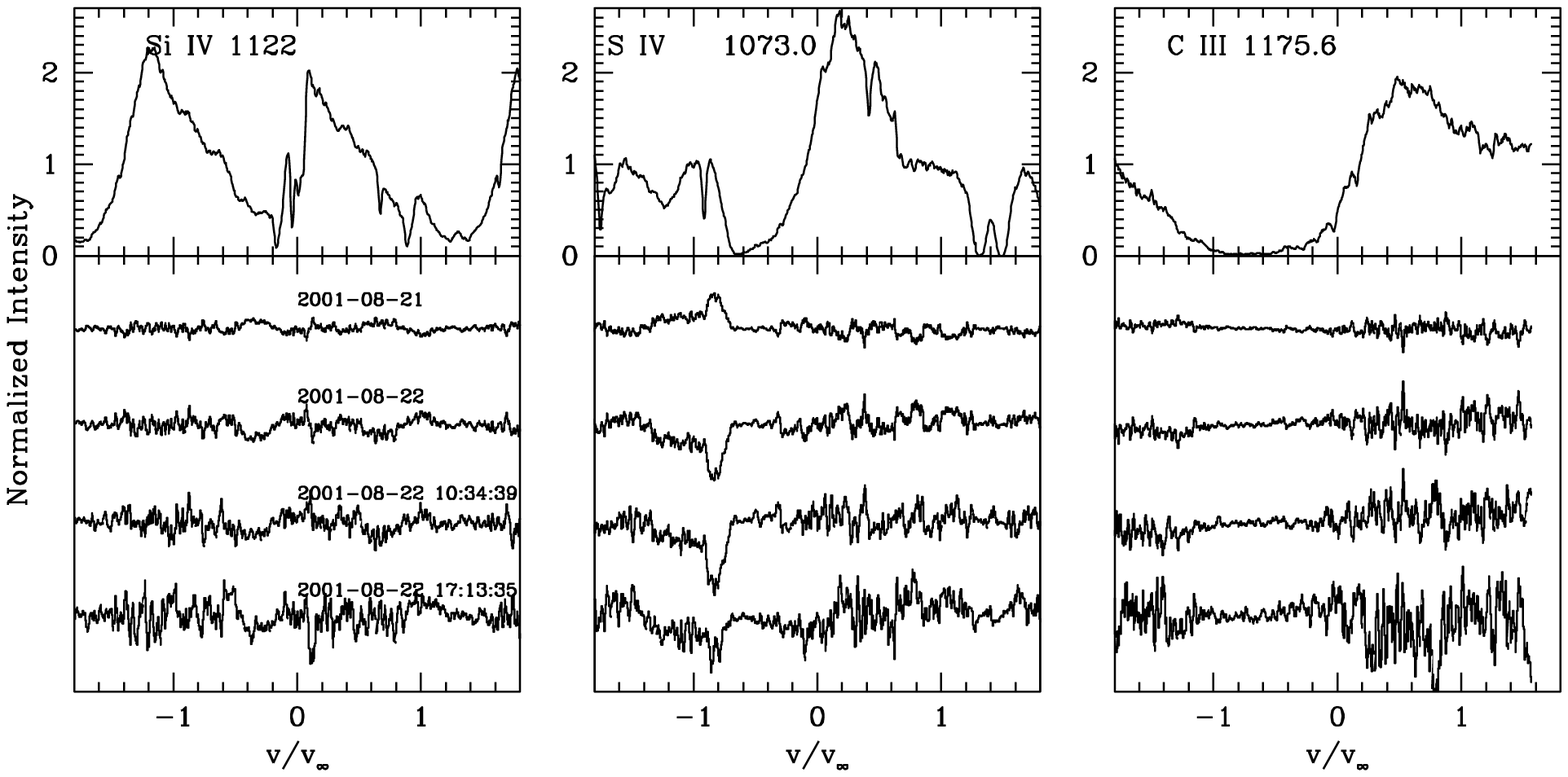}
\caption{
Same as Figure~\ref{spec_time_Hen2_131} for the Si~{\sc iv} $\lambda$1122, 
S~{\sc iv} $\lambda$1073, and C~{\sc iii} $\lambda$1176 P~Cygni profiles 
of SwSt\,1.  
}
\label{spec_time_SwSt1}
\end{figure*}

Figure~\ref{spec_time_SwSt1} plots profiles of the Si~{\sc iv} 
$\lambda$1122, S~{\sc iv} $\lambda$1073, and C~{\sc iii} 
$\lambda$1176 lines of SwSt\,1 (bottom-panels) normalized by 
the mean spectrum (top-panels).  
The two top normalized profiles correspond to the mean spectra 
acquired on 2001-08-21 and 2001-08-22, respectively.  
The latter spectrum has been expanded into two consecutive individual 
exposures obtained $\sim$6.6 hours apart.

The P~Cygni profiles of the three lines show very different types 
of variability.  
The Si~{\sc iv} line shows a ripple (R) overimposed on its trough, 
with some evidence that this ripple is traveling outwards in the 
wind from one exposure to the next.  
On the other hand, the S~{\sc iv} profile shows a narrow feature (N) at 
$\sim$0.85~$\times v_\infty$ with a shoulder extending bluewards of the 
terminal velocity (B).  
Finally, the changes of the C~{\sc iii} line manifest as variations 
of the slope across the whole trough from one spectrum to another.  
As for many other sources in this study, there is no obvious 
evidence of change in the emission region of the profile.


\section{Discussion}

\subsection{Occurrence of wind variability}

\begin{figure*}
\centerline{
\includegraphics[bb=58 184 552 678,width=0.95\columnwidth,angle=0]{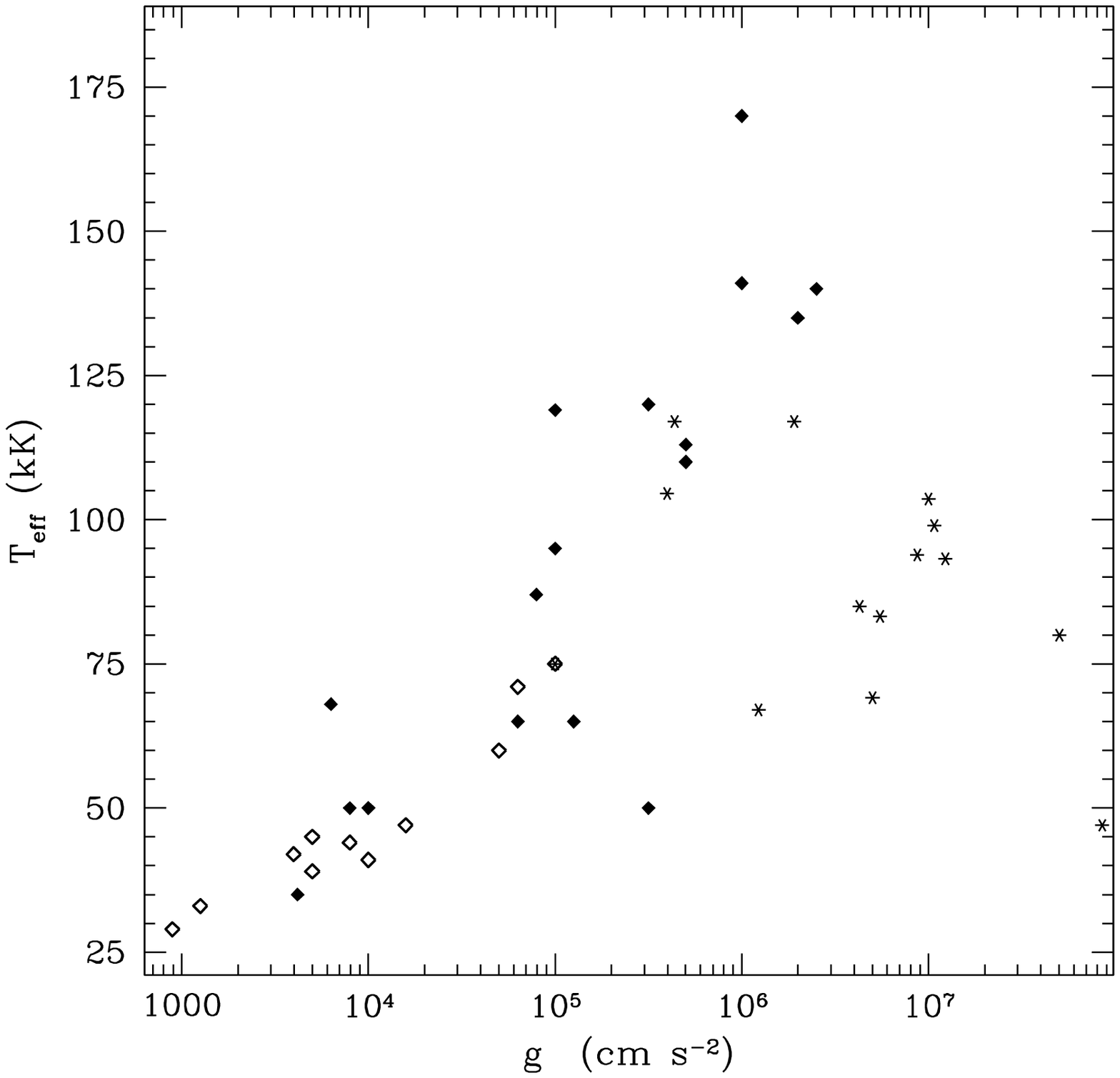}
\includegraphics[bb=58 184 552 678,width=0.95\columnwidth,angle=0]{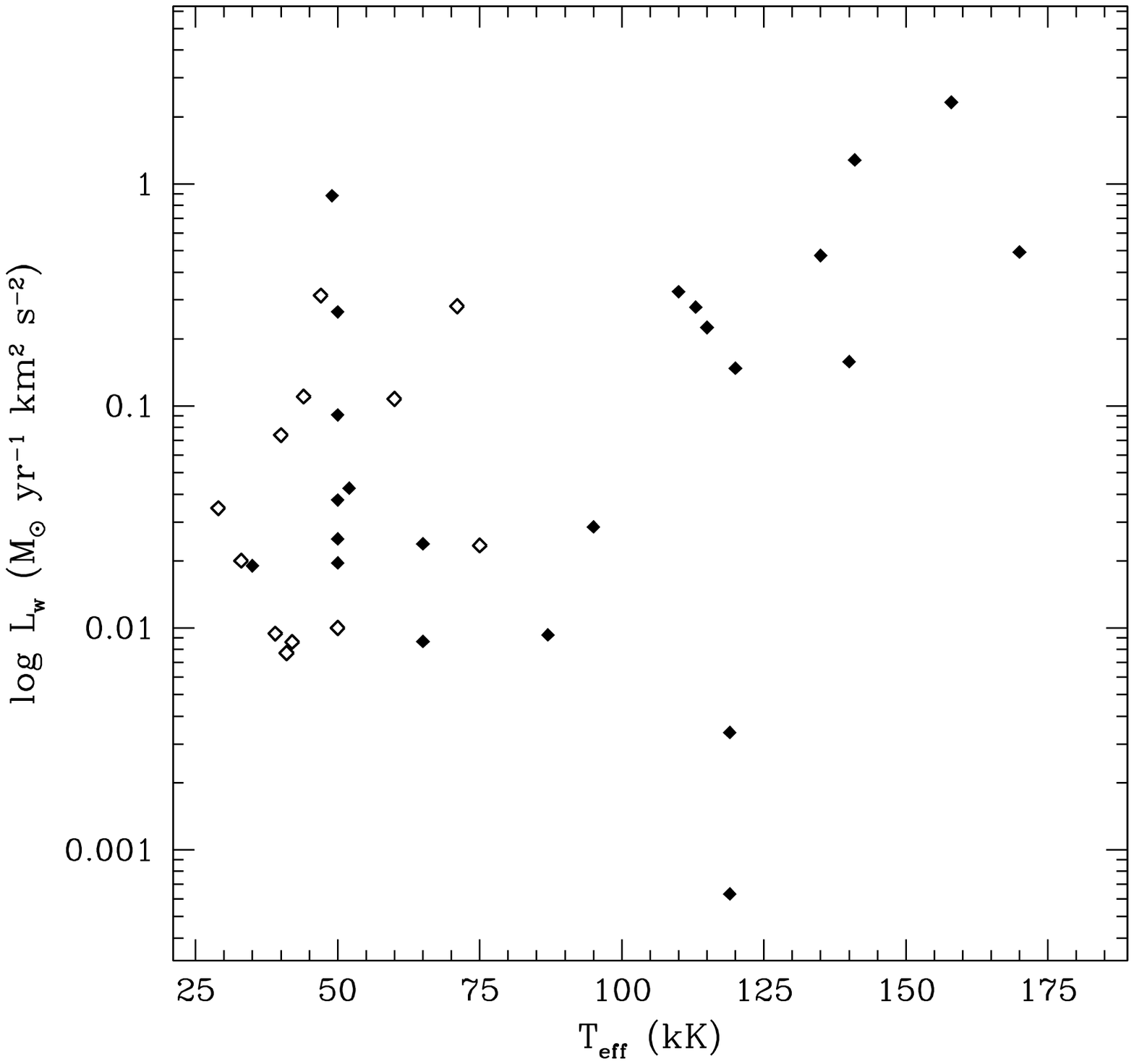}
}
\centerline{
\includegraphics[bb=58 184 552 678,width=0.95\columnwidth,angle=0]{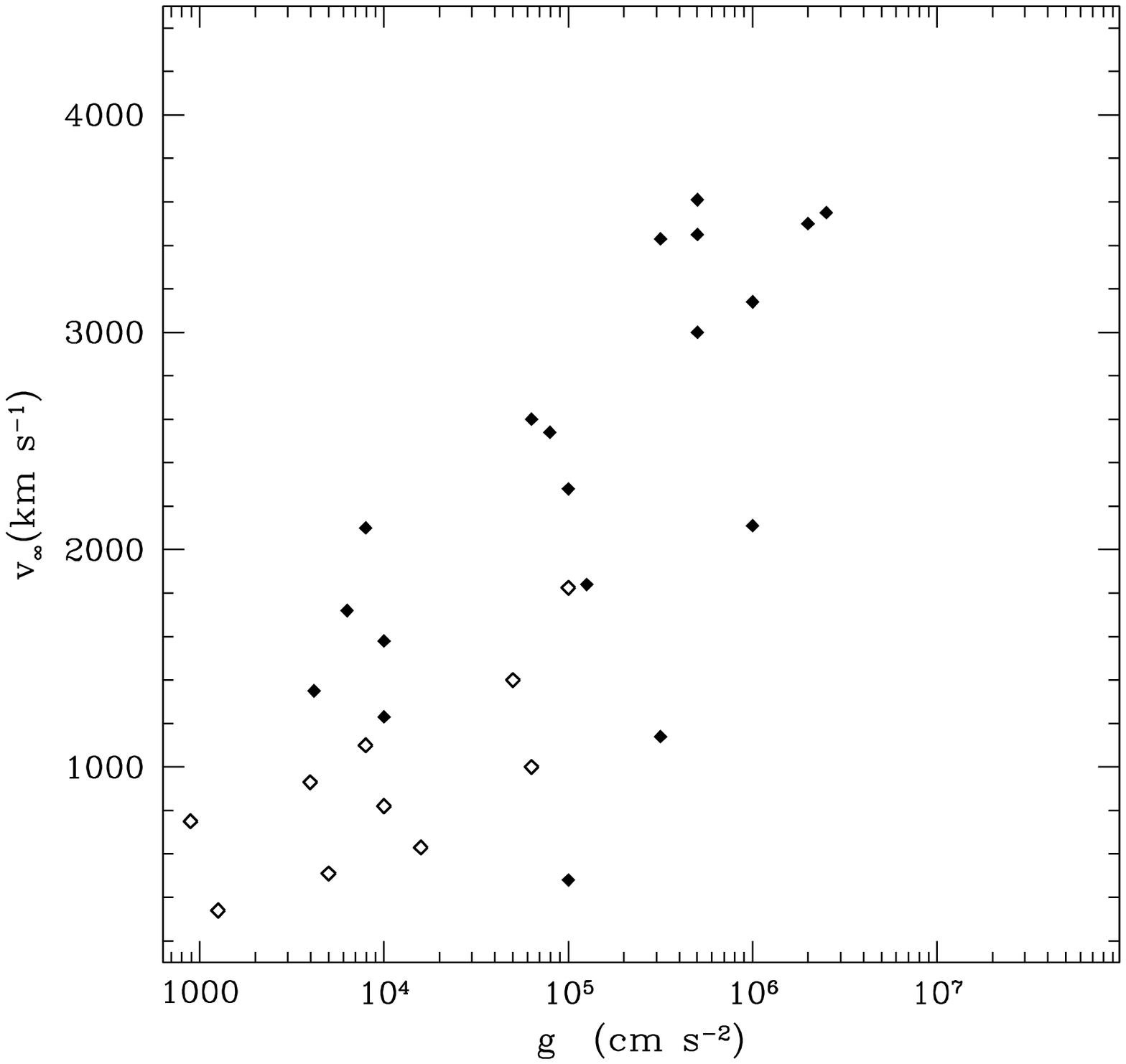}
\includegraphics[bb=58 184 552 678,width=0.95\columnwidth,angle=0]{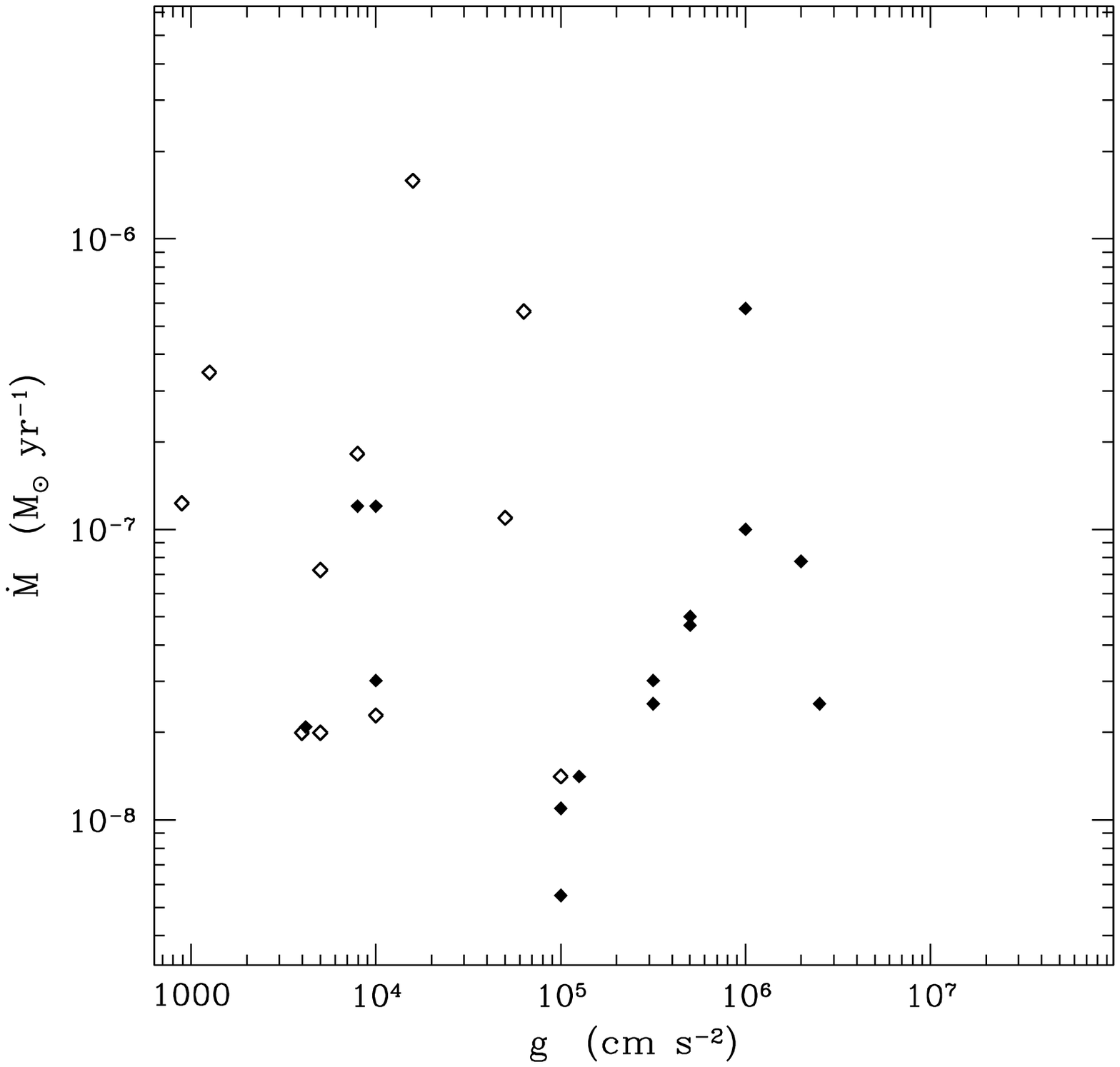}
}
\caption{
Distributions of the CSPNe with P~Cygni profile in \emph{FUSE} spectra in 
the $T_{\rm eff}$--$\log g$ ({\it top-left}), $L_{\rm wind}$--$T_{\rm eff}$ 
({\it top-right}), $v_\infty$--$\log g$ ({\it bottom-left}), and 
$\log \dot M$--$\log g$ ({\it bottom-right}) diagrams.  
The CSPNe with variable P~Cygni profiles are shown with open diamonds, 
while those without variable P~Cygni profiles are shown with filled 
diamonds.  
In addition, the $T_{\rm eff}$--$\log g$ diagram ({\it top-left}) 
shows the location of the CSPNe without P~Cygni profiles (stars).  
}
\label{stat}
\end{figure*}

In order to investigate whether the occurrence of wind variability 
correlates with stellar and wind parameters, we plot in Figure~\ref{stat} 
the location of the CSPNe with \emph{FUSE} observations in different 
diagrams comparing stellar ($T_{\rm eff}$, $\log g$) and wind ($\dot M$, 
$v_\infty$, $L_{\rm w} = \frac{1}{2} \dot M v_\infty^2$) properties.  
An inspection of Figure~\ref{stat} reveals that the CSPNe with 
variable \emph{FUSE} and/or \emph{IUE} P~Cygni profiles are to 
some extent clustered in all diagrams, as they tend to 
have low effective temperature ($T_{\rm eff}\leq$70,000~K), 
low gravity ($\log g\leq$5.3), and slow terminal velocities 
($v_\infty\leq$2000~km~s$^{-1}$).

It is, thus, tempting to conclude that the stellar winds of evolved CSPNe do 
not present variability.  
It must be kept in mind, however, that the P~Cygni profiles of many 
of the evolved CSPNe with the highest $T_{\rm eff}$ and $\log g$ are 
highly saturated (e.g., A\,78 or IC\,2448), as noted in 
Table~\ref{CSPN_FUSE_lines} (quality code N=2).  
It would be fair to state that the search for wind variability 
among the evolved CSPNe does not reach the same sensitivity as 
for the less evolved CSPNe, thus potentially introducing a bias 
in the detection rate among CSPNe at different evolutionary stages.

On the other hand, there are several CSPNe that, having very similar 
stellar and wind properties as those with variable P~Cygni profiles, 
do not show any variability.  
Very notable examples are Hb\,7, IC\,4776, NGC\,5882, NGC\,6058, NGC\,6210, 
NGC\,6891, PB\,8, and Tc\,1, as illustrated by Figure~\ref{stat} where they 
appear mixed among the CSPNe with variable P~Cygni profiles in one or 
several of the diagrams.  
In most cases, the quality of the available data, as noted in 
Table~\ref{CSPN_FUSE_lines} by N=1, may have hampered the 
detection of variability.  
The only CSPN that is definitely found to be non variable is NGC\,6210 
(N=3 in Table~\ref{CSPN_FUSE_lines}) based on 4 exposures with a total 
integration time of 6.6 ks obtained on August 2, 2000.  

Among the sample of CSPNe with \emph{FUSE} observations, there are 
several that are classified as [WR] stars.  
These stars are expected to have stronger winds and thus wind 
variability is more likely.  
BD+30$^\circ$3639 ([WC9]), NGC\,40 ([WC8]), and SwSt\,1 ([WC9pec]) show 
variable P~Cygni profiles, while the more evolved [WC]-PG\,1159 A\,30, 
A\,78, and NGC\,2371-72, and Hen\,2-99 ([WC9]), NGC\,2867 ([WO1-2]), 
PB\,6 ([WO1]), and PB\,8 ([WN/WC]) do not.  
As we have already discussed above, the quality of the observations does 
not allow us to make a definitive conclusion on the occurrence of wind 
variability among [WR] CSPNe.

To summarize, it can be concluded that the detection of variable P~Cygni 
profiles is more likely in less evolved CSPNe, either because the wind 
variability is associated to earlier evolutionary stages, its effects 
are more likely to be detected in denser winds, or the detection is 
more difficult in fast, tenuous winds resulting in saturated P~Cygni 
profiles.  
While a stronger wind might appear to be obviously promoted by higher 
luminosities occurring in younger CSPNe, it is not clear that 
variability bears the same relation to luminosity given the lack of 
variability in some CSPNe in the same luminosity group.  
Also note that the entire range of mass-loss rates has both variable 
and non variable winds.
The lack of detection of P~Cygni profile variability among a number 
of CSPNe with similar stellar and wind properties as those found to 
be variable suggests that the wind variability may be an intermitent 
phenomenom, or that variability and mass-loss may be not fully related.  
More sensitive and dedicated UV observations, such as those possible 
with the upcoming World Space Observatory-UV (\emph{WSO-UV}), are 
required to shed light into these questions.

\subsection{Characteristics of time-variable features}

The variability observed in the P~Cygni profiles of the CSPNe described 
in \S5 displays significant variety from broad ripples to narrow 
features.  
The latter can be ascribed to DACs typically detected in OB stars.  
The CSPNe that show variable narrow features are 
Hen\,2-131, NGC\,1535, NGC\,6826, and SwSt\,1.  
In all cases, the narrow features appear around the wind terminal velocity, 
with NGC\,1535 and NGC\,6826 having narrow features exceeding $v_\infty$ by 
$\sim$10\% and $\sim$20\%, respectively.

The fragmentary time coverage of the observations used in this analysis 
makes uncertain the assessment of the variations in velocity of the 
variable features seen in this sample of CSPNe.  
When velocity shifts are detected (e.g., broad ripples traveling 
bluewards, as in NGC\,6826), typical accelerations are in the 
range 0.01--0.05 km~s$^{-2}$, with features traveling a few tenths 
of the corresponding terminal velocities in time intervals of a 
few hours.  
The time-scales of the variable features of CSPNe to travel along the 
troughs of the P~Cygni profiles seem shorter than those of OB stars, 
typically in the range of a few days 
\citep[e.g., ][]{HPM95,Massa_etal95,Fullerton_etal97}. 
If we assume these velocity shifts are associated to the stellar rotation, 
as for OB stars, and thus that the rotation periods are of the order or 
greater than a few hours, then for typical radii (0.2--0.5 $R_\odot$) for 
the stars in our sample, the rotational velocities would be in the range 
50--150 km~s$^{-1}$.  
These rotational velocities compare favorably with those measured in stars 
in transition from CSPN to WD \citep[40--70 km~s$^{-1}$,][]{Rauch_etal04} 
and cover the range of rotational velocities recently derived for a 
sample of CSPNe also using \emph{FUSE} data 
\citep[50--110 km~s$^{-1}$,][]{Prinja_etal12,PMC12}.



%



\subsection{Correlation between hard X-rays and wind variability}

As for the strong stellar winds of OB stars  \citep{LW80,GO95}, the 
variability of the P~Cygni profiles of a CSPN may be associated with 
shocks in its stellar wind that can produce the hard X-ray emission 
detected, e.g., in NGC\,6543 \citep{Guerrero_etal01}.  
Therefore, a correlation between hard X-ray emission in CSPNe and wind 
variability may be expected.  
The search for such correlation, however, is hampered by the limited 
overlap between the samples of CSPNe with P~Cygni profiles in their 
\emph{FUSE} spectra and those observed by X-ray observatories.  
Among these CSPNe, three objects with variable \emph{FUSE} P~Cygni 
profiles are hard X-ray emitters (NGC\,2392, NGC\,6543, and NGC\,6826), 
while three others (BD+30$^\circ$3639, IC\,418, and NGC\,40) are not 
\citep[][Guerrero et al., in preparation]{Kastner_etal00,Guerrero_etal01,Montez_etal05,Kastner_etal12}.  
Among the CSPNe without P~Cygni profiles nor variable P~Cygni 
profiles, eight are not found to be hard X-ray emitters (Hen\,2-99, 
GJJC\,1, K\,1-16, NGC\,246, NGC\,1360, NGC\,2371-2, NGC\,3132, and 
NGC\,3587), whereas two (NGC\,7009 and NGC\,7094) are hard X-ray 
emitters \citep[][]{Kastner_etal12}, although their \emph{FUSE} 
spectra are not adequate to search for variability.  

%
%

The correlation between hard X-ray CSPNe emission and wind 
variability is not conclusive.  
A further clue on a possible shock wind origin of the hard X-ray 
emission from CSPNe may be provided by the comparison between 
the IP of a line with P~Cygni profile and its CSPN's $T_{\rm eff}$ 
(Figure~\ref{teff_line}).  
The CSPNe with the highest $T_{\rm eff}$ show P~Cygni 
profiles of species of the highest IP, such as Ne~{\sc vii}, while CSPNe 
of lower $T_{\rm eff}$ display only P~Cygni profiles of species of lower 
IP.  
This is the behavior expected for a photoionization origin of the lines.  
However, Figure~\ref{teff_line} also shows that the O~{\sc vi} 
line, while having a high IP, is found in stars with a wide range 
of effective temperatures (30,000~K$<T_{\rm eff}<$170,000~K).  
Interestingly, five out of the six CSPNe with $T_{\rm eff}<$45,000~K that 
show this line have variable P~Cygni profiles\footnote{
The sixth CSPN being Tc\,1, for which the available \emph{FUSE} 
observations may be insensitive to variability in the P~Cygni 
profiles of C~{\sc iii}, O~{\sc vi}, and S~{\sc iv} present in 
its spectrum due to low S/N ratio.  
}.  
This suggests that CSPNe with line variability tend to show lines 
with IP higher than expected for their $T_{\rm eff}$, implying that 
an additional source of ionization may be present.  
Certainly, Auger ionizations from X-ray photons produced in their 
stellar winds would naturally explain the additional ionization, 
as it occurs in OB stars \citep{CO79}.


We would also like to point that two of the three sources with variable 
winds but no hard X-ray emission, namely BD+30$^\circ$3639 and NGC\,40, 
are stars of the [WC] type.  
If X-ray are self-absorbed in thick winds, then these two objects are 
the most prone to be hard X-ray faint.  
Note also that stars of the [WC] type neither display variable O~{\sc vi} 
which may be associated to the production of X-ray-emitting hot gas.  
There are also two more objects, Hen\,2-99 and Tc\,1, that have very 
similar stellar properties to sources with wind variability, however 
they show neither wind variability nor P~Cygni profiles of the O~{\sc vi} 
lines.  


\section{Summary}

We have used the archive of \emph{FUSE} far-UV spectroscopic observations of 
CSPNe to search for the occurrence of P~Cygni profiles of high-excitation 
lines and to investigate their variability.  
For all spectral lines except O~{\sc vi}, there is a clear correlation 
between their occurrence, the IP of the ion, and the stellar effective 
temperature: the hottest CSPNe have only lines of the highest IP ions, 
while the coolest CSPNe have lines of low IP ions.  
This clearly implies the photoionization origin of the P~~Cygni profiles 
of these stars.  
Indeed, the detection of P~Cygni profiles of S~{\sc vi} and Ne~{\sc vii} in 
CSPNe with effective temperatures above 40,000~K and 100,000~K, respectively, 
shows that, contrary to OB stars, hot CSPNe can photoionize ``super-ions''.
On the other hand, the origin of the P~Cygni profiles of the O~{\sc vi} 
``super-ion'' in the spectra of CSPNe is ambiguous, as profiles of this 
line are found in CSPNe with high effective temperatures, but also in 
CSPNe too cold to photoionize it.

We have built a \emph{FUSE} atlas of P Cygni profiles of CSPNe.  
The P Cygni profiles of lines of the same CSPN may present very different 
shapes, making it difficult to estimate the terminal velocity of the wind.  
We have used the black velocity of saturated lines to estimate 
$v_\infty$, 
whereas for unsaturated P~Cygni profiles we introduce the grey velocity, 
$v_{\rm grey}$, the velocity of the blueward edge of the absorption region 
that has the lowest intensity, as a new indicator of $v_\infty$.  
A comparison with \emph{IUE} measurements available in the literature 
reveals that terminal velocities derived from \emph{IUE} data are 
usually overestimated as they rely on $v_{\rm edge}$.

P~Cygni profile variability in far-UV lines is reported for the first time 
for six CSPNe, namely Hen\,2-131, NGC\,40, NGC\,1535, NGC\,2392, Sp\,3, 
and SwSt\,1, adding them to a sample of another six CSPNe previously 
reported in \emph{FUSE} observations 
\citep[Hen\,2-138, IC\,418, IC\,2149, IC\,4593, NGC\,6543, and NGC\,6826, 
][]{Prinja_etal07,Prinja_etal10,Prinja_etal12}.  
Different types of variability are detected such as ripples and narrow 
features moving bluewards even at velocities above $v_\infty$, changes 
in the emission region of the profile and changes in the slope of the 
profile.  
Adding those CSPNe reported to have variable P~Cygni profiles in \emph{IUE} 
observations, the number of CSPNe with these characteristics amounts to 13.

Although there is not a tight correlation between the samples of hard 
X-ray CSPNe and those with variable stellar winds, we find evidence 
that shocks in the stellar winds of CSPNe produce additional ionization 
that can be associated to Auger ionizations from X-ray photons.  
In particular, all CSPNe with low effective temperature, $\lesssim$45,000 K, 
that show variability in P~Cygni profiles, present P~Cygni profile in the 
[O~{\sc vi}] line, even though the low effective temperature is not sufficient 
to photoionize this species.  
This strongly suggests a twofold origin of the P~Cygni profile in the 
[O~{\sc vi}] line, depending on the temperature of the CSPN and its 
variability.  
Therefore, CSPNe with variable P~Cygni profiles should be considered 
as prime candidates for future X-ray observations aimed at detecting 
hard X-ray emission.  


\begin{acknowledgements}
M.A.G. acknowledges support by grant AYA~2008-01934 of the Spanish 
Ministerio de Ciencia e Innovaci\'on (MICINN) cofunded by FEDER funds, 
and by a Distinguished Visitor grant of the Anglo-Australian Observatory 
(AAO).  
He also acknowledges with gratitude the hospitality of the Macquarie 
University where this paper was prepared while on sabbatical leave.  
Dr.\ Quentin A.\ Parker is kindly acknowledged for his support, as 
well as Dr.\ David Frew for his useful comments and suggestions. 
We appreciate the helpful comments of an anonymous referee which 
help improve this manuscript.

All of the data presented in this paper were obtained from the Mikulski 
Archive for Space Telescopes (MAST).  
MAST is located at the Space Telescope Science Institute (STScI) which 
is operated by the Association of Universities for Research in Astronomy, 
Inc., under NASA contract NAS5-26555. 
Support for MAST for non-HST data is provided by the NASA Office of Space 
Science via grant NNX09AF08G and by other grants and contracts.

\end{acknowledgements}

\end{document}